\documentclass[]{aa}
\usepackage{graphicx,deluxetable}
\usepackage{color}
\usepackage{natbib}
\begin{document}
\title{The $\beta$ Pictoris association:\\ Catalog\ of photometric rotational periods of low-mass members and candidate members\thanks{Tables 1--2 are also available in electronic form
at the CDS via anonymous ftp to cdsarc.u-strasbg.fr (130.79.128.5)
or via http://cdsweb.u-strasbg.fr/cgi-bin/qcat?J/A+A/. Table 3 is available only in electronic form. Figures A1--A73 are available as online material.}}
\author{S.\,Messina\inst{1}, M.\,Millward\inst{2}, A.\,Buccino\inst{3,4}\fnmsep\thanks{Visiting Astronomer, Complejo Astronomico El Leoncito operated under agreement between the Consejo Nacional de
Investigaciones Cientificas y Tecnicas de la Republica Argentina and the National Universities of La Plata, Cordoba and
                San Juan.}, L.\,Zhang\inst{5}, B.J.Medhi\inst{6},  E.\,Jofr\'e\inst{7,8}, R.\,Petrucci\inst{7,8}, Q.\,Pi\inst{5}, F.-J.\,Hambsch\inst{9,10},  P.\,Kehusmaa\inst{11}, C.\,Harlingten\inst{11}, S.\,Artemenko\inst{12}, I.\,Curtis\inst{13}, V.-P.\,Hentunen\inst{14}, L.\,Malo\inst{15}\fnmsep\thanks{Based on observations obtained at the Canada-France-Hawaii Telescope (CFHT) which is operated by the National Research Council of Canada, the Institut National des Sciences de l'Univers of the Centre National de la Recherche Scientique of France, and the University of Hawaii.}, P.\,Mauas\inst{3,4}, B.\,Monard\inst{16},  M. Muro Serrano\inst{17}, R.\,Naves\inst{18}, R.\,Santallo\inst{19},  A.\,Savuskin\inst{12}, T.G. Tan\inst{20}
}
\offprints{Sergio Messina}
\institute{INAF-Catania Astrophysical Observatory, via S.Sofia, 78 I-95123 Catania, Italy \\
\email{sergio.messina@oact.inaf.it}
\and   
York Creek Observatory, Georgetown, Tasmania, Australia \\
\email{mervyn.millward@yorkcreek.net}
\and
Instituto de Astronom\'ia y F\'isica del Espacio (IAFE-CONICET), Buenos Aires, Argentina\\
\email{abuccino@iafe.uba.ar; pablo@iafe.uba.ar}
\and
Departamento de F\'\i sica. FCEN-Universidad de Buenos Aires, Buenos Aires, Argentina
\and
Department of Physics, College of Science, Guizhou University, Guiyang 550025, P.R. China\\
\email{liy\_zhang@hotmail.com; piqingfeng@126.com}
\and
Aryabhatta Research Institute of Observational Sciences, Manora Peak, Nainital 263129, India\\
\email{biman@aries.res.in}
\and
Observatorio Astr\'onomico de C\'ordoba, Laprida 854, X5000BGR, C\'ordoba, Argentina\\
\email{jofre.emiliano@gmail.com; romina@oac.unc.edu.ar}
\and
Consejo National de Investigaciones Cient\'ificas y T\'ecnicas (CONICET),  Argentina
\and
Remote Observatory Atacama Desert (ROAD), Vereniging Voor Sterrenkunde (VVS), Oude Bleken 12, B-2400 Mol, Belgium\\
\email{hambsch@telenet.be}
\and
American Association of Variable Star Observers (AAVSO), Cambridge, MA, USA
\and
Harlingten Atacama Observatory, San Pedro de Atacama, Chile.\\
\email{petri@kehusmaa-astro.com}
\and
Research Institute Crimean Astrophysical Observatory, 298409, Nauchny, Crimea\\
\email{svetaartemenko@rambler.ru}
\and
 IAU Minor Planet Center code D79, 2 Yandra Street, Vale Park,
South Australia 5081
Australia \\
\email{Ivan.Curtis@keyworks.com.au}
\and
Taurus Hill Observatory,  Varkaus, Finland\\
\email{veli-pekka.hentunen@kassiopeia.net}
\and
Canada-France-Hawaii Telescope, 65-1238 Mamalahoa Hwy, Kamuela, HI 96743, USA\\
\email{malo@cfht.hawaii.edu}
\and
Klein Karoo Observatory, Western Cape, South Africa\\
\email{bmonard@mweb.co.za}
\and
Zeta UMa Observatory, Madrid, Spain\\
\email{mizarmms@hotmail.com}
\and
Montcabrer Observatory, C/Jaume Balmes, 24, Cabrils, Spain\\
\email{ramonnavesnogues@gmail.com}
\and
Southern Stars Observatory, Pamatai, Tahiti, French Polynesia\\
\email{santallo@southernstars-observatory.org}
\and
Perth Exoplanet Survey Telescope, Western Australia, Australia\\
\email{tgtan@bigpond.net.au}
}
\date{}
\titlerunning{Rotation among $\beta$ Pictoris members}
\authorrunning{S.\,Messina et al.}
\abstract {}
{We intended to compile the most complete catalog\ of bona fide members and candidate members of the $\beta$ Pictoris association, and to measure their rotation periods and basic properties from our own observations, public archives, and exploring the literature. } {We carried out a multi-observatories campaign to get our own photometric time series and collected all archived public photometric data time series for the stars in our catalog. Each time series was analyzed with the Lomb-Scargle and CLEAN periodograms to search for the stellar rotation periods. We complemented the measured rotational properties with detailed information on multiplicity, membership, and projected rotational velocity available in the literature and discussed star by star.} {We measured the rotation periods of 112 out of 117 among bona fide members and candidate members of the $\beta$ Pictoris association and, whenever possible, we also measured the luminosity, radius, and inclination of the stellar rotation axis. This represents to date the largest catalog of rotation periods of any young loose stellar association.} {We provided an extensive catalog of rotation periods together with other relevant basic properties useful to explore a number of open issues, such as the causes of spread of rotation periods among coeval stars, evolution of angular momentum, and lithium-rotation connection. 
}
\keywords{Stars: activity - Stars: late-type - Stars: rotation - 
Stars: starspots - Stars: open clusters and associations: individual:   \object{beta Pictoris} association}
\maketitle
\rm
\section{Introduction}
\object{$\beta$ Pictoris} is a nearby young loose stellar association. Its members have distances from the Sun in the range 10--80\,pc with an average value of about 40$\pm$17\,pc, and an age of about 25$\pm$3\,Myr (\citealt{Messina16a}; Paper II). 
Youth and proximity make this association a special  benchmark in stellar astrophysics studies.  Many studies have payed attention to this association (see, e.g., \citealt{Torres06}, \citeyear{Torres08}; \citealt{Lepine09}; \citealt{Kiss11}; \citealt{Schlieder10}, \citeyear{Schlieder12}; \citealt{Shkolnik12}; \citealt{Malo13}, \citeyear{Malo14a}, \citeyear{Malo14b}), thereby providing a significantly increased number of bona fide and new candidate members by about a factor 3 with respect to the association members detected in the discovery studies (e.g., \citealt{Zuckerman01}).\\
In particular, the stellar rotation period is a key basic parameter. Its knowledge, supplemented with information about other basic stellar properties, allows the exploration of a number of open issues related to the \object{$\beta$ Pictoris association}. Amongst others,  the impact of rotation on the lithium depletion (\citealt{Messina16a}; Paper II) and the origin of the rotation period spread among stars of similar mass and age (Messina et al. in preparation; Paper III). Moreover, the angular momentum evolution of young low-mass stars, the lifetime of primordial disks  and the timescale  of the star-disk locking phase can all be probed by means of the rotation of the star (see, e.g., \citealt{Messina14}, \citeyear{Messina15a}, \citeyear{Messina15b}). The rotation period of individual members is also relevant information in the radial velocity search for planetary companions. In fact, it helps to disentangle apparent radial velocity (RV) variations induced by the stellar magnetic activity from Doppler RV variations of Keplerian nature and to remove the activity noise in the RV time series.\\
The first comprehensive rotational investigation of the low-mass (spectral types F to M) members of  the $\beta$ Pictoris association was carried out by Messina et al. (\citeyear{Messina10}, \citeyear{Messina11}). They were able to measure 33 rotation periods on a total of 38 stars, among members and candidate members compiled from  \citet{Zuckerman01}, \citet{Song03}, \citet{Zuckerman04}, and Torres et al. (\citeyear{Torres06}, \citeyear{Torres08}).\\
Given the relevance of the rotation periods in several astrophysical open issues, in this work (Paper I) we have compiled the most complete catalog of low-mass members and candidate members of this association known to date and engaged a project to measure the rotation periods for all the stars in this catalog. \\
For the present work, we carried out a multi-observatories campaign dedicated to measure the rotation periods of 117 low-mass stars that are either bona fide members or candidate members of the $\beta$ Pictoris association. In Sect.\,2,  we present the  bibliographic sources to select our stellar sample. In Sect.\,3, we describe the main characteristics of the 13 observatories where we obtained our own observations. In Sect.\,4, we describe the method to perform the data reduction of these observations, and in Sect.\,5, we perform the periodogram analyses for all the time series. In Sect.\,6, we mention the method used to derive the basic stellar parameters, and in Sect.\,7 we detail the whole catalog. Finally in the Appendix, we present our results and the main characteristics mentioned in the literature for each star individually.

\section{Sample description}
We have carried out an extensive search in the literature to retrieve all members and candidate members of the $\beta$ Pictoris association.
We compiled a list of 117 stars, with spectral types later than about F3V, from the following major studies: Torres et al.  (\citeyear{Torres06}, \citeyear{Torres08}), \citet{Lepine09}; \citet{Kiss11}; Schlieder et al. (\citeyear{Schlieder10}, \citeyear{Schlieder12}); \citet{Shkolnik12}; Malo et al. (\citeyear{Malo13}, \citeyear{Malo14a}, \citeyear{Malo14b}), and others that are listed in the Appendix dedicated to the individual targets.
 Stars of earlier spectral types were excluded from our sample since the photometric rotation period requires the presence of magnetic activity (more specifically, of light rotational modulation by starspots) to be measured, which is possible in later spectral type stars with sufficiently deep convection zones.\\
We  measured the rotation periods of 112 out of 117 stars either from our own photometric monitoring or from photometric time series in public archives. For a few stars we adopted the rotation periods available in the  literature.\\
Information on the membership is not homogeneous for all the targets  either for the number of studies or for the 
methods used to determine this membership. For example, we found more than four membership studies for 52 targets, whereas we found only one membership study for 52 targets. In the present catalog we included all those stars that were considered members or candidate members in one membership study, at least. In a forthcoming paper, the rotation periods presented in this catalog, together with information on RV, proper motion, distance, and activity
indicators, will be used to assess the membership of each target.
\rm

\section{Observations}
The photometric rotation periods presented in this catalog were inferred from the analysis of new observations carried out by us and from photometric time series available in public archives.

\subsection{New photometry}

Our own photometric observations were carried out at 13 different observatories: seven located at southern latitudes and six at northern
latitudes.  In the following paragraphs, we give relevant information for each of them.

\subsubsection{Harlingten Atacama Observatory (HAO)}
The Harlingten Atacama Observatory is located in the Atacama Desert close to the town of San Pedro de Atacama, Chile ($-$22$^{\circ}$57$^{\prime}$10$^{\prime\prime}$;  68$^{\circ}$10$^{\prime}$49$^{\prime\prime}$W; 2450\,m a.s.l.). It is operated by P.\,Kehusmaa. The telescope is a 51 cm f/6.85 PlaneWave CDK with a  28$^{\prime}\times28^{\prime}$ field of view (FoV) and a plate scale of 0.83$^{\prime\prime}$/pixel. It is equipped with an Apogee Alta U42 CCD and Johnson/Cousins V, Rc filters.

\subsubsection{Yandra Street Vale Park Observatory (YSVP)}
The YSVP Observatory is located close to Vale Park, South Australia, Australia ($-$34$^{\circ}$53$^{\prime}$04$^{\prime\prime}$; 138$^{\circ}$37$^{\prime}$51$^{\prime\prime}$E; 44\,m a.s.l.). It is operated by I.\,Curtis. The telescope is a 23.5 cm f/10 (f6/3 with focal reducer) Schmidt-Cassegrain on a German Equatorial
mount with a 16.6$^{\prime}\times12.3^{\prime}$ FoV  and a plate scale of 0.6$^{\prime\prime}$/pixel. It is equipped with a cooled  1620$\times$1220 pixels Atik-320E CCD  with 4.4 $\mu$m pixel size,  Johnson UBVRI filters and the CBB blue-blocking filter for exoplanet observations.

\subsubsection{York Creek Observatory (YCO)}
The York Creek Observatory is located close to Launceston, Tasmania, Australia ($-$41$^{\circ}$06$^{\prime}$06$^{\prime\prime}$; 146$^{\circ}$50$^{\prime}$33$^{\prime\prime}$E; 28\,m a.s.l). It is operated by M.\,Millward. The telescope is a 25 cm f/10 Takahashi Mewlon reflector, equipped with a QSI 683ws-8 camera, and BVR standard Johnson-Cousins filters. The telescope has a 24.5$^{\prime}$$\times$18.5$^{\prime}$  FoV, and a plate scale of 0.44$^{\prime\prime}$/pixel. 

\subsubsection{Klein Karoo Observatory (KKO)}
The Klein Karoo Observatory  is located close to Klein Karoo, Western Cape, South Africa ($-$33$^{\circ}$45$^{\prime}$00$^{\prime\prime}$; 21$^{\circ}$00$^{\prime}$00$^{\prime\prime}$E). It is operated by B.\,Monard. The telescope is a  30 cm f/8 RCX-400 with a 21$\arcmin$$\times$14$\arcmin$ FoV, and a plate scale of 0.82$^{\prime\prime}$/pixel. It is equipped with a SBIG ST8-XME CCD camera and BV(RI)$_c$ filters.

\subsubsection{Montcabrer Observatory (MO)}
The Montcabrer Observatory is close to  Montcabrer,  Spain ($+$$41^{\circ}31^{\prime}11^{\prime\prime}$; $02^{\circ}23^{\prime}39^{\prime\prime}$E, 100\,m a.s.l.). It is operated by R.\,Naves.  The telescope is a 30 cm f/6 Meade LX-200  with a 26.3$^{\prime}\times17.5^{\prime}$ FoV and a plate scale of 1.03$^{\prime\prime}$/pixel. It is equipped with a SBIG ST-8 CCD camera, an AO-8T active optics unit, and Bessel BVRI filters.

\subsubsection{Crimean Astronomical Observatory (CrAO)}
The Crimean Astronomical Observatory  is located nearby  Nauchny, Crimea ($+$44$^{\circ}43^{\prime}37^{\prime\prime}$; 34$^{\circ}01^{\prime}02^{\prime\prime}$E; 600\,m a.s.l.). Photometric observations were performed by S.\,Artemenko and A.\,Savuskin. In this study we used the  0.5 m \rm Maksutov telescope with a  12.2$^{\prime}$$\times$12.2$^{\prime}$ FoV and a plate scale of 0.71$^{\prime\prime}$/pixel. It is equipped with a 1024$\times$1024 pixel  Apogee Alta U6 CCD camera and V, R Johnson filters. We also used the 1.25 m Ritchey-Chr\'etien telescope with 10.9$^{\prime}$$\times$10.9$^{\prime}$ FoV and a plate scale of  0.32$^{\prime\prime}$/pixel. It is equipped with a 2048$\times$2048 pixel FLI ProLine PL230 CCD camera and the same filters as the other telescope. 

\subsubsection{Xinglong station Observatory}
The Xinglong station Observatory is close to Xinglong, Yanshan, China ($+$40$^{\circ}$23$^{\prime}$39$^{\prime\prime}$; 117$^{\circ}$34$^{\prime}$30$^{\prime\prime}$E, 960\,m a.s.l.) and is a facility of  the National Astronomical Observatory of China. Photometric observations were collected by L.\,Zhang and Q.\,Pi. The telescope is  a 80 cm f/3.27 with a 16.5$^{\prime}$$\times$16.5$^{\prime}$ FoV  and a plate scale of 0.96$^{\prime\prime}$/pixel. It is equipped with a 1024$\times$1024 pixel  Marconi cooled CCD47-20 camera (13$\mu$m pixel size),  and standard Johnson-Cousin-Bessel  BVRI filters.

\subsubsection{ Aryabhatta Research Institute of Observational Sciences (ARIES)}
The Aryabhatta Research Institute of Observational Sciences is located close to Manora Peak,  Nainital, India  ($+$$29^{\circ}22^{\prime}49^{\prime\prime}$; $79^{\circ}27^{\prime}47^{\prime\prime}$E). Observations were collected by B.\,J.\,Medhi. The telescope  is a 104 cm f/13  Cassegrain with a 13$^{\prime}\times13^{\prime}$ FoV and a 
plate scale of 0.366$^{\prime\prime}$/pixel. It is equipped with a 2K$\times$2K CCD camera.

\subsubsection{Complejo Astron\'omico El Leoncito (CASLEO)}
The   Complejo Astron\'omico El Leoncito  Observatory is located close to San Juan, Argentina ($-$31$^{\circ}$47$^{\prime}$57$^{\prime\prime}$; 69$^{\circ}$18$^{\prime}$ 12$^{\prime\prime}$W; 2552\,m a.s.l.). Photometric observations were performed by A.\,Buccino, R.\,Petrucci, and E.\,Jofr\'e. For this study we used the Horacio Ghielmetti Telescope (THG),
which is a 40 cm f/8 remotely operated MEADE-RCX 400 Ritchey-Chretien telescope with a 49$^{\prime}\times$49$^{\prime}$ FoV and a plate scale of 0.57$^{\prime\prime}$/pixel. It is equipped with a $4096 \times
4096$ pixel Apogee Alta U16M camera (9 $\mu$m pixel size) and Johnson UBVRI and Clear filters (\citealt{Petrucci13}).

\subsubsection{Remote Observatory Atacama Desert Observatory (ROAD)  }
The Remote Observatory Atacama Desert is located in the Atacama Desert 
close to the town of San Pedro de Atacama, Chile ($-$22$^{\circ}$57$^{\prime}$10$^{\prime\prime}$;  68$^{\circ}$10$^{\prime}$49$^{\prime\prime}$W; 2450\,m a.s.l.). It is operated by F.-J.\,Hambsch. The telescope is housed at SPACE  (San  Pedro  de  Atacama  Celestial  Exploration\footnote{http://www.spaceobs.
com/index.html}).  The telescope is a 40 cm f/6.8 Optimized Dall-Kirkham (ODK) from Orion Optics. It is equipped with a 4K$\times$4K pixel FLI ML16803 CCD camera (9 $\mu$m pixel size) with  a  40$^{\prime}\times$40$^{\prime}$ FoV and BVI and Clear filters and a SA200 grating for low-resolution spectroscopy.

\subsubsection{Zeta UMa Observatory}
The Zeta UMa Observatory is located close to Madrid, Spain ($+$40$^{\circ}$25$^{\prime}$00$^{\prime\prime}$; 03$^{\circ}$42$^{\prime}$13$^{\prime\prime}$W; 709\,m a.s.l.). The telescope is operated by M. Muro Serrano. The telescope is a 13 cm f/5.7 Takahashi refractor
with a 80$^\prime$ $\times$ 60$^\prime$  FoV and a plate scale of 1.50$^{\prime\prime}$/pixel. It is equipped with 
a cooled QHY9 camera and  a set of  Johnson-Cousins V, R, and I  filters. 

\subsubsection{Taurus Hill Observatory}
The Taurus Hill Observatory is located close to Varkaus, Finland ($+$62$^{\circ}$18$^{\prime}$54$^{\prime\prime}$; 
28$^{\circ}$23$^{\prime}$21$^{\prime\prime}$E, 160\,m a.s.l.). The telescope is operated by V.-P. Hentunen.
The telescope is a  35 cm f/11 SC Celestron on a Paramount ME German equatorial mount with 24$^\prime$ $\times$ 16$^\prime$  FoV  and a plate scale of 0.95$^{\prime\prime}$/pixel. It is equipped with a 1530$\times$1020 pixel SBIG ST-8XME KAF-1603 CCD camera (9\,$\mu$m pixel size) and  
Johnson-Bessell BVR filters. 

\subsubsection{ Perth Exoplanet Survey Telescope  Observatory (PEST)}

The Perth Exoplanet Survey Telescope is located close to Perth, Australia ($-$31$^{\circ}$58$^{\prime}$; 
115$^{\circ}$47$^{\prime}$E; 24\,m a.s.l.). It is operated by T.-G. Tan. The telescope is a 30 cm f/10 Schmidt-Cassegrain (f/5 with focal reducer) with 31$^\prime$ $\times$ 21$^\prime$  FoV and a plate scale of 1.2$^{\prime\prime}$/pixel. It is equipped with a 1550$\times$1050 pixel SBIG ST-8XME CCD camera with a filter wheel loaded with BV(RI)$_c$ and Clear filters. Focusing is computer controlled with an Optec TCF-Si focuser.

\subsection{Archival data}
A large number of members and candidate members of the $\beta$ Pictoris association has photometric time series in one or more of the following public archives:
ASAS (All Sky Automated Survey; \citealt{Pojmanski97}), SuperWASP (Wide Angle Search for Planets; \citealt{Butters10}), INTEGRAL/OMC (\citealt{Domingo10}), Hipparcos (ESA 1997), NSVS (Northern Sky Variability Survey; \citealt{Wozniak04}), MEarth \citep{Berta12}, and CSS (Catalina Sky Survey; \citealt{Drake09}). We have retrieved and analyzed  all these available time series for the period search. In the Appendix, we indicate which archives were used for each star individually.\\

\section{Data reduction}
Although the new observations  we have collected come from different telescopes and instruments (see Sect.\,3.1), we adopted similar reduction procedures. Briefly, we used the tasks within IRAF\footnote{IRAF is distributed by the National Optical Astronomy Observatory, which is operated by the Association of the Universities for Research in Astronomy, inc. (AURA) under cooperative agreement with the National Science Foundation.}  for bias
correction and flat fielding. Then,  we used the technique of aperture photometry to extract magnitude time series for the targets and for other nearby stars detected in the frames to search for suitable comparison stars.
For a few targets, differential magnitude time series were computed with respect to one comparison star. However, whenever it was possible we preferred to compute differential values 
with respect to an ensemble comparison star. Generally, on each telescope pointing we collected five consecutive frames per filter. The corresponding differential magnitudes were subsequently averaged to get one average magnitude and its standard deviation, which is considered the measure of the photometric precision we obtained. After visual inspection of the light curves to identify possible flare events, we applied a 3$\sigma$ clipping to remove possible outliers.\\

\section{Period search methods}
\begin{figure*}
\begin{minipage}{20cm}
\includegraphics[scale = 0.5, trim = 0 0 0 0, clip, angle=90]{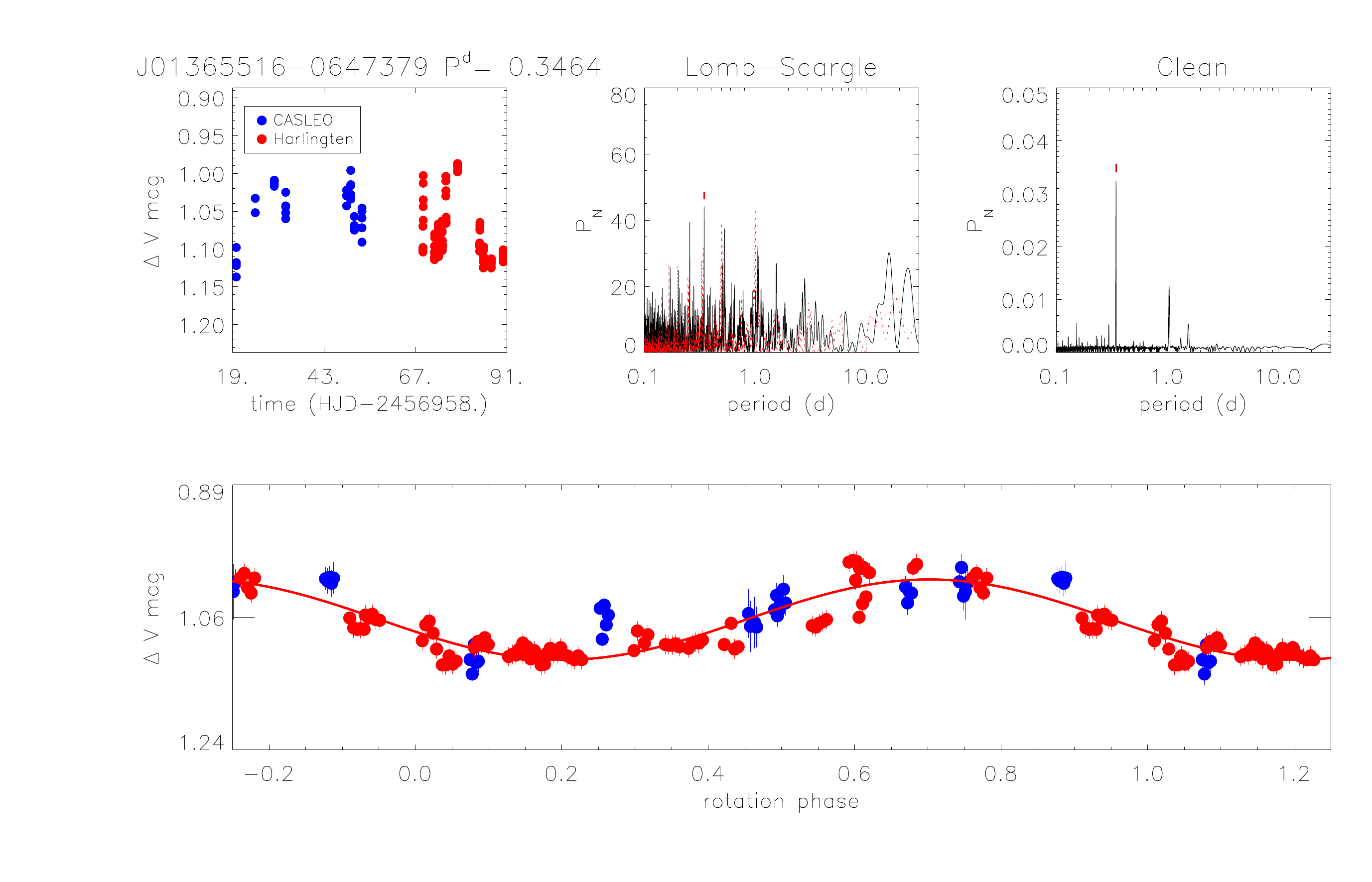}
\end{minipage}
\caption{\label{new} Results of periodogram analysis of 2MASS\, J01365516-0647379. In the top left panel we plot magnitudes vs. Heliocentric Julian Day. Different colors are used to distinguish data collected at CASLEO from data collected at HAO. In the top middle panel we plot the Lomb-Scargle periodogram with the spectral window function and power level corresponding to FAP = 1\% (horizontal dashed line) overplotted (red dotted line), and we indicate the peak corresponding to the rotation period. In the top right panel we plot the CLEAN periodogram. In the bottom panel we plot the light curve phased with the rotation period. The solid line represents the sinusoidal fit.}
\end{figure*}

\begin{figure*}
\begin{minipage}{20cm}
\includegraphics[scale = 0.5, trim = 0 0 0 0, clip, angle=90]{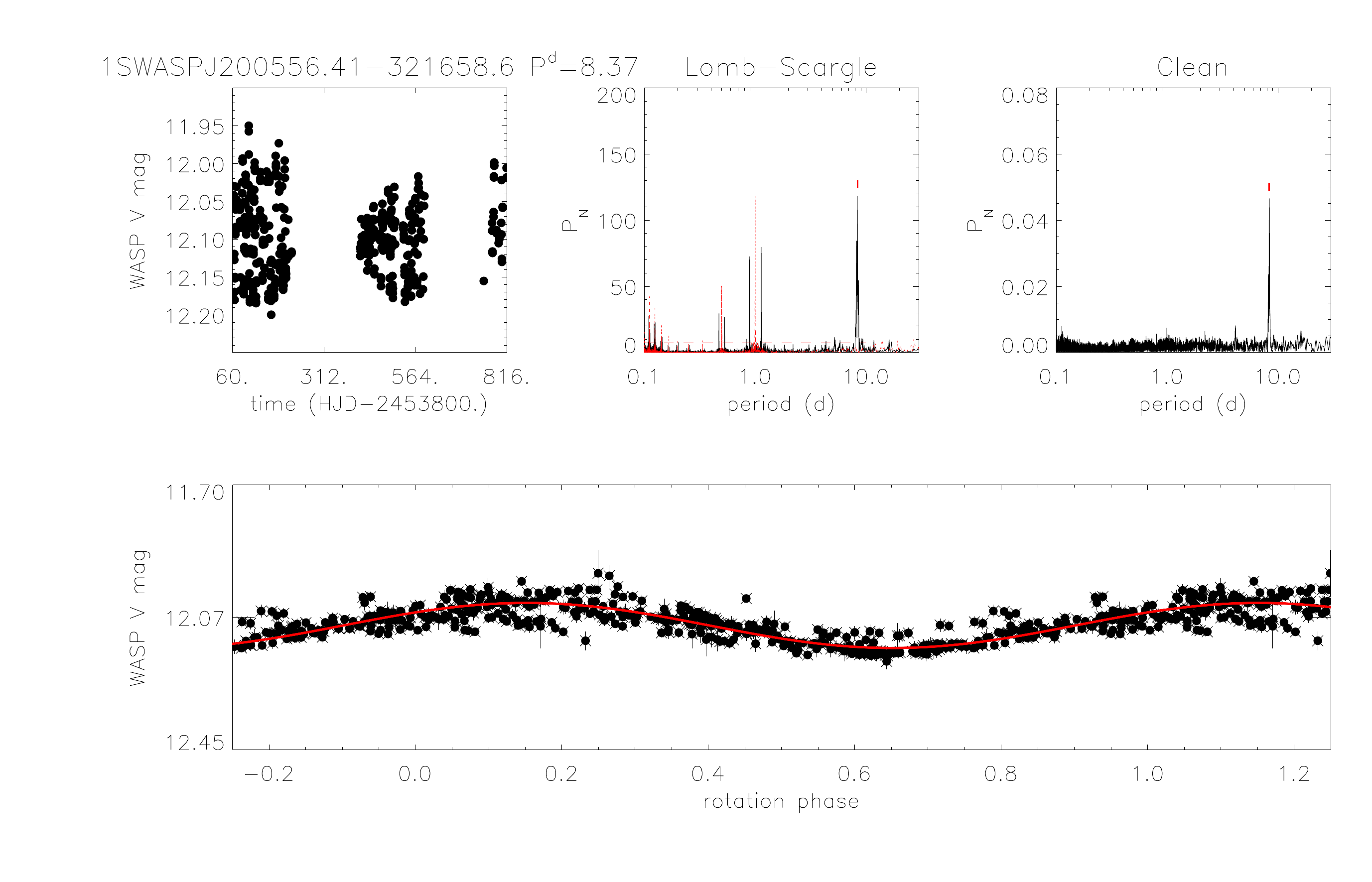}
\end{minipage}
\caption{\label{literature} Same as in Fig.\,\ref{new}, but for the SuperWASP data of 2MASS\,J20055640-3216591. }
\end{figure*}

To search for the stellar rotation periods of our targets, we followed an approach similar to that used in Messina et al. (\citeyear{Messina10}, \citeyear{Messina11}). Those papers provide a detailed description of the method.\\
Briefly, the period search was carried out by computing the Lomb$-$Scargle periodogram (LS; \citealt{Press02}; \citealt{Scargle82}; \citealt{Horne86}) and the  CLEAN periodogram \citep{Roberts87}.
The false  alarm probability (FAP) associated with our detected period, which is  the probability that a
peak of given height in the periodogram is caused simply by statistical variations, i.e., 
Gaussian noise, was computed through Monte Carlo simulations,  i.e., by generating 1000 artificial light curves obtained from the real light curve, keeping the date but permuting the magnitude values  (see, e.g., \citealt{Herbst02}). \rm For our following analysis, we considered only rotation periods that were measured with a confidence  level larger 
than 99\% (FAP$<0.01$).  \\
When data from more observation seasons were available, we computed LS and CLEAN periodograms for the complete series and for each season. The detection of the same periodicity in more (hopefully all) time sections further supports that the recurrent periodicity is the rotation period.  Three targets could be observed at more observatories. In these cases, we performed the periodogram analysis on each time series and on their combination. In the latter case, the shorter time series was aligned by applying a magnitude shift to the mean magnitude of the longer time series whenever necessary. This is a reasonable procedure since observations at different sites of the same target were performed almost contemporarily, ruling out effects from intrinsic long-term variations of the mean magnitude. \rm
 We followed the  method used by \citet{Lamm04} to compute the errors associated with the period determinations (see, e.g., Messina et al. \citeyear{Messina10} for details). 
 To derive the light curve amplitude, we fit the data with a sinusoid function whose period is equal to the stellar rotation period.  \\
  As result of our photometric analysis, we obtained the rotation periods of 112 out of 117 target stars. Specifically,
we collected new photometric time series for 31 targets. We measured 51 new rotation periods: 16  from our own photometry, 21 
from archived data, and 14 from both our own and archived data. We confirmed 28 previously known rotation periods: 6 using 
our own photometry and 22 using archived data. Finally, we adopted the rotation periods retrieved from the literature for 33 targets 
(of which 17 periods where retrieved from Messina et al. \citeyear{Messina10}, \citeyear{Messina11}).
We did not obtain the rotation period neither from our periodogram analysis nor from the literature  for the remaining 5 targets.\\
\rm
In Fig.\,\ref{new} and Fig.\,\ref{literature}, we give two examples that summarize the results of our period search carried out on our own photometric time series (the case of 2MASS\, J01365516-0647379) and on archive time series (the case of 2MASS\,J20055640-3216591).\\
We produced similar plots for all the photometric time series (either new or from archives) analyzed in this work and used to measure the rotation period. These plots are available as online material in Fig.\,A1--A73.

\section{Basic stellar parameters}
We use the V magnitudes and distances listed in Table 1, and the bolometric corrections for young stars tabulated by \citet{Pecaut13} to infer the absolute bolometric magnitudes that are transformed into luminosities. From the luminosities and effective temperatures, which are inferred from V$-$K$_s$ colors using the Teff - (V$-$K$_s$) relations from \citet{Pecaut13}, we derive the stellar radii. Uncertainties on luminosity and radius are computed according to the error propagation. Finally, we use the stellar radii and projected rotational velocities listed in Table 1 to infer the $\sin{i}$ values and hence the inclination of the stellar rotation axis.\\
These quantities are give in the Appendix and computed only for single stars and wide (resolved) components of multiple systems for which the V magnitude and the V$-$K$_s$ color are accurately measured. To minimize the impact of variability owing to magnetic activity, we use the brightest (presumably unspotted) V magnitude  ever observed.

\section{Catalog description}
Our catalog is presented in three tables. Table 1 lists  the target name,  coordinates, V magnitude, B$-$V, V$-$I, and V$-$K$_s$ colors,  distance to the Sun,  projected rotational velocity, spectral type classification, and separation between the components of multiple systems. The references  are given in the Appendix  in the 
discussion on each individual target.  Table 2 lists the target name, photometric rotation period and its uncertainty,   light curve amplitude,  photometric precision,  flag on the multiplicity, another designation name for the target,   the source of the photometric time series used to derive the mentioned period, the photometric filter, and the starting and ending Julian Day of the time series. \rm S,  B, \rm and T indicate a single star,  component of a binary system, \rm and a component of a triple system, respectively. The letter w and c indicate that the target is component of a  binary/multiple system \rm in a close orbit (projected separation between the components $<$ 60 \,AU) or in a wide orbit ($\ge$ 60 \,AU), according to Paper III (Messina et al. in preparation). The letter D indicates the presence of a debris disk. 
The question mark '?' indicates those targets whose single/binary nature is not known yet. Table 3 contains the photometric time series either collected by us or retrieved from the mentioned public archives and used to measure the rotation periods.  \\

 {\it Acknowledgements}. Research on stellar activity at INAF- Catania Astrophysical Observatory is supported by MIUR  (Ministero dell'Istruzione, dell'Universit\`a e della Ricerca).  This research has made use of the Simbad database, operated at CDS (Strasbourg, France).  We thank the referee Dr. Alexander Scholz for helpful comments. 
 We
used data from the WASP public archive in this research. The WASP consortium
comprises of the University of Cambridge, Keele University, University of
Leicester, The Open University, The QueenÕs University Belfast, St. Andrews
University, and the Isaac Newton Group. Funding for WASP comes from the
consortium universities and from the UKÕs Science and Technology Facilities
Council. This paper makes use of data from the MEarth Project, which is a collaboration between Harvard University and the Smithsonian Astrophysical Observatory. The MEarth Project acknowledges funding from the David and Lucile Packard Fellowship for Science and Engineering and the National Science Foundation under grants AST-0807690, AST-1109468, AST-1616624 and AST-1004488 (Alan T. Waterman Award), and a grant from the John Templeton Foundation.
LZ acknowledges  the support by the Joint Research Fund in Astronomy (U1431114 and U1631236) under cooperative agreement between the National Natural Science Foundation of China (NSFC) and Chinese Academy of Sciences (CAS). 
 This work made use of LAMOST data (the Large Sky Area Multi-Object Fiber Spectroscopic Telescope LAMOST), which is a National Major Scientific Project built by the Chinese Academy of Sciences. \\
 
\clearpage
\newpage

 \begin{deluxetable}{lrrrr@{\hspace{0.05cm}}r@{\hspace{0.05cm}}r@{\hspace{0.05cm}}r@{\hspace{0.05cm}}r@{\hspace{0.05cm}}c@{\hspace{0.05cm}}r@{\hspace{0.05cm}}}
\tablecolumns{11}
\tablecaption{\label{tab_period} List of $\beta$ Pictoris members with coordinates, V magnitude and colors, distance, projected rotational velocity, spectral type and separation between the components for multiple systems. 2MASSJ targets are abbreviated as J.
}
\tablewidth{0pt}
 \tablehead{\colhead{Target} & \colhead{RA } & \colhead{DEC} & \colhead{V} & \colhead{B$-$V} & \colhead{V$-$I} & \colhead{V$-$K$_s$} & \colhead{d} & \colhead{$v\sin{i}$} & \colhead{Sp.T} & \colhead{sep} \\
\colhead{    } & \colhead{(J2000)} & \colhead{(J2000)} & \colhead{(mag)} & \colhead{(mag)} & \colhead{(mag)} & \colhead{(mag)} & \colhead{(pc)} & \colhead{(km\,s$^{-1}$)} & \colhead{  } & \colhead{(AU)}\\
}
\startdata
  \hline
                       HIP\,560 &   00 06 50.08 & -23  06 27.20 &    6.15 &    0.39 &    0.45 &    0.91 &   39.1 &  170.0 &                  F3V &     --- \\
             J00172353-6645124 &  00 17 23.54 & -66  45 12.50 &   12.35 &    1.48 &    2.41 &    4.65 &   37.5 &    6.3 &                M2.5V &     --- \\
                TYC\,1186\,0706\,1 &  00 23 34.66 &  20  14 28.75 &   10.96 &    1.40 &    1.75 &    3.62 &   59.7 &    4.5 &             K7.5V+M5 &    101.5 \\
                       GJ\,2006A &  00 27 50.23 & -32  33 06.42 &   12.87 &    1.50 &    2.66 &    4.86 &   32.3 &    6.2 &               M3.5Ve &    577.3 \\
                       GJ\,2006B &  00 27 50.35 & -32  33 23.86 &   13.16 &    1.50 &    2.77 &    5.04 &   32.3 &    4.2 &               M3.5Ve &    577.3 \\
             J00323480+0729271A &  00 32 34.81 &  07  29 27.10 &   13.40 &    1.53 &    2.78 &    5.02 &   41.1 &   17.0 &                   M4V &     32.5 \\
             J00323480+0729271B &  00 32 34.81 &  07  29 27.10 &   12.62 &    1.53 &    3.05 &    5.68 &   41.1 &   15.0 &                  $>$M5 &     32.5 \\
                TYC\,5853\,1318\,1 &  01 07 11.94 & -19  35 36.00 &   11.41 &    1.40 &    1.80 &    4.16 &   43.0 &    9.1 &                   M1V &     --- \\
            J01112542+1526214A &  01 11 25.42 &  15  26 21.50 &   14.46 &    1.65 &    3.35 &    6.25 &   21.8 &   17.9 &                M5V &      8.9 \\
            J01112542+1526214B &  01 11 25.42 &  15  26 21.50 &   14.46 &    1.70 &    3.63 &    6.55 &   21.8 &   17.9 &                M6V &      8.9 \\
             J01132817-3821024 &  01 13 28.17 & -38  21 02.50 &   11.77 &    1.43 &    2.08 &    4.17 &   29.0 &    9.1 &           (M0V+M3V)+M1V &      0.0 \\
             J01351393-0712517 &  01 35 13.93 &  -07  12 51.77 &   13.42 &    1.54 &    2.87 &    5.50 &   37.9 &   55.1 &                M4.5V &      0.0 \\
             J01365516-0647379 &  01 36 55.16 &  -06  47 37.92 &   14.00 &    1.53 &    2.78 &    5.14 &   24.0 &   10.0 &               M4V+L0 &    134.2 \\
                TYC\,1208\,0468\,1 &  01 37 39.42 &  18  35 32.91 &    9.83 &    1.31 &    1.45 &    3.11 &   60.0 &   21.5 &                K3V+K5V &    103.8 \\
             J01535076-1459503 &  01 53 50.77 & -14  59 50.30 &   11.97 &    1.50 &    2.62 &    4.90 &   28.0 &   11.2 &              M3V+M3V &     80.7 \\
               J02014677+0117161 &  02 01 46.78 &  01  17 16.20 &   12.78 &    1.46 &    2.28 &    4.51 &   63.7 & --- &                    M &     --- \\
               RBS\,269 &  02 01 46.93 &  01  17 06.00 &   12.72 &    1.58 &    2.60 &    4.46 &   63.7 & --- &                    M &     --- \\
             J02175601+1225266 &  02 17 56.01 &  12  25 26.70 &   13.62 &    1.45 &    2.26 &    4.54 &   67.9 &   22.6 &                M3.5V &     --- \\
                      HIP\,10679 &  02 17 24.74 &  28  44 30.43 &    7.75 &    0.62 &    0.69 &    1.49 &   37.6 &    7.8 &                  G2V &    519.0 \\
                      HIP\,10680 &  02 17 25.28 &  28  44 42.16 &    6.95 &    0.52 &    0.59 &    1.16 &   37.6 &   37.5 &                  F5V &    519.0 \\
 		    HIP\,11152  & 02 23 26.64  &  +22 44 06.75  &  11.09  &  1.44 &  1.65  & 3.74   & 28.7 & 6.0   &		  M3V      & --- \\
                     HIP\,11437A &  02 27 29.25 &  30  58 24.60 &   10.12 &    1.18 &    1.32 &    3.04 &   42.3 &    5.0 &                   K4V &    930.8 \\
                     HIP\,11437B &  02 27 28.05 &  30  58 40.53 &   12.44 &    1.50 &    2.15 &    4.22 &   42.3 &    4.7 &                   M1V &    930.8 \\
                      HIP\,12545 &  02 41 25.90 &  05  59 18.00 &   10.37 &    1.21 &    1.46 &    3.30 &   42.0 &    9.0 &                 K6Ve &     --- \\
             J03350208+2342356 &  03 35 02.09 &  23  42 35.61 &   17.00 &    2.13 &    4.60 &    5.74 &   23.5 &   30.0 &                M8.5V &     --- \\
               J03461399+1709176 &  03 46 14.00 &  17  09 17.45 &   12.90 &    1.43 &    2.00 &    4.08 &   64.0 & --- &                 M0.5 &     --- \\
                        GJ\,3305 &  04 37 37.30 &  -02  29 28.00 &   10.59 &    1.45 &    2.00 &    4.18 &   29.8 &    5.3 &                M1+M? &      2.7 \\
             J04435686+3723033 &  04 43 56.87 &  37  23 03.30 &   12.98 &    1.46 &    2.30 &    4.18 &   59.0 &   10.6 &             M3Ve+M5? &    531.1 \\
                      HIP\,23200 &  04 59 34.83 &  01  47 00.68 &   10.05 &    1.39 &    1.84 &    3.99 &   24.0 &    8.7 &               M0.5Ve &      0.0 \\
                TYC\,1281\,1672\,1 &  05 00 49.28 &  15  27 00.71 &   10.75 &    1.24 &    1.40 &    3.15 &   54.0 &   17.2 &                 K2IV &     --- \\
                      HIP\,23309 &  05 00 47.10 & -57  15 25.00 &   10.00 &    1.40 &    1.79 &    3.76 &   26.2 &    5.8 &                 M0Ve &     --- \\
             J05015665+0108429 &  05 01 56.65 &  01  08 42.91 &   13.20 &    1.10 &    1.24 &    5.52 &   27.0 &    8.0 &                   M4V &     --- \\
                     HIP\,23418A &  05 01 58.80 &  09  59 00.00 &   11.45 &    1.54 &    2.81 &    4.78 &   24.6 &    7.7 &                  M3V &      0.0 \\
                     HIP\,23418B &  05 01 58.80 &  09  59 00.00 &   12.45 &    1.47 &    2.48 &    5.23 &   24.6 &   23.8 &                 $>$M3V &     16.7 \\
                    BD-211074A &  05 06 49.90 & -21  35 09.00 &   10.29 &    1.51 &    2.16 &    4.35 &   19.2 &    4.5 &                M1.5V &    157.9 \\
                    BD-211074B &  05 06 49.90 & -21  35 09.00 &   11.67 &    1.50 &    2.58 &    4.64 &   19.2 &    6.5 &                M2.5V &     15.4 \\
             J05082729-2101444 &  05 08 27.30 & -21  01 44.40 &   14.70 &    1.60 &    3.09 &    5.87 &   25.0 &   25.3 &                M5.6V &     --- \\
 TYC\,112\,1486\,1       &  05 20 31.83  &  +06 16 11.48  &  11.67  &  0.98 &  1.40 & 3.11 &  71    &  17.0   &  K4V & 30 \\
 TYC\,112\,917\,1        &  05 20 00.29  &  +06 13 03.57  &  11.58  &  1.26 &  1.38 & 3.00 &  67.8  &   8     &  K4V & --- \\
             J05241914-1601153 &  05 24 19.15 & -16  01 15.30 &   14.32 &    1.56 &    2.98 &    5.60 &   32.0 &   50.0 &              M4.5+M5 &     20.5 \\
                      HIP\,25486 &  05 27 04.76 & -11  54 03.47 &    6.22 &    0.55 &    0.63 &    1.29 &   27.0 &   53.0 &                  F7V &      0.0 \\
             J05294468-3239141 &  05 29 44.68 & -32  39 14.20 &   13.79 &    1.60 &    2.98 &    5.47 &   26.0 & --- &                 M4.5V &     --- \\
                TYC\,4770\,0797\,1 &  05 32 04.51 &  -03  05 29.38 &   11.32 &    1.45 &    2.22 &    4.31 &   42.0 &   12.0 &             M2V+M3.5V &      8.4 \\
             J05335981-0221325 &  05 33 59.81 &  -02  21 32.50 &   12.42 &    1.49 &    2.52 &    4.72 &   42.0 &    5.4 &                M2.9V &     --- \\
             J06131330-2742054 &  06 13 13.31 & -27  42 05.50 &   12.09 &    1.49 &    2.71 &    5.23 &   29.4 &    6.7 &                M3.V: &      0.0 \\
                      HIP\,29964 &  06 18 28.20 & -72  02 41.00 &    9.80 &    1.13 &    1.32 &    2.99 &   38.6 &   16.4 &                 K4Ve &     --- \\
           J07293108+3556003AB &  07 29 31.09 &  35  56 00.40 &   11.82 &    1.42 &    2.08 &    4.02 &   42.0 &   20.0 &                   M1+M3 &      8.4 \\
             J08173943-8243298 &  08 17 39.44 & -82  43 29.80 &   11.62 &    1.51 &    2.69 &    5.03 &   27.0 &   32.2 &                 M3.5V &     --- \\
             J08224744-5726530 &  08 22 47.45 & -57  26 53.00 &   13.37 &    1.58 &    3.08 &    5.57 &    8.0 &    6.4 &              M4.5+L0 &      5.1 \\
           J09361593+3731456AB &  09 36 15.91 &  37  31 45.50 &   11.09 &    1.44 &    2.06 &    4.10 &   33.7 &    6.0 &            M0.5+M0.5 &      0.0 \\
             J10015995+6651278 & 10 02 00.10 &  66  51 26.00 &   12.38 &    1.47 &    2.36 &    4.16 &   38.7 &   12.0 &                   M3 &     --- \\
                      HIP\,50156 & 10 14 19.17 &  21  04 29.55 &   10.08 &    1.37 &    1.83 &    3.82 &   20.0 &   23.1 &               M0.5V+? &      1.8 \\
                         TWA\,22 & 10 17 26.89 & -53  54 26.50 &   13.99 &    1.54 &    2.86 &    6.30 &   17.5 &    9.7 &                   M5+M6 &      1.8 \\
                    BD+262161A & 10 59 38.31 &  25  26 15.50 &    8.45 &    1.06 &    1.15 &    2.61 &   22.0 & --- &                   K2 &    115.7 \\
                    BD+262161B & 10 59 38.31 &  25  26 15.50 &    9.09 &    1.15 &    1.36 &    3.25 &   22.0 &    6.0 &                   K5 &    115.7 \\
             J11515681+0731262 & 11 51 56.81 &   07  31 26.25 &   12.38 &    1.40 &    2.26 &    4.61 &   33.2 & --- &             M2+M2+M8 &     16.6 \\
             J13545390-7121476 & 13 54 53.90 & -71  21 47.67 &   12.24 &    1.46 &    2.44 &    4.57 &   21.0 &    3.1 &                M2.5V &     --- \\
                     HIP\,69562A & 14 14 21.36 & -15  21 21.75 &   10.27 &    1.28 &    1.78 &    3.67 &   30.0 &  103.0 &               K5.5V+ &      9.0 \\
                     HIP\,69562B & 14 14 21.36 & -15  21 21.75 &   10.27 &    1.28 &    1.78 &    3.67 &   30.0 &   24.0 &               --- &     39.0 \\
                 TYC\,915\,1391\,1 & 14 25 55.93 &  14  12 10.14 &   10.89 &    1.08 &    1.20 &    3.60 &   51.8 & --- &                   K4V &     --- \\
                      HIP\,76629 & 15 38 57.50 & -57  42 27.00 &    7.97 &    0.82 &    0.88 &    2.12 &   39.7 &   16.6 &                  K0V &      0.0 \\
             J16430128-1754274 & 16 43 01.29 & -17  54 27.50 &   12.50 &    1.41 &    2.02 &    3.95 &   59.0 &    8.8 &                 M0.6 &     --- \\
             J16572029-5343316 & 16 57 20.30 & -53  43 31.70 &   12.44 &    1.47 &    2.47 &    4.65 &   51.0 &    3.5 &                  M3V &     --- \\
             J17150219-3333398 & 17 15 02.20 & -33  33 39.80 &   10.93 &    1.37 &    1.86 &    3.86 &   23.0 &   76.3 &                  M0V &     --- \\
                      HIP\,84586 & 17 17 25.50 & -66  57 04.00 &    7.23 &    0.76 &    0.84 &    2.53 &   31.4 &   31.0 &            G5IV+K5IV &      0.0 \\
                     HD\,155555C & 17 17 31.29 & -66  57 05.49 &   12.71 &    1.54 &    2.73 &    5.08 &   31.4 &    7.6 &               M3.5Ve &   1036.4 \\
                  TYC\,8728\,2262\,1 & 17 29 55.10 & -54  15 49.00 &    9.55 &    0.85 &    0.95 &    2.19 &   66.0 &   35.3 &                  K1V &     --- \\
                GSC\,08350-01924 & 17 29 20.67 & -50  14 53.00 &   13.47 &    1.46 &    2.56 &    4.77 &   64.0 &   23.5 &                  M3V &     51.2 \\
                      HD\,160305 & 17 41 49.03 & -50  43 28.00 &    8.35 &    0.51 &    0.65 &    1.36 &   72.0 & --- &                  F9V &     --- \\
                  TYC\,8742\,2065\,1 & 17 48 33.70 & -53  06 43.00 &    8.94 &    0.83 &    0.91 &    2.16 &   74.0 &   10.0 &                K0IV+ &      0.0 \\
                      HIP\,88399 & 18 03 03.41 & -51  38 56.43 &   12.50 &    0.46 &    0.53 &    4.23 &   47.0 & --- &               M2V+F6V &    302.7 \\
                      V4046\,Sgr & 18 14 10.50 & -32  47 33.00 &   10.44 &    1.18 &    1.43 &    3.19 &   72.0 &   14.2 &                K5V+K7V &      0.0 \\
                 UCAC2\,18035440 & 18 14 22.07 & -32  46 10.12 &   12.78 &    1.36 &    2.14 &    4.24 &   71.0 &    3.0 &                 M1Ve &      0.0 \\
             J18151564-4927472 & 18 15 15.64 & -49  27 47.20 &   12.86 &    1.47 &    2.48 &    4.78 &   61.0 &   76.7 &                  M3V &      0.0 \\
                      HIP\,89829 & 18 19 52.20 & -29  16 33.00 &    8.89 &    0.67 &    0.73 &    1.84 &   75.2 &  114.7 &                  G1V &     --- \\
            J18202275-1011131A & 18 20 22.74 & -10  11 13.62 &   10.63 &    1.18 &    1.44 &    3.35 &   61.0 & --- &            K5Ve &     81.1 \\
            J18202275-1011131B & 18 20 22.74 & -10  11 13.62 &   10.63 &    1.24 &    1.66 &    4.01 &   61.0 & --- &            K7Ve &     81.1 \\
             J18420694-5554254 & 18 42 06.95 & -55  54 25.50 &   13.53 &    1.53 &    2.81 &    4.95 &   54.0 &    8.6 &                M3.5V &     --- \\
                  TYC\,9077\,2489\,1 & 18 45 37.02 & -64  51 46.14 &    9.30 &    1.20 &    1.54 &    3.20 &   29.2 &  150.0 &                 K8Ve &      5.3 \\
                  TYC\,9073\,0762\,1 & 18 46 52.60 & -62  10 36.00 &   11.80 &    1.46 &    2.09 &    3.95 &   53.0 &    9.9 &                 M1Ve &     --- \\
                      HD\,173167 & 18 48 06.36 & -62  13 47.02 &    7.28 &    0.50 &    0.55 &    1.14 &   54.0 & --- &                  F5V &      0.0 \\
                  TYC\,7408\,0054\,1 & 18 50 44.50 & -31  47 47.00 &   11.20 &    1.35 &    1.82 &    3.66 &   50.0 &   49.7 &                 K8Ve &      0.0 \\
                      HIP\,92680 & 18 53 05.90 & -50  10 50.00 &    8.29 &    0.77 &    0.85 &    1.92 &   49.8 &   69.0 &                 K8Ve &     --- \\
                  TYC\,6872\,1011\,1 & 18 58 04.20 & -29  53 05.00 &   11.78 &    1.30 &    1.80 &    3.76 &   78.0 &   33.8 &                 M0Ve &     --- \\
             J19102820-2319486 & 19 10 28.21 & -23  19 48.60 &   13.20 &    1.50 &    2.63 &    4.99 &   67.0 &   12.2 &                  M4V &     --- \\
                  TYC\,6878\,0195\,1 & 19 11 44.70 & -26  04 09.00 &   10.27 &    1.05 &    1.18 &    2.90 &   79.0 &    9.8 &                 K4Ve &     86.9 \\
             J19233820-4606316 & 19 23 38.20 & -46  06 31.60 &   11.87 &    1.33 &    1.73 &    3.60 &   70.0 &   15.4 &                  M0V &     --- \\
             J19243494-3442392 & 19 24 34.95 & -34  42 39.30 &   14.28 &    1.56 &    2.99 &    5.50 &   54.0 &   10.9 &                  M4V &     --- \\
                  TYC\,7443\,1102\,1 & 19 56 04.37 & -32  07 37.71 &   11.80 &    1.40 &    1.80 &    3.95 &   55.0 &    6.0 &                M0.0V &   1298.2 \\
           J19560294-3207186AB & 19 56 02.94 & -32  07 18.70 &   13.30 &    1.48 &    2.45 &    5.12 &   55.0 &   35.0 &                  M4V &     11.0 \\
             J20013718-3313139 & 20 01 37.18 & -33  13 14.01 &   12.25 &    1.43 &    2.10 &    4.06 &   62.0 &    2.6 &                   M1V &     --- \\
             J20055640-3216591 & 20 05 56.41 & -32  16 59.15 &   11.96 &    1.20 &    1.95 &    4.02 &   52.0 & --- &                  M2V &     --- \\
                      HD\,191089 & 20 09 05.21 & -26  13 26.52 &    7.18 &    0.48 &    0.55 &    1.10 &   52.0 &   37.7 &                  F5V &     --- \\
           J20100002-2801410AB & 20 10 00.03 & -28  01 41.10 &   13.62 &    1.52 &    2.77 &    4.64 &   48.0 &   46.5 &            M2.5+M3.5 &     29.3 \\
             J20333759-2556521 & 20 33 37.59 & -25  56 52.20 &   14.87 &    1.65 &    3.30 &    5.99 &   48.3 &   21.0 &                 M4.5V &     --- \\
                    HIP\,102141A & 20 41 51.20 & -32  26 07.00 &   11.09 &    1.54 &    2.96 &    5.42 &   10.7 &   10.7 &                 M4Ve &     24.6 \\
                    HIP\,102141B & 20 41 51.10 & -32  26 10.00 &   11.13 &    1.55 &    2.96 &    5.42 &   10.7 &   10.7 &                 M4Ve &     24.6 \\
             J20434114-2433534 & 20 43 41.14 & -24  33 53.19 &   12.83 &    1.51 &    2.72 &    4.97 &   28.1 &   44.0 &            M3.7+M4.1 &     41.3 \\
                     HIP\,102409 & 20 45 09.50 & -31  20 27.00 &    8.73 &    1.47 &    2.10 &    4.20 &    9.9 &    9.3 &                 M1Ve &     --- \\
                     HIP\,103311 & 20 55 47.67 & -17  06 51.04 &    7.35 &    0.54 &    0.62 &    1.54 &   48.0 &  115.0 &                  F8V &     52.8 \\
                  TYC\,634902001 & 20 56 02.70 & -17  10 54.00 &   10.62 &    1.22 &    1.49 &    3.54 &   48.0 &   15.6 &              K6Ve+M2 &    105.6 \\
             J21100535-1919573 & 21 10 05.36 & -19  19 57.40 &   11.54 &    1.40 &    1.97 &    4.34 &   32.0 &    9.7 &                  M2V &     --- \\
             J21103147-2710578 & 21 10 31.48 & -27  10 57.80 &   15.20 &    1.59 &    3.04 &    5.60 &   41.0 &   15.8 &                M4.5V &    389.6 \\
             J21103096-2710513 & 21 10 30.96 & -27  10 51.30 &   15.72 &    1.66 &    3.31 &    5.60 &   41.0 &   14.6 &                  M5V &    389.6 \\
                     HIP\,105441 & 21 21 24.49 & -66  54 57.37 &    8.77 &    1.10 &    1.26 &    2.37 &   30.2 &    5.9 &                  K2V &    785.3 \\
                TYC\,9114\,1267\,1 & 21 21 28.72 & -66  55 06.30 &   10.59 &    1.40 &    1.76 &    3.58 &   30.2 &    4.5 &                  K7V &    785.3 \\
                 TYC\,9486\,927\,1 & 21 25 27.49 & -81  38 27.68 &   11.70 &    1.41 &    2.06 &    4.36 &   26.2 &   43.5 &                  M1V &      0.0 \\
           J21374019+0137137AB & 21 37 40.19 &  01  37 13.70 &   13.36 &    1.56 &    3.00 &    5.48 &   39.2 &   39.0 &                  M5V &     17.3 \\
               J21412662+2043107 & 21 41 26.63 &  20  43 10.70 &   13.50 &    1.50 &    2.52 &    4.89 &   52.7 & --- &                  M3V &     --- \\
                TYC\,2211\,1309\,1 & 22 00 41.59 &  27  15 13.60 &   11.39 &    1.40 &    1.80 &    3.67 &   45.6 &   30.0 &                  M0V &      0.0 \\
                  TYC\,9340\,0437\,1 & 22 42 48.90 & -71  42 21.00 &   10.60 &    1.35 &    1.73 &    3.71 &   36.0 &    7.5 &                 K7Ve &     --- \\
                     HIP\,112312 & 22 44 58.00 & -33  15 02.00 &   12.10 &    1.50 &    2.78 &    5.17 &   23.6 &   12.1 &                 M4Ve &    778.9 \\
                       TX\,Psa & 22 45 00.05 & -33  15 25.80 &   13.36 &    1.58 &    3.04 &    5.57 &   23.6 &   16.0 &               M4.5Ve &    778.9 \\
             J22571130+3639451 & 22 57 11.31 &  36  39 45.14 &   12.50 &    1.46 &    2.34 &    3.86 &   69.0 &   20.0 &                   M3V &     --- \\
                  TYC\,5832\,0666\,1 & 23 32 30.90 & -12  15 52.00 &   10.54 &    1.43 &    1.98 &    3.97 &   28.0 &    8.6 &                 M0Ve &     --- \\
             J23500639+2659519 & 23 50 06.39 &  26  59 51.93 &   14.26 &    1.53 &    2.57 &    4.96 &   25.0 & 36.0 &                  M3.5V &     --- \\
             J23512227+2344207 & 23 51 22.28 &  23  44 20.80 &   14.11 &    1.54 &    2.93 &    5.29 &   18.0 & 4.0 &                  M4V &    --- \\
\hline
\enddata
\end{deluxetable}

\clearpage

\begin{deluxetable}{lrrrrrcccc}
\tablecolumns{10}
\tablecaption{\label{tab_period} Results of the photometric period search. We list the target's name, the photometric rotation period, and its uncertainty, the peak-to-peak light curve amplitude, the photometric accuracy, the single/binary nature of the target, another name, the source of the photometric time series, the filter, and the observation JD interval used to compute the rotation period.
}
\tablewidth{0pt}
 \tablehead{\colhead{Target} & \colhead{P } & \colhead{$\Delta$P} & \colhead{$\Delta$mag} & \colhead{$\sigma$} & \colhead{Mult.} & \colhead{Other name} & \colhead{Source} & \colhead{Filter} & \colhead{JD$_{start}$--JD$_{end}$}\\
\colhead{    } & \colhead{(d)} & \colhead{(d)} & \colhead{(mag)} & \colhead{(mag)}  & \colhead{} & \colhead{} & \colhead{} & \colhead{} & \colhead{(JD-2400000)}\\
}
\startdata
  \hline
                      HIP\,560 &   0.224 &   0.005 &   0.008 &   0.007 &     S+D &          --- &  YSVP & I & 57297--57333 \\
             J00172353-6645124 &   6.644 &   0.027 &   0.100 &   0.012 &      S &          --- & INTEGRAL & V & 52856--56669  \\
                TYC\,1186\,0706\,1 &   7.9 &   0.1 &   0.070 &   0.028 &     Bw &      FK\,Psc &   literature & V & ---\\
                       GJ\,2006A &   3.99 &   0.05 &   0.170 &   0.006 &      Bw &          ---  & CASLEO & R & 56904--56952\\
                       GJ\,2006B &   4.91 &   0.05 &   0.120 &   0.005 &      Bw &          --- & CASLEO & R & 56904--56952 \\
             J00323480+0729271A &   3.355 &   0.005 &   0.045 &   0.006 &     Bc &      GJ3039A & MO+ARIES & R & 56948--57016 \\
             J00323480+0729271B &   0.925 &   0.008 &   0.045 &   0.006 &     Bc &      GJ3039B & MO+ARIES & R & 56948--57016 \\
                TYC\,5853\,1318\,1 &   7.26 &   0.07 &   0.10 &   0.03 &    S? &          --- & literature & V & ---\\
            J01112542+1526214A &   0.911 &   0.001 &   0.01 &   0.01 &      Bc &     GJ\,3076 & MEarth & RG715 & 55098--55495 \\
            J01112542+1526214B &   0.791 &   0.001 &   0.01 &   0.01 &      Bc &     GJ\,3076 & MEarth & RG715 & 55098--55495 \\
             J01132817-3821024 &   0.446 &   --- &   0.210 &   --- &     Tc &          --- & literature &  V & ---\\
             J01351393-0712517 &   0.703 &   --- &   0.080 &   --- &    SB2 &          ---  & literature &  V & ---\\
             J01365516-0647379 &   0.346 &   0.001 &   0.11 &   0.01 &    Bw &          --- & CASLEO+HAO & V & 56919--56989 \\
                TYC\,1208\,0468\,1 &   2.803 &   0.010 &   0.07 &   0.02 &      Bw &          --- & literature &  V & --- \\
             J01535076-1459503 &   1.515 &   --- &   0.110 &   --- &      BC &          --- & literature &  V & --- \\
               J02014677+0117161&   3.30/5.98 &   0.01 &   0.09 &   0.02 &      Bw &          ---  & NSVS & V & 51426--51576 \\
               RBS\,269 &   5.98/3.30 &   0.01 &   0.09 &   0.02 &      Bw &          ---  & NSVS & V & 51426--51576\\
             J02175601+1225266 &   1.995 &   0.005 &   0.05 &   0.005 &     S &          ---  & ARIES & I & 56959--57018\\
                      HIP\,10679 &   0.777 &   0.005 &   0.070 &   0.006 &    Bw+D &          ---  & literature & V & 56952--56962\\
                      HIP\,10680 &   0.240 &   0.001 &   0.030 &   0.006 &      Bw &          ---  & literature & V & 56952--56962\\
		  HIP\,11152    &  1.80/3.60    &  0.02     & 0.06    & 0.01      & S    &   --- & SW & V & 54056--54121 \\
                     HIP\,11437A &  12.5 &   0.5 &   0.20 &   0.01 &    Bw+D &      AG\,Tri & literature & V & 56634--57091\\
                     HIP\,11437B &   4.66 &   0.05 &   0.16 &   0.01 &      Bw &          --- & literature & V & 56634--57091 \\
                      HIP12545 &   4.83 &   0.03 &   0.180 &   0.032 &      S &          ---  & CrAO & V & 57321--57381\\
             J03350208+2342356 &   0.472 &   0.005 &   0.03 &   0.06 &    Bc? &          --- & NSVS & V & 51375--51620 \\
               J03461399+1709176 &   1.742 &   0.001 &   0.07 &   0.01 &     S &          --- & literature & $r$ & ---\\
                        GJ\,3305 &   4.89 &   0.01 &   0.05 &   0.05 &    Bc &          --- & ASAS & V & 52558--55145\\
             J04435686+3723033 &   4.288 &   --- &   --- &   --- &      Bw &    V962\_Per & literature & V & ---\\
                      HIP\,23200 &   4.430 &   0.030 &   0.150 &   0.032 &     SB1 &   V1005\,Ori  & literature & V & ---\\
                TYC\,1281\,1672\,1 &   2.76 &   0.01 &   0.12 &   0.01 &      S &   V1814\,Ori & INTEGRAL & V & 53460--56214\\
                      HIP\,23309 &   8.60 &   0.07 &   0.110 &   0.031 &       S &        --- & ASAS & V & 51868--54884\\
             J05015665+0108429 &   2.08 &   0.02 &   0.07 &   0.01 &      S? &          --- & INTEGRAL & V & 53592--56688\\
                     HIP\,23418A &   1.220 &   0.010 &   0.070 &   0.033 &     SB2 &          GJ3322A & ASAS & V & 51946--55164 \\
                     HIP\,23418B & --- &   --- &   --- &   --- &      Tc &          GJ3322B & --- & --- & --- \\
                    BD-211074A &   9.3 &   0.1 &   0.120 &   0.015 &      Tw &          GJ3331 & literature & R & --- \\
                    BD-211074B &   5.40 &   0.10 &   0.080 &   0.015 &      Tc &          GJ3332  & literature & R & ---\\
             J05082729-2101444 &   0.280 &   0.002 &   0.07 &   0.08 &       S &          --- & SW & V & 53993--54509\\
 TYC\,112\,1486\,1  & 2.18       & ---    &  0.09  & ---     & Tc & --- & literature & V & --- \\ 
 TYC\,112\,917\,1  & 3.51       & ---    &  0.08  &  ---    &  Tw & --- & literature & V & ---\\
             J05241914-1601153 &   0.401 &   0.001 &   0.15 &   0.05 &      Bc  &          --- & Catalina & V & 53598--56406 \\
                      HIP\,25486 &   0.966 &   0.002 &   0.10 &   0.01 &     SB2  &      AF\,Lep & literature & V & ---\\
             J05294468-3239141 &   1.532 &   0.005 &   0.03 &   0.05 &      S? &          --- & SW & V & 53993--54509\\
                TYC\,4770\,0797\,1 &   4.372 &   0.002 &   0.160 &   0.031 &    Bc &   V1311\,Ori & INTEGRAL & V & 54005--56685\\
             J05335981-0221325 &   7.250 &   --- &   0.170 &   --- &       S &          --- & INTEGRAL & V & 54005--56576\\
             J06131330-2742054 &  16.8 &   1.0 &   0.07 &   0.01 &      Tc &          --- & YCO & V & 57009--57140\\
                      HIP\,29964 &   2.670 &   0.010 &   0.120 &   0.032 &     S+D &      AO\,Men & literature & V & --- \\
           J07293108+3556003AB &   1.970 &   0.010 &   0.10 &   0.02 &      Bc &          --- & SW & V & 54056--54574\\
             J08173943-8243298 &   1.318 &   --- &   0.050 &   --- &     Bc? &          --- & literature & V & --- \\
             J08224744-5726530 & --- &   --- &   --- &   --- &    Tc &          --- & --- & --- & ---\\
           J09361593+3731456AB &  12.9 &   0.3 &   0.030 &   0.016 &    SB2 &       GJ\,9303 & SW & V & 54066--54604\\
             J10015995+6651278 &   2.49 &   0.02 &   0.060 &   0.012 &     Bc? &          --- & INTEGRAL & V & 55122--56846 \\
                      HIP\,50156 &   7.860 & --- &   0.050 & --- &    Bc &      DK\,Leo & literature & V & --- \\
                         TWA\,22 &   0.830 &   0.010 &   0.020 &   --- &      Bc &          --- & INTEGRAL & V & 52669--56988\\
                    BD+262161A &   2.022/0.974 &   0.005 &   0.010 &   0.006 &  Bw+D &          --- & SW & V & 54091--54225\\
                    BD+262161B &   0.974/2.022 &   0.005 &   0.010 &   0.006 &    Bw &          --- & SW & V & 54091--54225 \\
             J11515681+0731262 &   2.291 & --- &   0.130 & --- &    SB2 &          --- & literature & V & ---\\
             J13545390-7121476 &   3.65 &   0.02 &   0.020 &   0.008 &      S? &          --- & YCO & V & 56823--56921\\
                     HIP\,69562A &   0.298 &   0.005 &   0.17 &   0.02 &      Tc &      MV\,Vir & literature & V & ---\\
                     HIP\,69562B & --- & --- & --- & --- &      Tc &      MV\,Vir & --- & --- & --- \\
                 TYC\,915\,1391\,1 &   4.340 & --- &   0.360 & --- &      S &          --- & literature & V & ---\\
                      HIP\,76629 &   4.27 &   0.10 &   0.180 &   0.034 &     SB1 &    V343\,Nor  & literature & V & ---\\
             J16430128-1754274 &   5.14 &   0.04 &   0.140 &   0.031 &       S &          --- & literature & V & ---\\
             J16572029-5343316 &   7.15 &   0.05 &   0.020 &   0.008 &       S &          --- & YCO & V & 56823--56929\\
             J17150219-3333398 &   0.311 &   --- &   0.110 &   --- &     Bc? &          --- & INTEGRAL & V & 52698--56578 \\
                      HIP\,84586 &   1.680 &   0.010 &   0.120 &   0.034 &     SB2 &    V824\,Ara & YCO & V  & 56717--56788\\
                     HD\,155555C &   4.43 &   0.01 &   0.070 &   0.007 &      Tw &          --- & YCO & V  & 56717--56788\\
                  TYC\,8728\,2262\,1 &   1.775 &   0.005 &   0.150 &   0.036 &      S &          --- &  INTEGRAL & V & 52704--55984\\
                GSC\,08350-01924 &   1.906 &   0.005 &   0.05 &   0.01 &      Bc &          --- & YCO & V & 56950--56976\\
                      HD\,160305 &   1.336 &   0.008 &   0.060 &   0.036 &   S+D &          --- & literature & V & ---\\
                  TYC\,8742\,2065\,1 &   2.60/1.62 &   0.01 &   0.060 &   0.033 &    Tc &          --- & literature & V & --- \\
                      HIP\,88399 & --- &   --- &   --- &   --- &      Bw &          --- & --- & --- & ---\\
                      V4046\,Sgr &   2.42 &   0.01 &   0.090 &   0.033 &   SB2+D &   V4046\,Sgr & literature & V & --- \\
                 UCAC2\,18035440 &  12.05 &   0.5 &   0.14 &   0.03 &    SB &          --- & literature & V & --- \\
             J18151564-4927472 &   0.447 &   0.002 &   0.130 &   0.008 &    SB1 &          --- & YCO & V & 56800--56929 \\
                      HIP\,89829 &   0.571 &   0.001 &   0.140 &   0.037 &       S &          --- & INTEGRAL & V &  52730--56593\\
            J18202275-1011131 &   4.65/5.15 &   --- &   0.070 &   --- &    Bw+D &      FK\,Ser & ASAS & V & 51962--55092 \\
            J18202275-1011131B &  5.15/4.65 &   --- &   0.070 &   --- &      Bw &      FK\,Ser & ASAS & V & 51962--55092\\
             J18420694-5554254 &   5.403 &   --- &   0.070 &   --- &      S? &          --- & literature & V & ---\\
                  TYC\,9077\,2489\,1 &   0.345 &   0.005 &   0.160 &   0.033 &      Tc &          --- & literature & V & ---\\
                  TYC\,9073\,0762\,1 &   5.37 &   0.04 &   0.320 &   0.032 &      S&          --- & INTEGRAL & V & 54403--55984 \\
                      HD\,173167 &   0.290 &   0.005 &   0.220 &   0.033 &     SB1 &          --- & INTEGRAL & V & 54403--55984 \\
                  TYC\,7408\,0054\,1 &   1.075 &   0.005 &   0.150 &   0.033 &      EB &          --- & literature & V & ---\\
                      HIP\,92680 &   0.944 &   0.001 &   0.110 &   0.032 &      Bw &      PZ\,Tel & literature & V & ---\\
                  TYC\,6872\,1011\,1 &   0.503 &   0.004 &   0.060 &   0.032 &      Bw &          --- & literature & --- & ---\\
             J19102820-2319486 &   3.64 &   0.02 &   0.13 &   0.01 &       S &          --- & YCO & V & 56950--56972 \\
                  TYC\,687801951 &   5.70 &   0.05 &   0.090 &   0.032 &      Bw &          --- & literature & V & --- \\
             J19233820-4606316 &   3.237 &   --- &   0.110 &   --- &       S &          --- & SW & V & 53860--54574 \\
             J19243494-3442392 &   0.708 &   0.001 &   0.020 &   0.007 &     Bc? &          --- & ROAD & V & 57155--57199 \\
                  TYC\,7443\,1102\,1 &  11.3 &   0.2 &   0.09 &   0.03 &      Tw &          --- & literature & V & --- \\
           J19560294-3207186AB &   1.569 &   0.003 &   0.030 &   0.025 &      Tc &          --- & KKO & R & 56844--56914 \\
             J20013718-3313139 &  12.7 &   0.2 &   0.13 &   0.01 &      Tw &          --- & literature & V & --- \\
             J20055640-3216591 &   8.368 &   0.005 &   0.130 &   0.015 &       S &   V5663\,Sgr & SW & V & 53860--54614\\
                      HD\,191089 &   0.488 &   0.005 & --- & --- &     S+D &          --- & literature & V & ---  \\
           J20100002-2801410AB &   0.470 &   0.005 &   0.040 &   0.011 &      Bc &          --- & SW & V & 53860--54614\\
             J20333759-2556521 &   0.710 &   0.001 &   0.05 &   0.10 &       S &          --- & SW & V & 53958--54614\\
                    HIP\,102141A &   1.191 &   0.005 &   0.040 &   0.005 &      Bc &     AT\,MicA  & SW & V & 53860--53953\\
                    HIP\,102141B &   0.781 &   0.002 &   0.020 &   0.005 &      Bc &     AT\,MicB & SW & V & 53860--53953\\
             J20434114-2433534 &   1.610 &   0.010 &   0.03 &   0.04 &      Bc &          --- & SW & V & 53958--54388 \\
                     HIP\,102409 &   4.86 &   0.02 &   0.10 &   0.03 &     S+D &      AU\,Mic & SW & V & --- \\
                     HIP\,103311 &   0.356 &   0.004 &   0.06 &   0.02 &      Bc &          --- & literature & V & ---\\
                  TYC\,6349\,0200\,1 &   3.41 &   0.05 &   0.120 &   0.028 &      Bw &      AZ\,Cap & literature & V & --- \\
             J21100535-1919573 &   3.71 &   0.02 &   0.29 &   0.01 &       S &          --- & CASLEO & R & 56904--56952\\
             J21103147-2710578 &   1.867 &   0.008 &   0.04 &   0.14 &      Bw &          --- & SW & V & 53862--54614 \\
             J21103096-2710513 & --- &   --- &   --- &   --- &      Bw &          ---  & SW & V & 53862--54614\\
                     HIP\,105441 &   5.50 &   0.02 &   0.050 &   0.009 &    Bw &    V390\,Pav & KKO+CASLEO & R & 56904--56992 \\
                TYC\,9114\,1267\,1 &  20.5 &   1.0 &   0.015 &   0.010 &    Bw &          --- & KKO+CASLEO & R & 56904--56992\\
                 TYC\,9486\,927\,1 &   0.542 &   --- &   0.190 &   --- &     Bc &          --- & literature & V & ---\\
           J21374019+0137137AB &   0.202 &   0.001 &   0.130 &   0.004 &      Bc &          --- & CASLEO & R &  56904--56950\\
                J21412662+2043107 &   0.899 &   0.001 &   0.03 &   0.03 &   Bc? &          --- & SW & V & 53129--54410 \\
                TYC\,2211\,1309\,1 &   1.109 &   0.001 &   0.080 &   0.034 &     Bc &          --- & literature & V & --- \\
                  TYC\,9340\,0437\,1 &   4.46 &   0.03 &   0.16 &   0.03 &       S &          ---  & literature & V & --- \\
                     HIP\,112312 &   2.37 &   0.01 &   0.110 &   0.006 &      Bw &      WW\,Psa & literature & V & ---\\
                       TX\,Psa &   1.080 &   0.005 &   0.030 &   0.006 &      Bw &      TX\,Psa & literature & V & ---\\
             J22571130+3639451 &   1.220 &   0.020 &   0.04 &   0.01 &     S &          --- & SW & V &  53154--54666\\
                  TYC\,5832\,0666\,1 &   5.68 &   0.03 &   0.140 &   0.028 &       S &          --- & literature & V & ---\\
             J23500639+2659519 &   0.287 &   0.005 &   0.05 &   0.02 &    Bc? &          --- & MEarth & RG715 & 55857--56469 \\
             J23512227+2344207 &   3.208 &   0.004 &   0.060 &   0.003 &      S &          --- & MEarth & RG715 & 55851--56471 \\\hline
             \multicolumn{10}{l}{S: single; B:  binary system; T: triple system; w: wide orbit ($>$60\,AU); c: close orbit ($<$60\,AU)}\\
   \multicolumn{10}{l}{D: debris disc host star; ?: uncertain single/binary nature}\\            
\enddata
\end{deluxetable}

\clearpage

\bibliographystyle{aa.bst} 
\bibliography{mybib} 

\begin{thebibliography}{195}
\expandafter\ifx\csname natexlab\endcsname\relax\def\natexlab#1{#1}\fi

\bibitem[{{Abt} \& {Morrell}(1995)}]{Abt95}
{Abt}, H.~A. \& {Morrell}, N.~I. 1995, \apjs, 99, 135

\bibitem[{{Alcal{\'a}} {et~al.}(2000){Alcal{\'a}}, {Covino}, {Torres},
  {Sterzik}, {Pfeiffer}, \& {Neuh{\"a}user}}]{Alcala00}
{Alcal{\'a}}, J.~M., {Covino}, E., {Torres}, G., {et~al.} 2000, \aap, 353, 186

\bibitem[{{Alonso-Floriano} {et~al.}(2015){Alonso-Floriano}, {Caballero},
  {Cort{\'e}s-Contreras}, {Solano}, \& {Montes}}]{Alonso-Floriano15}
{Alonso-Floriano}, F.~J., {Caballero}, J.~A., {Cort{\'e}s-Contreras}, M.,
  {Solano}, E., \& {Montes}, D. 2015, \aap, 583, A85

\bibitem[{{Alonso-Floriano} {et~al.}(2011){Alonso-Floriano}, {Caballero}, \&
  {Montes}}]{Alonso-Floriano11}
{Alonso-Floriano}, F.~J., {Caballero}, J.~A., \& {Montes}, D. 2011, in Stellar
  Clusters \& Associations: A RIA Workshop on Gaia, 344--347

\bibitem[{{Amado} {et~al.}(2001){Amado}, {Zboril}, {Butler}, \&
  {Byrne}}]{Amado01}
{Amado}, P.~J., {Zboril}, M., {Butler}, C.~J., \& {Byrne}, P.~B. 2001,
  Contributions of the Astronomical Observatory Skalnate Pleso, 31, 13

\bibitem[{{Anthonioz} {et~al.}(2015){Anthonioz}, {M{\'e}nard}, {Pinte}, {Le
  Bouquin}, {Benisty}, {Thi}, {Absil}, {Duch{\^e}ne}, {Augereau}, {Berger},
  {Casassus}, {Duvert}, {Lazareff}, {Malbet}, {Millan-Gabet}, {Schreiber},
  {Traub}, \& {Zins}}]{Anthonioz15}
{Anthonioz}, F., {M{\'e}nard}, F., {Pinte}, C., {et~al.} 2015, \aap, 574, A41

\bibitem[{{Bailey} {et~al.}(2012){Bailey}, {White}, {Blake}, {Charbonneau},
  {Barman}, {Tanner}, \& {Torres}}]{Bailey12}
{Bailey}, III, J.~I., {White}, R.~J., {Blake}, C.~H., {et~al.} 2012, \apj, 749,
  16

\bibitem[{{Barnes} {et~al.}(2000){Barnes}, {Collier Cameron}, {James}, \&
  {Donati}}]{Barnes00}
{Barnes}, J.~R., {Collier Cameron}, A., {James}, D.~J., \& {Donati}, J.-F.
  2000, \mnras, 314, 162

\bibitem[{{Barrado y Navascu{\'e}s} {et~al.}(1999){Barrado y Navascu{\'e}s},
  {Stauffer}, {Song}, \& {Caillault}}]{Barrado99}
{Barrado y Navascu{\'e}s}, D., {Stauffer}, J.~R., {Song}, I., \& {Caillault},
  J.-P. 1999, \apjl, 520, L123

\bibitem[{{Batalha} {et~al.}(1998){Batalha}, {Quast}, {Torres}, {Pereira},
  {Terra}, {Jablonski}, {Schiavon}, {de La Reza}, \& {Sartori}}]{Batalha98}
{Batalha}, C.~C., {Quast}, G.~R., {Torres}, C.~A.~O., {et~al.} 1998, \aaps,
  128, 561

\bibitem[{{Bell} {et~al.}(2015){Bell}, {Mamajek}, \& {Naylor}}]{Bell15}
{Bell}, C.~P.~M., {Mamajek}, E.~E., \& {Naylor}, T. 2015, \mnras, 454, 593

\bibitem[{{Bennett} {et~al.}(1967){Bennett}, {Evans}, \& {Laing}}]{Bennett67}
{Bennett}, N.~W.~W., {Evans}, D.~S., \& {Laing}, J.~D. 1967, \mnras, 137

\bibitem[{{Berdnikov} \& {Pastukhova}(2008)}]{Berdnikov08}
{Berdnikov}, L.~N. \& {Pastukhova}, E.~N. 2008, Peremennye Zvezdy, 28

\bibitem[{{Bergfors} {et~al.}(2010){Bergfors}, {Brandner}, {Janson}, {Daemgen},
  {Geissler}, {Henning}, {Hippler}, {Hormuth}, {Joergens}, \&
  {K{\"o}hler}}]{Bergfors10}
{Bergfors}, C., {Brandner}, W., {Janson}, M., {et~al.} 2010, \aap, 520, A54

\bibitem[{{Berta} {et~al.}(2012){Berta}, {Irwin}, {Charbonneau}, {Burke}, \&
  {Falco}}]{Berta12}
{Berta}, Z.~K., {Irwin}, J., {Charbonneau}, D., {Burke}, C.~J., \& {Falco},
  E.~E. 2012, \aj, 144, 145

\bibitem[{{Beuzit} {et~al.}(2004){Beuzit}, {S{\'e}gransan}, {Forveille},
  {Udry}, {Delfosse}, {Mayor}, {Perrier}, {Hainaut}, {Roddier}, {Roddier}, \&
  {Mart{\'{\i}}n}}]{Beuzit04}
{Beuzit}, J.-L., {S{\'e}gransan}, D., {Forveille}, T., {et~al.} 2004, \aap,
  425, 997

\bibitem[{{Biazzo} {et~al.}(2005){Biazzo}, {Frasca}, {Henry}, {Catalano}, \&
  {Marilli}}]{Biazzo05}
{Biazzo}, K., {Frasca}, A., {Henry}, G.~W., {Catalano}, S., \& {Marilli}, E.
  2005, in ESA Special Publication, Vol. 560, 13th Cambridge Workshop on Cool
  Stars, Stellar Systems and the Sun, ed. F.~{Favata}, G.~A.~J. {Hussain}, \&
  B.~{Battrick}, 445

\bibitem[{{Biller} {et~al.}(2013){Biller}, {Liu}, {Wahhaj}, {Nielsen},
  {Hayward}, {Males}, {Skemer}, {Close}, {Chun}, {Ftaclas}, {Clarke}, {Thatte},
  {Shkolnik}, {Reid}, {Hartung}, {Boss}, {Lin}, {Alencar}, {de Gouveia Dal
  Pino}, {Gregorio-Hetem}, \& {Toomey}}]{Biller13}
{Biller}, B.~A., {Liu}, M.~C., {Wahhaj}, Z., {et~al.} 2013, \apj, 777, 160

\bibitem[{{Binks} \& {Jeffries}(2014)}]{Binks14}
{Binks}, A.~S. \& {Jeffries}, R.~D. 2014, \mnras, 438, L11

\bibitem[{{Binks} \& {Jeffries}(2016)}]{Binks16}
{Binks}, A.~S. \& {Jeffries}, R.~D. 2016, \mnras, 455, 3345

\bibitem[{{Bonnefoy} {et~al.}(2009){Bonnefoy}, {Chauvin}, {Dumas}, {Lagrange},
  {Beust}, {Desort}, {Teixeira}, {Ducourant}, {Beuzit}, \& {Song}}]{Bonnefoy09}
{Bonnefoy}, M., {Chauvin}, G., {Dumas}, C., {et~al.} 2009, \aap, 506, 799

\bibitem[{{Bopp} \& {Espenak}(1977)}]{Bopp77}
{Bopp}, B.~W. \& {Espenak}, F. 1977, \aj, 82, 916

\bibitem[{{Bopp} {et~al.}(1978){Bopp}, {Torres}, {Busko}, \& {Quast}}]{Bopp78}
{Bopp}, B.~W., {Torres}, C.~A.~O., {Busko}, I.~C., \& {Quast}, G.~R. 1978,
  Information Bulletin on Variable Stars, 1443

\bibitem[{{Bowler} {et~al.}(2015){Bowler}, {Liu}, {Shkolnik}, \&
  {Tamura}}]{Bowler15}
{Bowler}, B.~P., {Liu}, M.~C., {Shkolnik}, E.~L., \& {Tamura}, M. 2015, \apjs,
  216, 7

\bibitem[{{Brandt} {et~al.}(2014){Brandt}, {Kuzuhara}, {McElwain}, {Schlieder},
  {Wisniewski}, {Turner}, {Carson}, {Matsuo}, {Biller}, {Bonnefoy}, {Dressing},
  {Janson}, {Knapp}, {Moro-Mart{\'{\i}}n}, {Thalmann}, {Kudo}, {Kusakabe},
  {Hashimoto}, {Abe}, {Brandner}, {Currie}, {Egner}, {Feldt}, {Golota}, {Goto},
  {Grady}, {Guyon}, {Hayano}, {Hayashi}, {Hayashi}, {Henning}, {Hodapp},
  {Ishii}, {Iye}, {Kandori}, {Kwon}, {Mede}, {Miyama}, {Morino}, {Nishimura},
  {Pyo}, {Serabyn}, {Suenaga}, {Suto}, {Suzuki}, {Takami}, {Takahashi},
  {Takato}, {Terada}, {Tomono}, {Watanabe}, {Yamada}, {Takami}, {Usuda}, \&
  {Tamura}}]{Brandt14}
{Brandt}, T.~D., {Kuzuhara}, M., {McElwain}, M.~W., {et~al.} 2014, \apj, 786, 1

\bibitem[{{Browning} {et~al.}(2010){Browning}, {Basri}, {Marcy}, {West}, \&
  {Zhang}}]{Browning10}
{Browning}, M.~K., {Basri}, G., {Marcy}, G.~W., {West}, A.~A., \& {Zhang}, J.
  2010, \aj, 139, 504

\bibitem[{{Busko} {et~al.}(1980){Busko}, {Quast}, \& {Torres}}]{Busko80}
{Busko}, I.~C., {Quast}, G.~R., \& {Torres}, C.~A.~O. 1980, Information
  Bulletin on Variable Stars, 1898

\bibitem[{{Busko} \& {Torres}(1978)}]{Busko78}
{Busko}, I.~C. \& {Torres}, C.~A.~O. 1978, \aap, 64, 153

\bibitem[{{Butters} {et~al.}(2010){Butters}, {West}, {Anderson}, {Collier
  Cameron}, {Clarkson}, {Enoch}, {Haswell}, {Hellier}, {Horne}, {Joshi},
  {Kane}, {Lister}, {Maxted}, {Parley}, {Pollacco}, {Smalley}, {Street},
  {Todd}, {Wheatley}, \& {Wilson}}]{Butters10}
{Butters}, O.~W., {West}, R.~G., {Anderson}, D.~R., {et~al.} 2010, \aap, 520,
  L10

\bibitem[{{Byrne} {et~al.}(1984){Byrne}, {Doyle}, \& {Butler}}]{Byrne84}
{Byrne}, P.~B., {Doyle}, J.~G., \& {Butler}, C.~J. 1984, \mnras, 206, 907

\bibitem[{{Carpenter} {et~al.}(2005){Carpenter}, {Wolf}, {Schreyer},
  {Launhardt}, \& {Henning}}]{Carpenter05}
{Carpenter}, J.~M., {Wolf}, S., {Schreyer}, K., {Launhardt}, R., \& {Henning},
  T. 2005, \aj, 129, 1049

\bibitem[{{Chauvin} {et~al.}(2010){Chauvin}, {Lagrange}, {Bonavita},
  {Zuckerman}, {Dumas}, {Bessell}, {Beuzit}, {Bonnefoy}, {Desidera}, {Farihi},
  {Lowrance}, {Mouillet}, \& {Song}}]{Chauvin10}
{Chauvin}, G., {Lagrange}, A.-M., {Bonavita}, M., {et~al.} 2010, \aap, 509, A52

\bibitem[{{Chugainov}(1974)}]{Chugainov74}
{Chugainov}, P.~F. 1974, Izvestiya Ordena Trudovogo Krasnogo Znameni Krymskoj
  Astrofizicheskoj Observatorii, 52, 3

\bibitem[{{Churcher} {et~al.}(2011){Churcher}, {Wyatt}, \&
  {Smith}}]{Churcher11}
{Churcher}, L., {Wyatt}, M., \& {Smith}, R. 2011, \mnras, 410, 2

\bibitem[{{Close} {et~al.}(2003){Close}, {Siegler}, {Freed}, \&
  {Biller}}]{Close03}
{Close}, L.~M., {Siegler}, N., {Freed}, M., \& {Biller}, B. 2003, \apj, 587,
  407

\bibitem[{{Coates} {et~al.}(1980){Coates}, {Halprin}, {Sartori}, \&
  {Thompson}}]{Coates80}
{Coates}, D.~W., {Halprin}, L., {Sartori}, P., \& {Thompson}, K. 1980,
  Information Bulletin on Variable Stars, 1849

\bibitem[{{Coates} {et~al.}(1982){Coates}, {Innis}, {Moon}, \&
  {Thompson}}]{Coates82}
{Coates}, D.~W., {Innis}, J.~L., {Moon}, T.~T., \& {Thompson}, K. 1982,
  Information Bulletin on Variable Stars, 2248

\bibitem[{{Cutispoto}(1993)}]{Cutispoto93}
{Cutispoto}, G. 1993, \aaps, 102, 655

\bibitem[{{Cutispoto}(1996)}]{Cutispoto96}
{Cutispoto}, G. 1996, \aaps, 119, 281

\bibitem[{{Cutispoto}(1998{\natexlab{a}})}]{Cutispoto98b}
{Cutispoto}, G. 1998{\natexlab{a}}, \aaps, 127, 207

\bibitem[{{Cutispoto}(1998{\natexlab{b}})}]{Cutispoto98a}
{Cutispoto}, G. 1998{\natexlab{b}}, \aaps, 131, 321

\bibitem[{{Cutispoto} {et~al.}(2000){Cutispoto}, {Pastori}, {Guerrero},
  {Tagliaferri}, {Messina}, {Rodon{\`o}}, \& {de Medeiros}}]{Cutispoto00}
{Cutispoto}, G., {Pastori}, L., {Guerrero}, A., {et~al.} 2000, \aap, 364, 205

\bibitem[{{Cutispoto} {et~al.}(2002){Cutispoto}, {Pastori}, {Pasquini}, {de
  Medeiros}, {Tagliaferri}, \& {Andersen}}]{Cutispoto02}
{Cutispoto}, G., {Pastori}, L., {Pasquini}, L., {et~al.} 2002, \aap, 384, 491

\bibitem[{{Cutispoto} {et~al.}(1999){Cutispoto}, {Pastori}, {Tagliaferri},
  {Messina}, \& {Pallavicini}}]{Cutispoto99}
{Cutispoto}, G., {Pastori}, L., {Tagliaferri}, G., {Messina}, S., \&
  {Pallavicini}, R. 1999, \aaps, 138, 87

\bibitem[{{Cutri} {et~al.}(2003){Cutri}, {Skrutskie}, {van Dyk}, {Beichman},
  {Carpenter}, {Chester}, {Cambresy}, {Evans}, {Fowler}, {Gizis}, {Howard},
  {Huchra}, {Jarrett}, {Kopan}, {Kirkpatrick}, {Light}, {Marsh}, {McCallon},
  {Schneider}, {Stiening}, {Sykes}, {Weinberg}, {Wheaton}, {Wheelock}, \&
  {Zacarias}}]{Cutri03}
{Cutri}, R.~M., {Skrutskie}, M.~F., {van Dyk}, S., {et~al.} 2003, {2MASS All
  Sky Catalog of point sources.}

\bibitem[{{da Silva} {et~al.}(2009){da Silva}, {Torres}, {de La Reza}, {Quast},
  {Melo}, \& {Sterzik}}]{daSilva09}
{da Silva}, L., {Torres}, C.~A.~O., {de La Reza}, R., {et~al.} 2009, \aap, 508,
  833

\bibitem[{{Dal} \& {Evren}(2011)}]{Dal11}
{Dal}, H.~A. \& {Evren}, S. 2011, \pasj, 63, 427

\bibitem[{{de la Reza} \& {Pinz{\'o}n}(2004)}]{delaReza04}
{de la Reza}, R. \& {Pinz{\'o}n}, G. 2004, \aj, 128, 1812

\bibitem[{{de La Reza} {et~al.}(1986){de La Reza}, {Quast}, {Torres}, {Mayor},
  {Meylan}, \& {Llorente de Andres}}]{delaReza86}
{de La Reza}, R., {Quast}, G., {Torres}, C.~A.~O., {et~al.} 1986, in ESA
  Special Publication, Vol. 263, New Insights in Astrophysics. Eight Years of
  UV Astronomy with IUE, ed. E.~J. {Rolfe}

\bibitem[{{Deacon} {et~al.}(2016){Deacon}, {Schlieder}, \& {Murphy}}]{Deacon16}
{Deacon}, N.~R., {Schlieder}, J.~E., \& {Murphy}, S.~J. 2016, \mnras, 457, 3191

\bibitem[{{Delfosse} {et~al.}(1998){Delfosse}, {Forveille}, {Perrier}, \&
  {Mayor}}]{Delfosse98}
{Delfosse}, X., {Forveille}, T., {Perrier}, C., \& {Mayor}, M. 1998, \aap, 331,
  581

\bibitem[{{Delorme} {et~al.}(2012){Delorme}, {Lagrange}, {Chauvin}, {Bonavita},
  {Lacour}, {Bonnefoy}, {Ehrenreich}, \& {Beust}}]{Delorme12}
{Delorme}, P., {Lagrange}, A.~M., {Chauvin}, G., {et~al.} 2012, \aap, 539, A72

\bibitem[{{Desidera} {et~al.}(2015){Desidera}, {Covino}, {Messina}, {Carson},
  {Hagelberg}, {Schlieder}, {Biazzo}, {Alcal{\'a}}, {Chauvin}, {Vigan},
  {Beuzit}, {Bonavita}, {Bonnefoy}, {Delorme}, {D'Orazi}, {Esposito}, {Feldt},
  {Girardi}, {Gratton}, {Henning}, {Lagrange}, {Lanzafame}, {Launhardt},
  {Marmier}, {Melo}, {Meyer}, {Mouillet}, {Moutou}, {Segransan}, {Udry}, \&
  {Zaidi}}]{Desidera15}
{Desidera}, S., {Covino}, E., {Messina}, S., {et~al.} 2015, \aap, 573, A126

\bibitem[{{Domingo} {et~al.}(2010){Domingo}, {Guti{\'e}rrez-S{\'a}nchez},
  {R{\'{\i}}squez}, {Caballero-Garc{\'{\i}}a}, {Mas-Hesse}, \&
  {Solano}}]{Domingo10}
{Domingo}, A., {Guti{\'e}rrez-S{\'a}nchez}, R., {R{\'{\i}}squez}, D., {et~al.}
  2010, Astrophysics and Space Science Proceedings, 14, 493

\bibitem[{{Drake} {et~al.}(2009){Drake}, {Djorgovski}, {Mahabal}, {Beshore},
  {Larson}, {Graham}, {Williams}, {Christensen}, {Catelan}, {Boattini},
  {Gibbs}, {Hill}, \& {Kowalski}}]{Drake09}
{Drake}, A.~J., {Djorgovski}, S.~G., {Mahabal}, A., {et~al.} 2009, \apj, 696,
  870

\bibitem[{{Dunstone} {et~al.}(2008){Dunstone}, {Hussain}, {Collier Cameron},
  {Marsden}, {Jardine}, {Stempels}, {Ramirez Velez}, \& {Donati}}]{Dunstone08}
{Dunstone}, N.~J., {Hussain}, G.~A.~J., {Collier Cameron}, A., {et~al.} 2008,
  \mnras, 387, 481

\bibitem[{{Elliott} {et~al.}(2014){Elliott}, {Bayo}, {Melo}, {Torres},
  {Sterzik}, \& {Quast}}]{Elliott14}
{Elliott}, P., {Bayo}, A., {Melo}, C.~H.~F., {et~al.} 2014, \aap, 568, A26

\bibitem[{{Elliott} {et~al.}(2016){Elliott}, {Bayo}, {Melo}, {Torres},
  {Sterzik}, {Quast}, {Montes}, \& {Brahm}}]{Elliott16}
{Elliott}, P., {Bayo}, A., {Melo}, C.~H.~F., {et~al.} 2016, \aap, 590, A13

\bibitem[{{Elliott} {et~al.}(2015){Elliott}, {Hu{\'e}lamo}, {Bouy}, {Bayo},
  {Melo}, {Torres}, {Sterzik}, {Quast}, {Chauvin}, \& {Barrado}}]{Elliott15}
{Elliott}, P., {Hu{\'e}lamo}, N., {Bouy}, H., {et~al.} 2015, \aap, 580, A88

\bibitem[{{Favata} {et~al.}(1995){Favata}, {Barbera}, {Micela}, \&
  {Sciortino}}]{Favata95}
{Favata}, F., {Barbera}, M., {Micela}, G., \& {Sciortino}, S. 1995, \aap, 295,
  147

\bibitem[{{Feigelson} {et~al.}(2006){Feigelson}, {Lawson}, {Stark}, {Townsley},
  \& {Garmire}}]{Feigelson06}
{Feigelson}, E.~D., {Lawson}, W.~A., {Stark}, M., {Townsley}, L., \& {Garmire},
  G.~P. 2006, \aj, 131, 1730

\bibitem[{{Fern{\'a}ndez} {et~al.}(2008){Fern{\'a}ndez}, {Figueras}, \&
  {Torra}}]{Fernandez08}
{Fern{\'a}ndez}, D., {Figueras}, F., \& {Torra}, J. 2008, \aap, 480, 735

\bibitem[{{Frith} {et~al.}(2013){Frith}, {Pinfield}, {Jones}, {Barnes},
  {Pavlenko}, {Martin}, {Brown}, {Kuznetsov}, {Marocco}, {Tata}, \&
  {Cappetta}}]{Frith13}
{Frith}, J., {Pinfield}, D.~J., {Jones}, H.~R.~A., {et~al.} 2013, \mnras, 435,
  2161

\bibitem[{{Gahm} {et~al.}(1993){Gahm}, {Gullbring}, {Fischerstrom}, {Lindroos},
  \& {Loden}}]{Gahm93}
{Gahm}, G.~F., {Gullbring}, E., {Fischerstrom}, C., {Lindroos}, K.~P., \&
  {Loden}, K. 1993, \aaps, 100, 371

\bibitem[{{Gaidos} {et~al.}(2000){Gaidos}, {Henry}, \& {Henry}}]{Gaidos00}
{Gaidos}, E.~J., {Henry}, G.~W., \& {Henry}, S.~M. 2000, \aj, 120, 1006

\bibitem[{{Garc{\'{\i}}a-Alvarez} {et~al.}(2011){Garc{\'{\i}}a-Alvarez},
  {Lanza}, {Messina}, {Drake}, {van Wyk}, {Shobbrook}, {Butler}, {Kilkenny},
  {Doyle}, \& {Kashyap}}]{Garcia-Alvarez11}
{Garc{\'{\i}}a-Alvarez}, D., {Lanza}, A.~F., {Messina}, S., {et~al.} 2011,
  \aap, 533, A30

\bibitem[{{Gershberg} {et~al.}(1999){Gershberg}, {Katsova}, {Lovkaya},
  {Terebizh}, \& {Shakhovskaya}}]{Gershberg99}
{Gershberg}, R.~E., {Katsova}, M.~M., {Lovkaya}, M.~N., {Terebizh}, A.~V., \&
  {Shakhovskaya}, N.~I. 1999, \aaps, 139, 555

\bibitem[{{Gizis} {et~al.}(2000){Gizis}, {Monet}, {Reid}, {Kirkpatrick}, \&
  {Burgasser}}]{Gizis00}
{Gizis}, J.~E., {Monet}, D.~G., {Reid}, I.~N., {Kirkpatrick}, J.~D., \&
  {Burgasser}, A.~J. 2000, \mnras, 311, 385

\bibitem[{{Haakonsen} \& {Rutledge}(2009)}]{Haakonsen09}
{Haakonsen}, C.~B. \& {Rutledge}, R.~E. 2009, \apjs, 184, 138

\bibitem[{{Hartman} {et~al.}(2010){Hartman}, {Bakos}, {Kov{\'a}cs}, \&
  {Noyes}}]{Hartman10}
{Hartman}, J.~D., {Bakos}, G.~{\'A}., {Kov{\'a}cs}, G., \& {Noyes}, R.~W. 2010,
  \mnras, 408, 475

\bibitem[{{Hartmann} {et~al.}(1986){Hartmann}, {Hewett}, {Stahler}, \&
  {Mathieu}}]{Hartmann86}
{Hartmann}, L., {Hewett}, R., {Stahler}, S., \& {Mathieu}, R.~D. 1986, \apj,
  309, 275

\bibitem[{{Hebb} {et~al.}(2007){Hebb}, {Petro}, {Ford}, {Ardila}, {Toledo},
  {Minniti}, {Golimowski}, \& {Clampin}}]{Hebb07}
{Hebb}, L., {Petro}, L., {Ford}, H.~C., {et~al.} 2007, \mnras, 379, 63

\bibitem[{{He{\l}miniak} {et~al.}(2012){He{\l}miniak}, {Konacki},
  {R{\'o}{\.Z}yczka}, {Ka{\l}u{\.Z}ny}, {Ratajczak}, {Borkowski}, {Sybilski},
  {Muterspaugh}, {Reichart}, {Ivarsen}, {Haislip}, {Crain}, {Foster},
  {Nysewander}, \& {LaCluyze}}]{Helminiak12}
{He{\l}miniak}, K.~G., {Konacki}, M., {R{\'o}{\.Z}yczka}, M., {et~al.} 2012,
  \mnras, 425, 1245

\bibitem[{{Herbst} {et~al.}(2002){Herbst}, {Bailer-Jones}, {Mundt},
  {Meisenheimer}, \& {Wackermann}}]{Herbst02}
{Herbst}, W., {Bailer-Jones}, C.~A.~L., {Mundt}, R., {Meisenheimer}, K., \&
  {Wackermann}, R. 2002, \aap, 396, 513

\bibitem[{{Hoffman} {et~al.}(2009){Hoffman}, {Harrison}, \&
  {McNamara}}]{Hoffman09}
{Hoffman}, D.~I., {Harrison}, T.~E., \& {McNamara}, B.~J. 2009, \aj, 138, 466

\bibitem[{{Horne} \& {Baliunas}(1986)}]{Horne86}
{Horne}, J.~H. \& {Baliunas}, S.~L. 1986, \apj, 302, 757

\bibitem[{{Hosey} {et~al.}(2015){Hosey}, {Henry}, {Jao}, {Dieterich},
  {Winters}, {Lurie}, {Riedel}, \& {Subasavage}}]{Hosey15}
{Hosey}, A.~D., {Henry}, T.~J., {Jao}, W.-C., {et~al.} 2015, \aj, 150, 6

\bibitem[{{Innis} {et~al.}(2007){Innis}, {Coates}, \& {Kaye}}]{Innis07}
{Innis}, J.~L., {Coates}, D.~W., \& {Kaye}, T.~G. 2007, Peremennye Zvezdy, 27

\bibitem[{{Innis} {et~al.}(1990){Innis}, {Coates}, {Thompson}, \& {Lloyd
  Evans}}]{Innis90}
{Innis}, J.~L., {Coates}, D.~W., {Thompson}, K., \& {Lloyd Evans}, T. 1990,
  \mnras, 242, 306

\bibitem[{{Janson} {et~al.}(2014){Janson}, {Bergfors}, {Brandner}, {Bonnefoy},
  {Schlieder}, {K{\"o}hler}, {Hormuth}, {Henning}, \& {Hippler}}]{Janson14}
{Janson}, M., {Bergfors}, C., {Brandner}, W., {et~al.} 2014, \apjs, 214, 17

\bibitem[{{Janson} {et~al.}(2012){Janson}, {Hormuth}, {Bergfors}, {Brandner},
  {Hippler}, {Daemgen}, {Kudryavtseva}, {Schmalzl}, {Schnupp}, \&
  {Henning}}]{Janson12}
{Janson}, M., {Hormuth}, F., {Bergfors}, C., {et~al.} 2012, \apj, 754, 44

\bibitem[{{Jao} {et~al.}(2003){Jao}, {Henry}, {Subasavage}, {Bean}, {Costa},
  {Ianna}, \& {M{\'e}ndez}}]{Jao03}
{Jao}, W.-C., {Henry}, T.~J., {Subasavage}, J.~P., {et~al.} 2003, \aj, 125, 332

\bibitem[{{J{\"a}rvinen} {et~al.}(2015){J{\"a}rvinen}, {Arlt}, {Hackman},
  {Marsden}, {K{\"u}ker}, {Ilyin}, {Berdyugina}, {Strassmeier}, \&
  {Waite}}]{Jarvinen15}
{J{\"a}rvinen}, S.~P., {Arlt}, R., {Hackman}, T., {et~al.} 2015, \aap, 574, A25

\bibitem[{{Jayawardhana} {et~al.}(2006){Jayawardhana}, {Coffey}, {Scholz},
  {Brandeker}, \& {van Kerkwijk}}]{Jayawardhana06}
{Jayawardhana}, R., {Coffey}, J., {Scholz}, A., {Brandeker}, A., \& {van
  Kerkwijk}, M.~H. 2006, \apj, 648, 1206

\bibitem[{{Jensen} \& {Mathieu}(1997)}]{Jensen97}
{Jensen}, E.~L.~N. \& {Mathieu}, R.~D. 1997, \aj, 114, 301

\bibitem[{{Kaisler} {et~al.}(2004){Kaisler}, {Zuckerman}, {Song}, {Macintosh},
  {Weinberger}, {Becklin}, {Konopacky}, \& {Patience}}]{Kaisler04}
{Kaisler}, D., {Zuckerman}, B., {Song}, I., {et~al.} 2004, \aap, 414, 175

\bibitem[{{Kalas} {et~al.}(2004){Kalas}, {Liu}, \& {Matthews}}]{Kalas04}
{Kalas}, P., {Liu}, M.~C., \& {Matthews}, B.~C. 2004, Science, 303, 1990

\bibitem[{{Kasper} {et~al.}(2007){Kasper}, {Apai}, {Janson}, \&
  {Brandner}}]{Kasper07}
{Kasper}, M., {Apai}, D., {Janson}, M., \& {Brandner}, W. 2007, \aap, 472, 321

\bibitem[{{Kastner} {et~al.}(2011){Kastner}, {Sacco}, {Montez}, {Huenemoerder},
  {Shi}, {Alecian}, {Argiroffi}, {Audard}, {Bouvier}, {Damiani}, {Donati},
  {Gregory}, {G{\"u}del}, {Hussain}, {Maggio}, \& {Montmerle}}]{Kastner11}
{Kastner}, J.~H., {Sacco}, G.~G., {Montez}, R., {et~al.} 2011, \apjl, 740, L17

\bibitem[{{Kazarovets} {et~al.}(1999){Kazarovets}, {Samus}, {Durlevich},
  {Frolov}, {Antipin}, {Kireeva}, \& {Pastukhova}}]{Kazarovets99}
{Kazarovets}, E.~V., {Samus}, N.~N., {Durlevich}, O.~V., {et~al.} 1999,
  Information Bulletin on Variable Stars, 4659

\bibitem[{{Kiraga}(2012)}]{Kiraga12}
{Kiraga}, M. 2012, \actaa, 62, 67

\bibitem[{{Kiraga} \& {Stepien}(2007)}]{Kiraga07}
{Kiraga}, M. \& {Stepien}, K. 2007, \actaa, 57, 149

\bibitem[{{Kiss} {et~al.}(2011){Kiss}, {Mo{\'o}r}, {Szalai}, {Kov{\'a}cs},
  {Bayliss}, {Gilmore}, {Bienaym{\'e}}, {Binney}, {Bland-Hawthorn}, {Campbell},
  {Freeman}, {Fulbright}, {Gibson}, {Grebel}, {Helmi}, {Munari}, {Navarro},
  {Parker}, {Reid}, {Seabroke}, {Siebert}, {Siviero}, {Steinmetz}, {Watson},
  {Williams}, {Wyse}, \& {Zwitter}}]{Kiss11}
{Kiss}, L.~L., {Mo{\'o}r}, A., {Szalai}, T., {et~al.} 2011, \mnras, 411, 117

\bibitem[{{Klutsch} {et~al.}(2014){Klutsch}, {Freire Ferrero}, {Guillout},
  {Frasca}, {Marilli}, \& {Montes}}]{Klutsch14}
{Klutsch}, A., {Freire Ferrero}, R., {Guillout}, P., {et~al.} 2014, \aap, 567,
  A52

\bibitem[{{Koen} \& {Eyer}(2002)}]{Koen02}
{Koen}, C. \& {Eyer}, L. 2002, \mnras, 331, 45

\bibitem[{{Kraus} {et~al.}(2014){Kraus}, {Shkolnik}, {Allers}, \&
  {Liu}}]{Kraus14}
{Kraus}, A.~L., {Shkolnik}, E.~L., {Allers}, K.~N., \& {Liu}, M.~C. 2014, \aj,
  147, 146

\bibitem[{{Lamm} {et~al.}(2004){Lamm}, {Bailer-Jones}, {Mundt}, {Herbst}, \&
  {Scholz}}]{Lamm04}
{Lamm}, M.~H., {Bailer-Jones}, C.~A.~L., {Mundt}, R., {Herbst}, W., \&
  {Scholz}, A. 2004, \aap, 417, 557

\bibitem[{{L{\'e}pine} \& {Simon}(2009)}]{Lepine09}
{L{\'e}pine}, S. \& {Simon}, M. 2009, \aj, 137, 3632

\bibitem[{{Lloyd Evans} \& {Koen}(1987)}]{LloydEvans87}
{Lloyd Evans}, T. \& {Koen}, M.~C.~J. 1987, South African Astronomical
  Observatory Circular, 11, 21

\bibitem[{{L{\'o}pez-Santiago} {et~al.}(2010){L{\'o}pez-Santiago}, {Montes},
  {G{\'a}lvez-Ortiz}, {Crespo-Chac{\'o}n}, {Mart{\'{\i}}nez-Arn{\'a}iz},
  {Fern{\'a}ndez-Figueroa}, {de Castro}, \& {Cornide}}]{Lopez-Santiago10}
{L{\'o}pez-Santiago}, J., {Montes}, D., {G{\'a}lvez-Ortiz}, M.~C., {et~al.}
  2010, \aap, 514, A97

\bibitem[{{Maire} {et~al.}(2016){Maire}, {Bonnefoy}, {Ginski}, {Vigan},
  {Messina}, {Mesa}, {Galicher}, {Gratton}, {Desidera}, {Kopytova}, {Millward},
  {Thalmann}, {Claudi}, {Ehrenreich}, {Zurlo}, {Chauvin}, {Antichi},
  {Baruffolo}, {Bazzon}, {Beuzit}, {Blanchard}, {Boccaletti}, {de Boer},
  {Carle}, {Cascone}, {Costille}, {De Caprio}, {Delboulb{\'e}}, {Dohlen},
  {Dominik}, {Feldt}, {Fusco}, {Girard}, {Giro}, {Gisler}, {Gluck}, {Gry},
  {Henning}, {Hubin}, {Hugot}, {Jaquet}, {Kasper}, {Lagrange}, {Langlois}, {Le
  Mignant}, {Llored}, {Madec}, {Martinez}, {Mawet}, {Milli},
  {M{\"o}ller-Nilsson}, {Mouillet}, {Moulin}, {Moutou}, {Orign{\'e}}, {Pavlov},
  {Petit}, {Pragt}, {Puget}, {Ramos}, {Rochat}, {Roelfsema}, {Salasnich},
  {Sauvage}, {Schmid}, {Turatto}, {Udry}, {Vakili}, {Wahhaj}, {Weber}, \&
  {Wildi}}]{Maire16}
{Maire}, A.-L., {Bonnefoy}, M., {Ginski}, C., {et~al.} 2016, \aap, 587, A56

\bibitem[{{Makarov}(2007)}]{Makarov07}
{Makarov}, V.~V. 2007, \apjs, 169, 105

\bibitem[{{Maldonado} {et~al.}(2015){Maldonado}, {Affer}, {Micela},
  {Scandariato}, {Damasso}, {Stelzer}, {Barbieri}, {Bedin}, {Biazzo},
  {Bignamini}, {Borsa}, {Claudi}, {Covino}, {Desidera}, {Esposito}, {Gratton},
  {Gonz{\'a}lez Hern{\'a}ndez}, {Lanza}, {Maggio}, {Molinari}, {Pagano},
  {Perger}, {Pillitteri}, {Piotto}, {Poretti}, {Prisinzano}, {Rebolo}, {Ribas},
  {Shkolnik}, {Southworth}, {Sozzetti}, \& {Su{\'a}rez
  Mascare{\~n}o}}]{Maldonado15}
{Maldonado}, J., {Affer}, L., {Micela}, G., {et~al.} 2015, \aap, 577, A132

\bibitem[{{Malo} {et~al.}(2014{\natexlab{a}}){Malo}, {Artigau}, {Doyon},
  {Lafreni{\`e}re}, {Albert}, \& {Gagn{\'e}}}]{Malo14a}
{Malo}, L., {Artigau}, {\'E}., {Doyon}, R., {et~al.} 2014{\natexlab{a}}, \apj,
  788, 81

\bibitem[{{Malo} {et~al.}(2014{\natexlab{b}}){Malo}, {Doyon}, {Feiden},
  {Albert}, {Lafreni{\`e}re}, {Artigau}, {Gagn{\'e}}, \& {Riedel}}]{Malo14b}
{Malo}, L., {Doyon}, R., {Feiden}, G.~A., {et~al.} 2014{\natexlab{b}}, \apj,
  792, 37

\bibitem[{{Malo} {et~al.}(2013){Malo}, {Doyon}, {Lafreni{\`e}re}, {Artigau},
  {Gagn{\'e}}, {Baron}, \& {Riedel}}]{Malo13}
{Malo}, L., {Doyon}, R., {Lafreni{\`e}re}, D., {et~al.} 2013, \apj, 762, 88

\bibitem[{{Mamajek} \& {Bell}(2014)}]{Mamajek14}
{Mamajek}, E.~E. \& {Bell}, C.~P.~M. 2014, \mnras, 445, 2169

\bibitem[{{Mason} {et~al.}(2001){Mason}, {Wycoff}, {Hartkopf}, {Douglass}, \&
  {Worley}}]{Mason01}
{Mason}, B.~D., {Wycoff}, G.~L., {Hartkopf}, W.~I., {Douglass}, G.~G., \&
  {Worley}, C.~E. 2001, \aj, 122, 3466

\bibitem[{{McCarthy} {et~al.}(2001){McCarthy}, {Zuckerman}, \&
  {Becklin}}]{McCarthy01}
{McCarthy}, C., {Zuckerman}, B., \& {Becklin}, E.~E. 2001, \aj, 121, 3259

\bibitem[{{McCarthy} \& {White}(2012)}]{McCarthy12}
{McCarthy}, K. \& {White}, R.~J. 2012, \aj, 143, 134

\bibitem[{{Mekkaden} \& {Sinachopoulos}(1991)}]{Mekkaden91}
{Mekkaden}, M.~V. \& {Sinachopoulos}, D. 1991, in Astronomische Gesellschaft
  Abstract Series, Vol.~6, Astronomische Gesellschaft Abstract Series, ed.
  G.~{Klare}, 63

\bibitem[{{Mentuch} {et~al.}(2008){Mentuch}, {Brandeker}, {van Kerkwijk},
  {Jayawardhana}, \& {Hauschildt}}]{Mentuch08}
{Mentuch}, E., {Brandeker}, A., {van Kerkwijk}, M.~H., {Jayawardhana}, R., \&
  {Hauschildt}, P.~H. 2008, \apj, 689, 1127

\bibitem[{{Messina} {et~al.}(2011){Messina}, {Desidera}, {Lanzafame},
  {Turatto}, \& {Guinan}}]{Messina11}
{Messina}, S., {Desidera}, S., {Lanzafame}, A.~C., {Turatto}, M., \& {Guinan},
  E.~F. 2011, \aap, 532, A10

\bibitem[{{Messina} {et~al.}(2010){Messina}, {Desidera}, {Turatto},
  {Lanzafame}, \& {Guinan}}]{Messina10}
{Messina}, S., {Desidera}, S., {Turatto}, M., {Lanzafame}, A.~C., \& {Guinan},
  E.~F. 2010, \aap, 520, A15

\bibitem[{{Messina} {et~al.}(2015{\natexlab{a}}){Messina}, {Hentunen}, \&
  {Zambelli}}]{Messina15a}
{Messina}, S., {Hentunen}, V.-P., \& {Zambelli}, R. 2015{\natexlab{a}},
  Information Bulletin on Variable Stars, 6145

\bibitem[{{Messina} {et~al.}(2016{\natexlab{a}}){Messina}, {Lanzafame},
  {Feiden}, {Millward}, {Desidera}, {Buccino}, {Curtis}, {Jofre'}, {Kehusmaa},
  {Medhi}, {Monard}, \& {Petrucci}}]{Messina16a}
{Messina}, S., {Lanzafame}, A.~C., {Feiden}, G.~A., {et~al.}
  2016{\natexlab{a}}, ArXiv e-prints: 1607.06634

\bibitem[{{Messina} {et~al.}(2016{\natexlab{b}}){Messina}, {Leto}, \&
  {Pagano}}]{2016c}
{Messina}, S., {Leto}, G., \& {Pagano}, I. 2016{\natexlab{b}}, \apss, 361, 291

\bibitem[{{Messina} {et~al.}(2015{\natexlab{b}}){Messina}, {Millward}, \&
  {Bradstreet}}]{Messina15c}
{Messina}, S., {Millward}, M., \& {Bradstreet}, D.~H. 2015{\natexlab{b}}, \na,
  37, 105

\bibitem[{{Messina} {et~al.}(2014){Messina}, {Monard}, {Biazzo}, {Melo}, \&
  {Frasca}}]{Messina14}
{Messina}, S., {Monard}, B., {Biazzo}, K., {Melo}, C.~H.~F., \& {Frasca}, A.
  2014, \aap, 570, A19

\bibitem[{{Messina} {et~al.}(2015{\natexlab{c}}){Messina}, {Muro Serrano},
  {Artemenko}, {Bailey}, {Savushkin}, \& {Nelson}}]{Messina15b}
{Messina}, S., {Muro Serrano}, M., {Artemenko}, S., {et~al.}
  2015{\natexlab{c}}, \apss, 360, 17

\bibitem[{{Messina} {et~al.}(2016{\natexlab{c}}){Messina}, {Naves}, \&
  {Medhi}}]{2016b}
{Messina}, S., {Naves}, R., \& {Medhi}, B.~J. 2016{\natexlab{c}}, \na, 48, 5

\bibitem[{{Messina} {et~al.}(2016{\natexlab{d}}){Messina}, {Santallo}, {Tan},
  \& et~al.}]{2016d}
{Messina}, S., {Santallo}, R., {Tan}, T.-G., \& et~al. 2016{\natexlab{d}},
  \aap\,\,\,\, submitted

\bibitem[{{Mochnacki} {et~al.}(2002){Mochnacki}, {Gladders}, {Thomson}, {Lu},
  {Ehlers}, {Guler}, {Hussain}, {Kameda}, {King}, {Mitchell}, {Rowe},
  {Schindler}, \& {Scott}}]{Mochnacki02}
{Mochnacki}, S.~W., {Gladders}, M.~D., {Thomson}, J.~R., {et~al.} 2002, \aj,
  124, 2868

\bibitem[{{Montet} {et~al.}(2015){Montet}, {Bowler}, {Shkolnik}, {Deck},
  {Wang}, {Horch}, {Liu}, {Hillenbrand}, {Kraus}, \& {Charbonneau}}]{Montet15}
{Montet}, B.~T., {Bowler}, B.~P., {Shkolnik}, E.~L., {et~al.} 2015, \apjl, 813,
  L11

\bibitem[{{Mo{\'o}r} {et~al.}(2006){Mo{\'o}r}, {{\'A}brah{\'a}m}, {Derekas},
  {Kiss}, {Kiss}, {Apai}, {Grady}, \& {Henning}}]{Moor06}
{Mo{\'o}r}, A., {{\'A}brah{\'a}m}, P., {Derekas}, A., {et~al.} 2006, \apj, 644,
  525

\bibitem[{{Mo{\'o}r} {et~al.}(2013){Mo{\'o}r}, {Szab{\'o}}, {Kiss}, {Kiss},
  {{\'A}brah{\'a}m}, {Szul{\'a}gyi}, {K{\'o}sp{\'a}l}, \& {Szalai}}]{Moor13}
{Mo{\'o}r}, A., {Szab{\'o}}, G.~M., {Kiss}, L.~L., {et~al.} 2013, \mnras, 435,
  1376

\bibitem[{{Morlet} {et~al.}(2000){Morlet}, {Salaman}, \& {Gili}}]{Morlet00}
{Morlet}, G., {Salaman}, M., \& {Gili}, R. 2000, \aaps, 145, 67

\bibitem[{{Mugrauer} {et~al.}(2010){Mugrauer}, {Vogt}, {Neuh{\"a}user}, \&
  {Schmidt}}]{Mugrauer10}
{Mugrauer}, M., {Vogt}, N., {Neuh{\"a}user}, R., \& {Schmidt}, T.~O.~B. 2010,
  \aap, 523, L1

\bibitem[{{Nakajima} {et~al.}(2010){Nakajima}, {Morino}, \&
  {Fukagawa}}]{Nakajima10}
{Nakajima}, T., {Morino}, J.-I., \& {Fukagawa}, M. 2010, \aj, 140, 713

\bibitem[{{Nataf} {et~al.}(2010){Nataf}, {Stanek}, \& {Bakos}}]{Nataf10}
{Nataf}, D.~M., {Stanek}, K.~Z., \& {Bakos}, G.~{\'A}. 2010, \actaa, 60, 261

\bibitem[{{Naz{\'e}} {et~al.}(2010){Naz{\'e}}, {Rauw}, \& {Ud-Doula}}]{Naze10}
{Naz{\'e}}, Y., {Rauw}, G., \& {Ud-Doula}, A. 2010, \aap, 510, A59

\bibitem[{{Neuh{\"a}user} {et~al.}(2002){Neuh{\"a}user}, {Guenther},
  {Mugrauer}, {Ott}, \& {Eckart}}]{Neuhauser02}
{Neuh{\"a}user}, R., {Guenther}, E., {Mugrauer}, M., {Ott}, T., \& {Eckart}, A.
  2002, \aap, 395, 877

\bibitem[{{Neuh{\"a}user} {et~al.}(2003){Neuh{\"a}user}, {Guenther}, {Alves},
  {Hu{\'e}lamo}, {Ott}, \& {Eckart}}]{Neuhauser03}
{Neuh{\"a}user}, R., {Guenther}, E.~W., {Alves}, J., {et~al.} 2003,
  Astronomische Nachrichten, 324, 535

\bibitem[{{Newton} {et~al.}(2014){Newton}, {Charbonneau}, {Irwin},
  {Berta-Thompson}, {Rojas-Ayala}, {Covey}, \& {Lloyd}}]{Newton14}
{Newton}, E.~R., {Charbonneau}, D., {Irwin}, J., {et~al.} 2014, \aj, 147, 20

\bibitem[{{Nordstr{\"o}m} {et~al.}(2004){Nordstr{\"o}m}, {Mayor}, {Andersen},
  {Holmberg}, {Pont}, {J{\o}rgensen}, {Olsen}, {Udry}, \&
  {Mowlavi}}]{Nordstrom04}
{Nordstr{\"o}m}, B., {Mayor}, M., {Andersen}, J., {et~al.} 2004, \aap, 418, 989

\bibitem[{{Norton} {et~al.}(2007){Norton}, {Wheatley}, {West}, {Haswell},
  {Street}, {Collier Cameron}, {Christian}, {Clarkson}, {Enoch}, {Gallaway},
  {Hellier}, {Horne}, {Irwin}, {Kane}, {Lister}, {Nicholas}, {Parley},
  {Pollacco}, {Ryans}, {Skillen}, \& {Wilson}}]{Norton07}
{Norton}, A.~J., {Wheatley}, P.~J., {West}, R.~G., {et~al.} 2007, \aap, 467,
  785

\bibitem[{{Ortega} {et~al.}(2009){Ortega}, {Jilinski}, {de la Reza}, \&
  {Bazzanella}}]{Ortega09}
{Ortega}, V.~G., {Jilinski}, E., {de la Reza}, R., \& {Bazzanella}, B. 2009,
  \aj, 137, 3922

\bibitem[{{Parihar} {et~al.}(2009){Parihar}, {Messina}, {Distefano},
  {Shantikumar}, \& {Medhi}}]{Parihar09}
{Parihar}, P., {Messina}, S., {Distefano}, E., {Shantikumar}, N.~S., \&
  {Medhi}, B.~J. 2009, \mnras, 400, 603

\bibitem[{{Pasquini} {et~al.}(1991){Pasquini}, {Cutispoto}, {Gratton}, \&
  {Mayor}}]{Pasquini91}
{Pasquini}, L., {Cutispoto}, G., {Gratton}, R., \& {Mayor}, M. 1991, \aap, 248,
  72

\bibitem[{{Patel} {et~al.}(2014){Patel}, {Metchev}, \& {Heinze}}]{Patel14}
{Patel}, R.~I., {Metchev}, S.~A., \& {Heinze}, A. 2014, \apjs, 212, 10

\bibitem[{{Pecaut} \& {Mamajek}(2013)}]{Pecaut13}
{Pecaut}, M.~J. \& {Mamajek}, E.~E. 2013, \apjs, 208, 9

\bibitem[{{Petrucci} {et~al.}(2013){Petrucci}, {Jofr{\'e}}, {Schwartz},
  {C{\'u}neo}, {Mart{\'{\i}}nez}, {G{\'o}mez}, {Buccino}, \&
  {Mauas}}]{Petrucci13}
{Petrucci}, R., {Jofr{\'e}}, E., {Schwartz}, M., {et~al.} 2013, \apjl, 779, L23

\bibitem[{{Pojmanski}(1997)}]{Pojmanski97}
{Pojmanski}, G. 1997, \actaa, 47, 467

\bibitem[{{Pojmanski}(2002)}]{Pojmanski02}
{Pojmanski}, G. 2002, \actaa, 52, 397

\bibitem[{{Press} {et~al.}(2002){Press}, {Teukolsky}, {Vetterling}, \&
  {Flannery}}]{Press02}
{Press}, W.~H., {Teukolsky}, S.~A., {Vetterling}, W.~T., \& {Flannery}, B.~P.
  2002, {Numerical recipes in C++ : the art of scientific computing}

\bibitem[{{Quast} {et~al.}(2000){Quast}, {Torres}, {de La Reza}, {da Silva}, \&
  {Mayor}}]{Quast00}
{Quast}, G.~R., {Torres}, C.~A.~O., {de La Reza}, R., {da Silva}, L., \&
  {Mayor}, M. 2000, in IAU Symposium, Vol. 200, IAU Symposium, 28

\bibitem[{{Rameau} {et~al.}(2013){Rameau}, {Chauvin}, {Lagrange}, {Klahr},
  {Bonnefoy}, {Mordasini}, {Bonavita}, {Desidera}, {Dumas}, \&
  {Girard}}]{Rameau13}
{Rameau}, J., {Chauvin}, G., {Lagrange}, A.-M., {et~al.} 2013, \aap, 553, A60

\bibitem[{{Randich} {et~al.}(1993){Randich}, {Gratton}, \&
  {Pallavicini}}]{Randich93}
{Randich}, S., {Gratton}, R., \& {Pallavicini}, R. 1993, \aap, 273, 194

\bibitem[{{Rebull} {et~al.}(2008){Rebull}, {Stapelfeldt}, {Werner}, {Mannings},
  {Chen}, {Stauffer}, {Smith}, {Song}, {Hines}, \& {Low}}]{Rebull08}
{Rebull}, L.~M., {Stapelfeldt}, K.~R., {Werner}, M.~W., {et~al.} 2008, \apj,
  681, 1484

\bibitem[{{Reid} {et~al.}(2007){Reid}, {Cruz}, \& {Allen}}]{Reid07}
{Reid}, I.~N., {Cruz}, K.~L., \& {Allen}, P.~R. 2007, \aj, 133, 2825

\bibitem[{{Reid} {et~al.}(1995){Reid}, {Hawley}, \& {Gizis}}]{Reid95}
{Reid}, I.~N., {Hawley}, S.~L., \& {Gizis}, J.~E. 1995, \aj, 110, 1838

\bibitem[{{Reid} {et~al.}(2002){Reid}, {Kirkpatrick}, {Liebert}, {Gizis},
  {Dahn}, \& {Monet}}]{Reid02}
{Reid}, I.~N., {Kirkpatrick}, J.~D., {Liebert}, J., {et~al.} 2002, \aj, 124,
  519

\bibitem[{{Reiners} {et~al.}(2012){Reiners}, {Joshi}, \& {Goldman}}]{Reiners12}
{Reiners}, A., {Joshi}, N., \& {Goldman}, B. 2012, \aj, 143, 93

\bibitem[{{Rhee} {et~al.}(2007){Rhee}, {Song}, {Zuckerman}, \&
  {McElwain}}]{Rhee07}
{Rhee}, J.~H., {Song}, I., {Zuckerman}, B., \& {McElwain}, M. 2007, \apj, 660,
  1556

\bibitem[{{Riaz} {et~al.}(2006){Riaz}, {Gizis}, \& {Harvin}}]{Riaz06}
{Riaz}, B., {Gizis}, J.~E., \& {Harvin}, J. 2006, \aj, 132, 866

\bibitem[{{Ribas}(2003)}]{Ribas03}
{Ribas}, I. 2003, \aap, 400, 297

\bibitem[{{Riedel}(2012)}]{Riedel12}
{Riedel}, A.~R. 2012, PhD thesis, Georgia State University

\bibitem[{{Riedel} {et~al.}(2014){Riedel}, {Finch}, {Henry}, {Subasavage},
  {Jao}, {Malo}, {Rodriguez}, {White}, {Gies}, {Dieterich}, {Winters},
  {Davison}, {Nelan}, {Blunt}, {Cruz}, {Rice}, \& {Ianna}}]{Riedel14}
{Riedel}, A.~R., {Finch}, C.~T., {Henry}, T.~J., {et~al.} 2014, \aj, 147, 85

\bibitem[{{Riviere-Marichalar} {et~al.}(2014){Riviere-Marichalar}, {Barrado},
  {Montesinos}, {Duch{\^e}ne}, {Bouy}, {Pinte}, {Menard}, {Donaldson}, {Eiroa},
  {Krivov}, {Kamp}, {Mendigut{\'{\i}}a}, {Dent}, \&
  {Lillo-Box}}]{Riviere-Marichalar14}
{Riviere-Marichalar}, P., {Barrado}, D., {Montesinos}, B., {et~al.} 2014, \aap,
  565, A68

\bibitem[{{Roberts} {et~al.}(1987){Roberts}, {Lehar}, \& {Dreher}}]{Roberts87}
{Roberts}, D.~H., {Lehar}, J., \& {Dreher}, J.~W. 1987, \aj, 93, 968

\bibitem[{{Rodriguez} {et~al.}(2010){Rodriguez}, {Kastner}, {Wilner}, \&
  {Qi}}]{Rodriguez10}
{Rodriguez}, D.~R., {Kastner}, J.~H., {Wilner}, D., \& {Qi}, C. 2010, \apj,
  720, 1684

\bibitem[{{Rodriguez} \& {Zuckerman}(2012)}]{Rodriguez12}
{Rodriguez}, D.~R. \& {Zuckerman}, B. 2012, \apj, 745, 147

\bibitem[{{Royer} {et~al.}(2002){Royer}, {Grenier}, {Baylac}, {G{\'o}mez}, \&
  {Zorec}}]{Royer02}
{Royer}, F., {Grenier}, S., {Baylac}, M.-O., {G{\'o}mez}, A.~E., \& {Zorec}, J.
  2002, \aap, 393, 897

\bibitem[{{Scargle}(1982)}]{Scargle82}
{Scargle}, J.~D. 1982, \apj, 263, 835

\bibitem[{{Schlieder} {et~al.}(2010){Schlieder}, {L{\'e}pine}, \&
  {Simon}}]{Schlieder10}
{Schlieder}, J.~E., {L{\'e}pine}, S., \& {Simon}, M. 2010, \aj, 140, 119

\bibitem[{{Schlieder} {et~al.}(2012){Schlieder}, {L{\'e}pine}, \&
  {Simon}}]{Schlieder12}
{Schlieder}, J.~E., {L{\'e}pine}, S., \& {Simon}, M. 2012, \aj, 143, 80

\bibitem[{{Scholz} {et~al.}(2007){Scholz}, {Coffey}, {Brandeker}, \&
  {Jayawardhana}}]{Scholz07}
{Scholz}, A., {Coffey}, J., {Brandeker}, A., \& {Jayawardhana}, R. 2007, \apj,
  662, 1254

\bibitem[{{Schr{\"o}der} {et~al.}(2009){Schr{\"o}der}, {Reiners}, \&
  {Schmitt}}]{Schroder09}
{Schr{\"o}der}, C., {Reiners}, A., \& {Schmitt}, J.~H.~M.~M. 2009, \aap, 493,
  1099

\bibitem[{{Shkolnik} {et~al.}(2009){Shkolnik}, {Liu}, \& {Reid}}]{Shkolnik09}
{Shkolnik}, E., {Liu}, M.~C., \& {Reid}, I.~N. 2009, \apj, 699, 649

\bibitem[{{Shkolnik} {et~al.}(2012){Shkolnik}, {Anglada-Escud{\'e}}, {Liu},
  {Bowler}, {Weinberger}, {Boss}, {Reid}, \& {Tamura}}]{Shkolnik12}
{Shkolnik}, E.~L., {Anglada-Escud{\'e}}, G., {Liu}, M.~C., {et~al.} 2012, \apj,
  758, 56

\bibitem[{{Shulyak} {et~al.}(2011){Shulyak}, {Seifahrt}, {Reiners},
  {Kochukhov}, \& {Piskunov}}]{Shulyak11}
{Shulyak}, D., {Seifahrt}, A., {Reiners}, A., {Kochukhov}, O., \& {Piskunov},
  N. 2011, \mnras, 418, 2548

\bibitem[{{Siebert} {et~al.}(2011){Siebert}, {Williams}, {Siviero}, {Reid},
  {Boeche}, {Steinmetz}, {Fulbright}, {Munari}, {Zwitter}, {Watson}, {Wyse},
  {de Jong}, {Enke}, {Anguiano}, {Burton}, {Cass}, {Fiegert}, {Hartley},
  {Ritter}, {Russel}, {Stupar}, {Bienaym{\'e}}, {Freeman}, {Gilmore}, {Grebel},
  {Helmi}, {Navarro}, {Binney}, {Bland-Hawthorn}, {Campbell}, {Famaey},
  {Gerhard}, {Gibson}, {Matijevi{\v c}}, {Parker}, {Seabroke}, {Sharma},
  {Smith}, \& {Wylie-de Boer}}]{Siebert11}
{Siebert}, A., {Williams}, M.~E.~K., {Siviero}, A., {et~al.} 2011, \aj, 141,
  187

\bibitem[{{Siess} {et~al.}(2000){Siess}, {Dufour}, \& {Forestini}}]{Siess00}
{Siess}, L., {Dufour}, E., \& {Forestini}, M. 2000, \aap, 358, 593

\bibitem[{{Soderblom} {et~al.}(1998){Soderblom}, {King}, \&
  {Henry}}]{Soderblom98}
{Soderblom}, D.~R., {King}, J.~R., \& {Henry}, T.~J. 1998, \aj, 116, 396

\bibitem[{{Song} {et~al.}(2003){Song}, {Zuckerman}, \& {Bessell}}]{Song03}
{Song}, I., {Zuckerman}, B., \& {Bessell}, M.~S. 2003, \apj, 599, 342

\bibitem[{{Song} {et~al.}(2012){Song}, {Zuckerman}, \& {Bessell}}]{Song12}
{Song}, I., {Zuckerman}, B., \& {Bessell}, M.~S. 2012, \aj, 144, 8

\bibitem[{{Stempels} \& {Gahm}(2004)}]{Stempels04}
{Stempels}, H.~C. \& {Gahm}, G.~F. 2004, \aap, 421, 1159

\bibitem[{{Stephenson}(1986)}]{Stephenson86}
{Stephenson}, C.~B. 1986, \aj, 91, 144

\bibitem[{{Teixeira} {et~al.}(2009){Teixeira}, {Ducourant}, {Chauvin},
  {Krone-Martins}, {Bonnefoy}, \& {Song}}]{Teixeira09}
{Teixeira}, R., {Ducourant}, C., {Chauvin}, G., {et~al.} 2009, \aap, 503, 281

\bibitem[{{Thalmann} {et~al.}(2014){Thalmann}, {Desidera}, {Bonavita},
  {Janson}, {Usuda}, {Henning}, {K{\"o}hler}, {Carson}, {Boccaletti},
  {Bergfors}, {Brandner}, {Feldt}, {Goto}, {Klahr}, {Marzari}, \&
  {Mordasini}}]{Thalmann14}
{Thalmann}, C., {Desidera}, S., {Bonavita}, M., {et~al.} 2014, \aap, 572, A91

\bibitem[{{Torres} \& {Ferraz Mello}(1973)}]{Torres73}
{Torres}, C.~A.~O. \& {Ferraz Mello}, S. 1973, \aap, 27, 231

\bibitem[{{Torres} {et~al.}(2006){Torres}, {Quast}, {da Silva}, {de La Reza},
  {Melo}, \& {Sterzik}}]{Torres06}
{Torres}, C.~A.~O., {Quast}, G.~R., {da Silva}, L., {et~al.} 2006, \aap, 460,
  695

\bibitem[{{Torres} {et~al.}(2008){Torres}, {Quast}, {Melo}, \&
  {Sterzik}}]{Torres08}
{Torres}, C.~A.~O., {Quast}, G.~R., {Melo}, C.~H.~F., \& {Sterzik}, M.~F. 2008,
  {Young Nearby Loose Associations}

\bibitem[{{Udalski} \& {Geyer}(1985)}]{Udalski85}
{Udalski}, A. \& {Geyer}, E.~H. 1985, Information Bulletin on Variable Stars,
  2692

\bibitem[{{Valenti} \& {Fischer}(2005)}]{Valenti05}
{Valenti}, J.~A. \& {Fischer}, D.~A. 2005, \apjs, 159, 141

\bibitem[{{van den Ancker} {et~al.}(2000){van den Ancker}, {P{\'e}rez}, {de
  Winter}, \& {McCollum}}]{vandenAncker00}
{van den Ancker}, M.~E., {P{\'e}rez}, M.~R., {de Winter}, D., \& {McCollum}, B.
  2000, \aap, 363, L25

\bibitem[{{van Leeuwen}(2007)}]{vanLeeuwen07}
{van Leeuwen}, F. 2007, \aap, 474, 653

\bibitem[{{Vogt} {et~al.}(1983){Vogt}, {Penrod}, \& {Soderblom}}]{Vogt83}
{Vogt}, S.~S., {Penrod}, G.~D., \& {Soderblom}, D.~R. 1983, \apj, 269, 250

\bibitem[{{Waite} {et~al.}(2011){Waite}, {Marsden}, {Carter}, {Al{\'e}cian},
  {Brown}, {Burton}, \& {Hart}}]{Waite11}
{Waite}, I.~A., {Marsden}, S.~C., {Carter}, B.~D., {et~al.} 2011, \pasa, 28,
  323

\bibitem[{{Weise} {et~al.}(2010){Weise}, {Launhardt}, {Setiawan}, \&
  {Henning}}]{Weise10}
{Weise}, P., {Launhardt}, R., {Setiawan}, J., \& {Henning}, T. 2010, \aap, 517,
  A88

\bibitem[{{White} {et~al.}(2007){White}, {Gabor}, \& {Hillenbrand}}]{White07}
{White}, R.~J., {Gabor}, J.~M., \& {Hillenbrand}, L.~A. 2007, \aj, 133, 2524

\bibitem[{{Wo{\'z}niak} {et~al.}(2004){Wo{\'z}niak}, {Vestrand}, {Akerlof},
  {Balsano}, {Bloch}, {Casperson}, {Fletcher}, {Gisler}, {Kehoe}, {Kinemuchi},
  {Lee}, {Marshall}, {McGowan}, {McKay}, {Rykoff}, {Smith}, {Szymanski}, \&
  {Wren}}]{Wozniak04}
{Wo{\'z}niak}, P.~R., {Vestrand}, W.~T., {Akerlof}, C.~W., {et~al.} 2004, \aj,
  127, 2436

\bibitem[{{Zhang} {et~al.}(2016){Zhang}, {Pi}, {Han}, {Shi}, {Wang}, {Luo},
  {Zhang}, {Hou}, \& {Wang}}]{Zhang16}
{Zhang}, L., {Pi}, Q., {Han}, X.~L., {et~al.} 2016, \na, 44, 66

\bibitem[{{Zuckerman} \& {Song}(2004)}]{Zuckerman04}
{Zuckerman}, B. \& {Song}, I. 2004, \araa, 42, 685

\bibitem[{{Zuckerman} {et~al.}(2001){Zuckerman}, {Song}, {Bessell}, \&
  {Webb}}]{Zuckerman01}
{Zuckerman}, B., {Song}, I., {Bessell}, M.~S., \& {Webb}, R.~A. 2001, \apjl,
  562, L87

\end{thebibliography}

\clearpage

\begin{appendix}

\section{Discussion of individual targets}
For each target, we provide the code to identify its single/binary nature,  according to the notation presented in Sect.\,7, and the V$-$K$_s$ color and rotation period.
\begin{itemize}

\item \bf \object{HIP 560} \rm\\

 S+D; V$-$K$_s$ = 0.91\,mag; P = 0.224\,d\\

 HIP\,560 is a F3V  single star (\citealt{Torres06}; \citealt{Pecaut13}; \citealt{Rodriguez12}).
Its membership to the $\beta$ Pictoris association was first suggested by \citet{Zuckerman01} and, 
subsequently, confirmed by \citet{Torres06}, \citet{Lepine09}, \citet{Schlieder12}, and \citet{Malo13}.
The measured projected rotational velocities are $v \sin{i} = $ 155\,km\,s$^{-1}$ (\citealt{Abt95});  $v \sin{i} = $ 170\,km\,s$^{-1}$ \citep{Royer02}; $v \sin{i} = $ 155\,km\,s$^{-1}$ \citep{delaReza04}; and  $v \sin{i} = $ 170.7\,km\,s$^{-1}$ \citep{Torres06}. Infrared excess was first measured by \citet{Rebull08}. Then, \citet{Patel14} and \citet{Riviere-Marichalar14} reported the discovery of a debris disk.\\
HIP\,560 was observed by Hipparcos and also included in the ASAS survey (ASAS 000650-2306.5). Our periodogram analysis of the ASAS photometric time series revealed a photometric rotation period P = 0.20$\pm$0.02\,d and an amplitude of the light curve of $\Delta$V = 0.02\,mag.
However the photometric precision,  $\sigma$ = 0.036\,mag, is very poor. Our periodogram analysis of the Hipparcos photometric time series, with  good photometric precision ($\sigma$ = 0.007\,mag)
but  very poor observation sampling, revealed two major peaks at P = 0.14\,d and P = 0.27\,d. \\
We observed  HIP\,560 at YSVP  from October 1 to November 6, 2015 for a total of seven nights. We were able to collect 2951 frames in the I filter using 10\,s integration and operating the telescope in defocused mode, owing to the brightness of the target. The target was observed  up to ten consecutive hours during   the nights with the longer observation sequence. These magnitudes were computed with respect to an ensemble comparison consisting of two nearby stars, HD\,242 and CD-236. The differential magnitude time series  was binned using a bin width of 15 min and then analyzed for the period search. The achieved photometric precision was $\sigma_I$ = 0.007\,mag. The LS periodogram found the most significant period  P = 0.224$\pm$0.005\,d. The CLEAN periodogram detected the same period P = 0.22\,d (see online Fig.\,\ref{A1}), also in agreement with the results based on the ASAS photometry, within the uncertainties. Taking all this into account, we assumed that P = 0.224\,d is the stellar rotation period. The light curve amplitude is $\Delta$I = 0.01\,mag.\\
We derived a luminosity L = 4.18$\pm$1.50\,L$_\odot$, a radius R =1.42$\pm$0.40\,R$_\odot$, and an inclination  for the spin axis of the star of $i$ = 32$\pm$5$^{\circ}$, when the average projected rotational velocity is adopted.\\

\item \bf \object{2MASS J00172353-6645124} \rm\\

S; V$-$K$_s$ = 4.647\,mag; P = 6.644\,d\\

2MASS\,J00172353-6645124 is a M2.5V star \citep{Riaz06}.
Its membership to the $\beta$ Pictoris association was first proposed by \citet{Riedel12}, by Malo et al. (\citeyear{Malo13}, \citeyear{Malo14a}, \citeyear{Malo14b}), and, subsequently, confirmed by \citet{Riedel14}, who found no evidence of multiplicity. These latter authors noted, however, that the absolute magnitude is 0.4 mag above the isochrone best fitting the $\beta$ Pictoris members. \citet{Riedel14}  measured a trigonometric distance d = 37.5\,pc. \citet{Malo14a} measured a projected rotational velocity $v \sin{i} =$ 6.3\,km\,s$^{-1}$. \\
Photometric time series of this star are available in the ASAS (ASAS J001723-6645.1), INTEGRAL/OMC (ID 8846000060), and SuperWASP (1SWASPJ001723.46-664512.1) archives.
\citet{Kiraga12} reported a rotation period P = 6.645\,d derived from the ASAS photometric time series, which is also confirmed by 
our own analysis of the same ASAS data. We also found a similar rotation period (P = 6.644$\pm$0.027\,d) in the LS and CLEAN 
periodograms of the INTEGRAL/OMC time series with an amplitude of the light rotational modulation $\Delta$V = 0.10 mag (see online Fig.\,\ref{A2}).
A similar period P = 6.61$\pm$0.02\,d is retrieved from the SuperWASP photometric time series with a light curve amplitude $\Delta$V = 0.12\,mag (see online Fig.\,\ref{A3}). \\
 We derived a luminosity L = 0.074$\pm$0.020\,L$_\odot$ and a radius R = 0.78$\pm$0.026\,R$_\odot$.
 From these values we inferred $\sin{i} =$ 1.06$\pm$0.1, which is slightly larger than $\sin{i} =$ 1, which corresponds to a star viewed equator-on. Taking the uncertainties into account, we consider the inclination of the rotation axis to be $i$ $\simeq$ 90$^{\circ}$. 
 The observed relatively large light curve amplitude $\Delta$V = 0.12\,mag (inferred from the SuperWASP data) is consistent with this configuration, which maximizes the amplitude of the light rotational modulation.\\

\item \bf \object{TYC 1186 0706 1} (FK Psc) \rm \\

Bw; V$-$K$_s$ =  3.623\,mag; P =  7.90\,d\\

TYC\,1186\,0706\,1  consists of two components with a separation $\rho$ = 1.7$^{\prime\prime}$ ($\sim$102\,AU) and a flux ratio  $\sim$2 in the H band \citep{Brandt14}. 
The primary component is a K7.5V star \citep{Lepine09}.  On the basis of the magnitude difference between
the two components, we estimate  a $\sim$M5 spectral type for the secondary component (using a 23 Myr isochrone from
\citealt{Siess00}). TYC\,1186\,0706\,1 was  proposed by \citet{Lepine09} as a highly probable member of the association. Membership is also supported by \citet{McCarthy12}, who measured a Li equivalent width (EW) comparable with those of similar spectral type members.
However, subsequent analysis by Malo et al. (\citeyear{Malo13}, \citeyear{Malo14a}), who determined  a projected rotational velocity $v \sin{i}=$  4.5$\pm$1\,km\,s$^{-1}$, found a membership probability $<$90\%. On the other hand,  \citet{Brandt14} did not consider TYC\,1186\,0706\,1 to be reliably associated with the $\beta$ Pictoris association.\\
A  rotation period P = 7.9165\,d was first measured by \citet{Norton07} using the first season of SuperWASP data. Similar rotation periods,
P = 7.70\,d and P = 7.90\,d, were found by Messina et al. (\citeyear{Messina10}, \citeyear{Messina11}) analyzing the ASAS time series and two SuperWASP seasons, respectively.\\

\item \bf \object{GJ 2006}AB \rm\\

 Bw; (V$-$K$_s$)$_A$ = 4.858\,mag; P$_A$ =  3.99\,d; (V$-$K$_s$)$_B$ = 5.044\,mag;   P$_B$ = 4.91\,d\\

GJ\,2006AB  is a visual wide binary consisting of M3.5V + M3.5V components.
\citet{Jao03} measured a separation $\rho$ = 17.87$\arcsec$ (PA = 175.02$^{\circ}$) corresponding to $\sim$ 580\,AU and 
a magnitude difference $\Delta$V = 0.26\,mag.  Moreover, they found that  the V-band magnitude difference ranged from  $\Delta$V = 0.17\,mag to 0.39\, mag, thus inferring that  one component, at least, is a variable star. 
The WDS catalog (The Washington Double Star Catalog, \citealt{Mason01}) also reported a separation $\rho$ = 18.0$\arcsec$ (PA = 180$^{\circ}$) and a magnitude difference $\Delta$V = 0.29\,mag at the epoch 1920. 
GJ\,2006AB was proposed by \citet{Riedel14} and  \citet{Malo13}  as 
candidate member of the $\beta$ Pictoris association. Its membership to the Tucana/Horologium association was investigated and then rejected by \citet{Kraus14}. Finally, 
\citet{Malo14a} concluded that GJ\,2006AB can be considered a bona fide member of the $\beta$ Pictoris  association.
\citet{Kraus14} measured a $v \sin{i}_A =$ 5.5$\pm$0.4\,km\,s$^{-1}$, whereas \citet{Malo14a} computed  $v \sin{i}_A =$ 6.2$\pm$1.4\,km\,s$^{-1}$ and $v \sin{i}_B <$ 4.2\,km\,s$^{-1}$.\\
Our analysis of the unresolved ASAS photometry (ASAS  002750-3233.1) yielded a rotation period P = 3.97\,d. However, to obtain the rotation periods of both components we planned our own observations.
We observed GJ\,2006AB at CASLEO from September 3 to October 21, 2014 for a total of 12 nights, using the R filter. Differential magnitudes were obtained with respect to the nearby nonvariable star 2MASS\,00281856-3233255.    The components of GJ\,2006 were resolved and we determined the rotation periods of
the A component, P$_{\rm A}$ = 3.99$\pm$0.05\,d, (see online Fig.\,\ref{A4}), in good agreement with the previous results based on the unresolved ASAS photometry, and of the B component, P$_{\rm B}$ = 4.91$\pm$0.05\,d (see online Fig.\,\ref{A5}), with a confidence level
$> 99\%$. The A component exhibited a light curve amplitude $\Delta$R = 0.19 mag, whereas the component B has an amplitude of $\Delta$R = 0.12 mag.\\
For the primary component, adopting an average $v\sin{i}$, we derived a luminosity L$_{\rm A}$ = 0.040$\pm$0.01\,L$_\odot$, a radius R$_{\rm A}$ = 0.61$\pm$0.20\,R$_\odot$, and an inclination $i_{\rm A}$ = 55$\pm$7$^\circ$.
We derived L$_{\rm B}$ = 0.036$\pm$0.010\,L$_\odot$, a radius R$_{\rm B}$ = 0.60$\pm$0.20\,R$_\odot$ for the secondary component  and an inclination $i_{\rm B}$ $<$ 42$\pm$7$^\circ$.\\

\item \bf \object{2MASS J00323480+072927}AB (GJ\,3039) \rm \\

Bc; (V$-$K$_s$)$_A$ = 5.02\,mag; P$_A$ =  3.355\,d; (V$-$K$_s$)$_B$ = 5.68\,mag;   P$_B$ = 0.925\,d\\

 2MASS\,J00323480+072927AB is a visual close binary consisting of  M4 dwarf and a companion of later spectral type with an angular separation  $\rho = 0.{\small \arcsec}79$ \citep{McCarthy01} ($\sim$30\,AU). 
This system was found by \citet{Schlieder12} to be likely member of the association. However, the membership has been recently rejected by \citet{Binks16}.\\
 It was photometrically monitored  by the ASAS 
(ASAS  003235+0729.5) and NSVS  (ID  9141792). A detailed investigation on the photometric variability of this system was carried out by  \citet{2016b} who collected  high-precision, multi-band photometric time series at the MO  and at ARIES  from October to December 2014; this work provides more detailed information. Briefly, Messina et al. measured the rotation periods of the two components, and found that the brighter M4 component has a rotation period P = 3.355\,d and the
 fainter period P = 0.925\,d. They also detected, for the first time for this system, two flare events in the R filter in correspondence of the light curve phase of minimum.  As discussed in  \citet{2016b}, the rotation periods of GJ\,3039A and B fit well into the distribution of either single stars or wide binaries of the $\beta$ Pictoris association. Since the component A has an inclination  $i_{\rm A}$ $\simeq$90$^\circ$, the system may be seen edge on.  In this case,  the projected separation ($\sim$30\,AU) may be  significantly smaller than the real separation, making this system a wide binary. In fact, the two components have rotation periods comparable to those measured  in wide binaries. Moreover,  \citet{2016b} raised the suspect that GJ\,3039B is not physically bounded to GJ\,3039A but may still be a member of the $\beta$ Pictoris association. \\

\item \bf \object{TYC 5853 1318 1} \rm \\

S?; V$-$K$_s$ = 4.158\,mag; P =   7.26\,d\\

TYC\,5853\,1318\,1 is a M1V star \citep{Riaz06} proposed as likely member of the $\beta$ Pictoris association by \citet{Kiss11}. However,  they noted that its spatial location slightly differs  from the location of the bona fide members.
\citet{Malo13} found a significant membership probability ($\sim85\%$)  of the $\beta$ Pictoris association, when the binary hypothesis is considered. However, the same membership probability was found of the Tucana/Horologium  association (THA), when the star is supposed as single. \citet{Kraus14} considered it a likely member of the older Tucana/Horologium association on the basis of its lithium content, which is smaller than in other bona fide members of the $\beta$ Pictoris association. However, this star is inside the Li depletion gap (see, \citealt{Messina16a}), a circumstance that would explain the lower Li content.  \citet{Malo14a} measured a projected rotational velocity  $v \sin{i}=$  11.5$\pm$1.4\,km\,s$^{-1}$, whereas \citet{Kraus14} measured  $v \sin{i}=$  6.8$\pm$0.6\,km\,s$^{-1}$.\\
\citet{Messina11} measured a rotation period  P = 7.26\,d derived from SuperWASP data. Our present analysis of ASAS data (ASAS  010712-1935.6) confirms the period P = 7.3\,d, which is found in six out of eight seasons and in the complete time series from the LS periodogram. On the other hand, the CLEAN algorithm also detected the same period in two seasons and in the complete series.
If we consider the membership of THA, and a statistical distance d = 52$\pm$3\,pc, we derive a luminosity  L = 0.25$\pm$0.06\,L$_\odot$, a radius R = 1.25$\pm$0.26\,R$_\odot$, and an inclination $i$ $\sim$ 90$^\circ$ for the spin axis of the star adopting $<v \sin{i}>  = 9.1$\,k\,ms$^{-1}$. However, if we consider the membership of the $\beta$ Pictoris association (i.e., supposing the binary hypothesis),  and a kinematic distance d = 43$\pm$4\,pc, we derive a radius R = 1.00$\pm$0.13\,R$_\odot$ that gives $\sin{i} \sim$ 1.3. To conciliate radius, rotation period, and $v \sin{i}$ to get $\sin{i} \sim$ 1, we must invoke a significant radius inflation.\\
We note that TYC\,5853\,1318\,1 is the only bona fide member in our catalog whose single/binary nature is not known yet.\\

\item \bf \object{2MASS J01112542+1526214} \rm (GJ\,3076) \\

 Bc; (V$-$K$_s$)$_A$ = 6.252\,mag; P$_A$ =  0.9107\,d;
(V$-$K$_s$)$_B$ 6.552\,mag; P$_B$ =  0.7909\,d\\

2MASS\,J01112542+1526214 is a visual close binary discovered by \citet{Beuzit04} consisting of two components separated by $\rho$ = 0.41$^{\prime\prime}$ ($\sim$8\,AU) with a K-band magnitude difference $\Delta$K = 0.69\,mag (and an estimated V magnitude difference $\Delta$V = 1\,mag) at a trigomometric  distance  d = 21.8\,pc \citep{Riedel14}. \citet{Reid95} determined that the combined spectral classification for this system is M5.0, whereas \citet{Janson12} provided the classification M5V + M6V. The membership of the association was proposed by \citet{Malo13} and confirmed by \citet{Riedel14}. Later, \citet{Malo14b} concluded that it can be considered a bona fide member of the association. This system is also included by \citet{Elliott16} in the list of $\beta$ Pictoris members.
\citet{Hosey15} found evidence of long-term photometric  variability with a possible cycle of about 5 yr. 
According to \citet{Delfosse98}, the projected rotational velocity is $v \sin{i}_{\rm AB} =$  15.2\,km\,s$^{-1}$ . \citet{Malo14a} measured a value of $v \sin{i}_{\rm AB} =$ 17.9\,km\,s$^{-1}$ and a Li EW$_{\rm AB}$ = 593\,m\AA. \\
This system was observed photometrically (but unresolved) as part of the MEarth project \citep{Berta12}. Our LS and CLEAN analyses of the MEarth archival photometric time series allowed us to measure two highly significant (FAP $<$ 0.01) photometric rotation periods. The most powerful is  P = 0.9107$\pm$0.001\,d that we attribute to the brighter component, and P = 0.7909$\pm$0.001\,d, which is still observed after filtering out the primary period, that we attribute to the secondary fainter component (see online Fig.\,\ref{A6}).\\

\item  \bf   \object{2MASS J01132817-3821024} \rm \\

Tc; V$-$K$_s$ = 4.174\,mag;  P = 0.4456\,d\\

2MASS\,J01132817-3821024 is a triple system consisting of a M1 + M3 eclipsing binary system  and a close M1 visual companion.  
The eclipsing binary is detached  and was discovered within the ASAS survey \citep{Pojmanski02}.  
\citet{Parihar09} classified its components as M1V + M3V and measured an orbital period P = 0.4456\,d. 
The visual close companion was detected by \citet{Bergfors10}  at a distance $\rho$ = 1.415$\arcsec$ (PA = 27.2$^{\circ}$), with a magnitude difference $\Delta$z$^\prime$ = $\Delta$i$^\prime$ = 0.17 mag with respect to the eclipsing binary. 
\citet{Janson12} derived the following parameters assuming the eclipsing binary as a unique component: 
SpT$_{\rm AB}$ = M0.5, SpT$_{\rm C}$ = M1.0, M$_{\rm AB}$ = 0.56\,M$_\odot$, M$_{\rm C}$ = 0.54\,M$_\odot$, and $\rho$ = 71.4 AU,  
P$_{\rm orb}$ = 812\,yr. The most extensive investigation was carried out by \citet{Helminiak12} who derived the 
physical/orbital parameters of the component A + B of the eclipsing binary, as well as hints of binarity 
for the component B, which is likely composed by two equal mass (M $\simeq$ 0.45-0.50\,M$_\odot$) stars. 
\citet{Malo13} measured an upper limit for the rotational velocity $v \sin{i} < 9$\, km\,s$^{-1}$, and found that this system might be a member either of the $\beta$ Pictoris (22\%) or the Tucana/Horologium (59\%), or the Columba (19\%) associations.\\

\item  \bf   \object{2MASS J01351393-0712517} \rm \\

SB2; V$-$K$_s$ =  5.502\,mag;  P =  0.7031\,d\\

        2MASS\,J01351393-0712517 is a M4.5V SB2 spectroscopic binary \citep{Malo13}. It was proposed by \citet{Shkolnik12} as member of the $\beta$ Pictoris association with membership quality AAB. \citet{Malo13} found that this system might be also a member 
of the Columba association.  They measured a projected rotational velocity $v \sin{i}$  = 55.1$\pm$4.8\,km\,s$^{-1}$. The membership of the $\beta$ Pictoris association was confirmed by  \citet{Binks14} who measured $v \sin{i}$  = 50.7$\pm$5.9\,km\,s$^{-1}$, and more recently by \citet{Binks16}, and \citet{Elliott16}.\\
The stellar rotation period P = 0.7031\,d was measured by \citet{Kiraga12},  based on the ASAS time series. \\

\item  \bf   \object{2MASS J01365516-0647379} \rm  \\

Bw; V$-$K$_s$ =   5.138\,mag; P = 0.3464\,d\\

 This is a visual wide binary consisting of M4  dwarf and a stellar companion with spectral type later than L0   \citep{Bergfors10}. 
 The components have a separation $\rho$ = 5.587$\arcsec$ (PA = 179.9$^{\circ}$; $\sim$134\,AU), and a magnitude difference $\Delta$z$^\prime$ = 5.07\,mag. 
 The faint component was also detected by \citet{Bowler15} with a slightly different separation $\rho$ =  6.662$\pm$0.003$\arcsec$ and $\Delta$K = 13.8\,mag.
  It is at a trigonometric distance  d = 24$\pm$0.4\,pc \citep{vanLeeuwen07}. \citet{Shkolnik12}  proposed a membership (with quality match ABA) of the Pleiades cluster, 
 whereas Malo et al. (\citeyear{Malo13}; \citeyear{Malo14b}) found a high probability that this M4 star is a candidate member of the $\beta$ Pictoris association. 
   \citet{Malo14a} also measured a projected rotational velocity $v \sin{i}$ = 10\,km\,s$^{-1}$ and inferred  T$_{\rm eff}$ = 3500\,K, which they noticed to be too high for its M4 spectral type. On the other hand, Malo et al.  did not detect any Li line, noticing that it is detectable in similar T$_{\rm eff}$ members. 
This system is found by \citet{Elliott16} to have the star \object{2MASS J01373545-0645375} as common proper motion companion at a separation of 14635\,AU.  Interestingly, this system has been identified by \citet{Alonso-Floriano11} as the wide proper-motion companion ($\rho$ = 612$\arcsec$) of \object{EX Cet},  which is a BY Dra-type star in the Her-Lyr moving group with an age of about 200 Myr at the same distance d = 24\,pc.
\citet{Gaidos00} measured Li EW = 88.7$\pm$2.3\,m\AA\,\, and $v \sin{i}$  = 2.9\,km\,s$^{-1}$ for EX Cet.   \\
 We observed this target  at CASLEO  in September--October 2014 for a total of eight nights in the R filter and at HAO  in 
November 2014 for a total of ten nights in the V filter. Differential magnitudes were obtained with respect to an ensemble of  seven nearby nonvariable stars. The two time series were combined, first by adding to the R data a magnitude offset such that their mean magnitude was equal to the V mean magnitude, and subsequently, by tuning that magnitude offset by minimizing the dispersion in the phased light curve between V and R magnitudes.
 We  measured a rotation period P = 0.3464$\pm$0.0005\,d with high confidence level (FAP $<$ 0.01) with both LS and CLEAN periodograms.
 The light curve exhibits an amplitude $\Delta$V = 0.11 mag (see online Fig.\,\ref{A7}).
 We inferred from our own photometry  magnitudes V = 14.00$\pm$0.05\,mag and R = 13.32$\pm$0.05\,mag.
 We derived for the primary component a luminosity L = 0.010$\pm$0.003\,L$_\odot$, a radius R = 0.32$\pm$0.10\,R$_\odot$, and an inclination $i$ $\sim$ 12$\pm$2$^\circ$.\\

\item  \bf   \object{TYC 1208 0468 1} \rm \\

 Bw; V$-$K$_s$ =   3.114\,mag; P = 2.803\,d\\

        TYC\,1208\,0468\,1 is a wide visual binary consisting of  K3V + K5V components \citep{Malo14b}  separated by $\rho$ = 1.73$^{\prime\prime}$ \citep{Morlet00} ($\sim$100\,AU) at a trigonometric distance d = 60\,pc. The membership of the $\beta$ Pictoris association was suggested by \citet{Schlieder10} and later by \citet{Bell15}; whereas \citet{Malo14a} indicated a possible membership of the  \object{Columba association}. The projected rotational velocity is 
$v \sin{i} = $  16.6\,km\,s$^{-1}$ \citep{Malo14a}. \\
A photometric rotation period P = 2.803\,d and an amplitude of the light curve $\Delta$V = 0.073\,mag were measured by \citet{Kiraga12} based on ASAS photometry. \\

\item  \bf   \object{2MASS J01535076-1459503} \rm \\

Bw; V$-$K$_s$ =   4.897\,mag; P = 1.515\,d\\

 2MASS\,J01535076-1459503 is a visual binary consisting of two M3V + M3V components with an angular separation $\rho$ = 2.876$\arcsec$ (PA = 291.9$^{\circ}$; $\sim$80\,AU), 
 magnitude differences $\Delta$z$^\prime$ = 0.01\,mag,  $\Delta$i$^\prime$ = 0.13\,mag, and at a trigonometric distance d = 18\,pc \citep{Bergfors10}. 
It was proposed by Malo et al. (2014a) as candidate member of the $\beta$ Pictoris association, who measured a 
projected rotational velocity  $v \sin{i} < $  11.2 km\,s$^{-1}$. \\
 \citet{Kiraga12} determined a rotation period P = 1.515\,d from the ASAS photometric time series. \\

\item \bf \object{2MASS J02014677+0117161}  and \object{RBS 269} \rm \\

Bw; (V$-$K$_s$)$_S$ = 4.515\,mag; (V$-$K$_s$)$_N$ = 4.462\,mag;  P =  5.98\,d and P = 3.30\,d\\

2MASS\,J02014677+0117161 (southern component) and RBS\,269 (northern component)  form a visual binary consisting of two M dwarfs with a separation $\rho$ = 10.48$^{\prime\prime}$ ($\sim$ 670\,AU) and a position angle PA = 167$^{\circ}$ at a distance d = 63.70\,pc \citep{Alonso-Floriano15}. This system is associated with the X-ray source RX J0201.7+0117 \citep{Haakonsen09}.  These stars are listed by \citet{Schlieder12}, \citet{Binks14}, and Malo et al. (\citeyear{Malo14a}, \citeyear{Malo14b}) among the members of the $\beta$ Pictoris association. \\
\citet{Kiraga12} measured a photometric period P = 6.006\,d and an  amplitude for the rotational modulation  $\Delta$V = 0.086\,mag derived from unresolved ASAS time series. We have re-analyzed the ASAS data and found two significant power peaks in both LS and CLEAN periodograms at P = 6.00\,d and P = 3.41\,d. We retrieved another photometric time series from the INTEGRAL/OMC archive (ID 0037000047). The LS and CLEAN periodograms
found a rotation period  P = 5.87$\pm$0.22\,d (see online Fig.\,\ref{A8}) in agreement, within the uncertainties, with the longer  period derived from ASAS data.
This system was also observed by NSVS (ID 12053205). Our analysis considering these data revealed the periods P = 3.30$\pm$0.01\,d and P = 5.98$\pm$0.02\,d (see online Fig.\,\ref{A9}). These two  values likely represent the rotation periods of the two components, although we are not capable to establish of which component they are.\\

\item  \bf   \object{2MASS J02175601+1225266} \rm \\

S; V$-$K$_s$ =  4.537\,mag; P = 1.995\,d\\

\citet{Riaz06} inferred a M3.5V spectral type for this object from TiO-band strength at a spectroscopic distance d = 43\,pc.
They associated this star with the X-ray source 1RXS\,J021756.5+122532, which exhibits strong X-ray to bolometric luminosity 
L$_{\rm X}$/L$_{\rm bol}$ = $-$2.89. \citet{Schlieder12} listed it among the $\beta$ Pictoris candidate members. Its membership was
subsequently confirmed by Binks \& Jeffries (\citeyear{Binks14}, \citeyear{Binks16}) who also inferred a kinematic distance  d = 67.9$\pm$6\,pc, assuming the $\beta$ Pictoris membership, and measured $v \sin{i}$ = 22.6$\pm$3.0\,km\,s$^{-1}$.
No nearby visual companion was found by the AstraLux Large M-dwarf Survey \citep{Janson12}. Two RV measurements are available RV = +7.0$\pm$1.4\,km\,s$^{-1}$ from \citet{Binks14} and RV = +8.0$\pm$0.3\,km\,s$^{-1}$ from Malo et al. (in preparation), who also measured $v \sin{i}$ = 17\,km\,s$^{-1}$ and detected no sign of binarity in their  ESPaDOnS@CFHT spectrum. \\
We carried out multi-band photometric observations of this target at ARIES  in October and December 2014 for a total of five nights. Differential magnitudes were obtained with respect to the nearby nonvariable star 2MASS 02173354+1228171.          
The LS and CLEAN analyses revealed a rotation period P = 1.995$\pm$0.005\,d in all filters with FAP = 1\%.
We observed this star at the phases of  maximum and minimum light, but were unable to cover the other rotation phases (see online Fig.\,\ref{A10}). Moreover, we retrieved a photometric time series for this target from the CSS (MLS\_J021756.0+122526)  totaling 67 measurements from  2006 to late 2013.
Despite the low photometric precision of this data ($\sigma_{\rm V}$ = 0.05 mag), our LS analysis revealed a significant (FAP $<$ 1\%) power peak at P = 2.14$\pm$0.05\,d (see online Fig.\,\ref{A11}), 
in agreement with the rotation period independently derived  from our own observations.\\
We derive a luminosity L = 0.12$\pm$0.01\,L$_\odot$, a radius R = 1.16$\pm$0.12\,R$_\odot$, and an inclination $i$ = 55$^{\circ}$.\\

\item  \bf  \object{HIP 10680}/\object{HIP 10679} (=  HD\,14082AB) \rm \\

Bw+D; (V$-$K$_s$)$_A$ = 1.163\,mag; P$_A$ =  0.2396\,d;  (V$-$K$_s$)$_B$ = 1.488\,mag; P$_B$ = 0.777\,d\\

This is a visual binary consisting of  F5V + G2V components with an angular separation $\rho$ $\sim$ 13.8$\arcsec$ \citep{Mason01} and a distance  d = 37.62\,pc \citep{Pecaut13}; ($\sim$520\,AU). 
The cooler component hosts a debris disk \citep{Rebull08}.
The membership of the $\beta$ Pictoris association was first proposed by \citet{Barrado99} and then by \citet{Song03} and \citet{Zuckerman04}. Then, its membership was recovered by \citet{Makarov07}, \citet{Torres08}, \citet{Lepine09}, and by \citet{McCarthy12} on the basis of Li EW.
This system was searched by SEEDS (Strategic Exploration of Exoplanets and Disks with Subaru), but no companion candidates were detected within a projected separation of 7.5$\arcsec$ ($\sim$210 AU) from each component \citep{Brandt14}. We retrieved one spectrum for each component in the 3800-9000\,\AA \,\,region from the LAMOST archive. From our analysis, we inferred a F2V spectral type for HIP\,10680, measured a Li EW = 254$\pm$7\,m\AA,\,\,and detected the H$\alpha$ line in absorption with EW = 3.81$\pm$0.23\,\AA. We inferred  a G3V spectral type, measured a Li EW = 341$\pm$8\,m\AA,\,\,and detected the H$\alpha$ line in absorption with EW = 3.46$\pm$0.33\,\AA\ for HIP\,10679.\\
HIP\,10680 and HIP\,10679 have projected equatorial velocities  $v \sin{i}$ = 37.6 km\,s$^{-1}$  and $v \sin{i}$ = 7.8 km\,s$^{-1}$, respectively \citep{Valenti05}. HIP\,10680 is reported in the Hipparcos catalog  as  a likely algol-type eclipsing binary with a period P = 7.06\,d. However, a note in the catalog reports the possibility that this photometry has been contaminated at some epochs by the presence of the close companion generating a spurious variability. \\
The most recent photometric investigation was carried out by \citet{Messina15a} and provides a more detailed information. Briefly, we collected a photometric time series at the Taurus Hill Observatory from October 2014 to January 2015.
We measured the rotation periods of both components. The rotation period of HIP\,10680 is P = 0.2396\,d
whereas the rotation period of HIP\,10679 is P = 0.777\,d. For HIP\,10680, we derived a luminosity L = 1.88$\pm$0.17\,L$_\odot$ and a radius R = 1.11$\pm$0.10\,R$_\odot$. Combining radius and projected stellar velocity, we estimated an inclination for the stellar rotation axis $i$ = 10$^{\circ}$ .
We derived a luminosity L = 0.96$\pm$0.09\,L$_\odot$ and a radius R = 0.95$\pm$0.09\,R$_\odot$ for HIP\,10679. Combining radius and projected rotational velocity, we found the same inclination for the stellar rotation axis $i$ = 10$^{\circ}$ .\\

\item \bf \object{HIP 11152} \rm \\

S; V$-$K$_s$ = 3.74\,mag; P = 3.6\,d\\

HIP\,11152 is a M3V star studied by \citet{Schlieder10} who proposed its membership
to the $\beta$ Pictoris association and measured a projected rotational velocity $v\sin{i}$ = 6$\pm$2\,km\,s$^{-1}$. \citet{McCarthy12} and Malo et al. (\citeyear{Malo13}, \citeyear{Malo14b}) considered it a bona fide member.\\
Our LS and CLEAN analyses of the SuperWASP photometric time series allowed us to infer
a rotation period P = 1.80$\pm$0.01\,d and a light curve amplitude of $\Delta$V = 0.06\,mag (see online Fig.\,\ref{A12}).  We derived  a luminosity L = 0.073$\pm$0.020\,L$_\odot$, radius R = 0.59$\pm$0.19\,R$_\odot$, and an inclination $i$ = 21$\pm$7$^\circ$. Such  highly inclined stars generally exhibit low-amplitude light curves that are smaller than the observed light curve (see, e.g., Messina  et al., Paper III). Therefore, we suspect that this star may have two major spot groups on opposite hemispheres that modulate the stellar flux with half the rotation period. Therefore, the correct rotation period would be P = 3.6\,d, which is twice as long as that detected by the periodogram (to which corresponds an inclination $i$ $\sim$ 45$\pm$7$^\circ$). \\

\item \bf \object{HIP 11437}AB (AG Tri) \rm  \\

Bw+D; (V$-$K$_s$)$_A$ =  3.040\,mag; P$_A$ = 12.5\,d; (V$-$K$_s$)$_B$ =   4.219\,mag; P$_B$ = 4.66\,d\\

HIP\,11437AB  is a wide visual binary  consisting of K4V + M1V  components \citep{Messina15b} at a distance d = 42.3\,pc \citep{Rodriguez12}. 
The components have an angular separation $\rho$ = 22$^{\prime\prime}$ ($\sim$930\,AU) \citep{Mason01}. 
This system was first proposed as a member of the $\beta$ Pictoris association by \citet{Song03} and \citet{Zuckerman04}. This membership was subsequently 
confirmed by \citet{Makarov07}, \citet{Torres06}, and \citet{Lepine09}. The measured projected rotational velocities for this system are $v \sin{i}_A$ =  5\,km\,s$^{-1}$ \citep{Cutispoto00}; $v \sin{i}_A$ =  4.7\,km\,s$^{-1}$ 
and $v \sin{i}_B$ =  5\,km\,s$^{-1}$ \citep{Bailey12}, and more recently,  $v \sin{i}_A$ =  5.25$\pm$0.75\,km\,s$^{-1}$ and $v \sin{i}_B$ =  7.46$\pm$0.75\,km\,s$^{-1}$ \citep{Messina15b}.
High-contrast H-band imaging with the Subaru telescope by \citet{Brandt14}
 did not detect any companion candidates within 7$^{\prime\prime}$ ($\sim$ 300 AU projected) from both components. 
The presence of a debris disk around the A component was first detected by \citet{Rebull08} using MIPS observations,
and recently by \citet{Riviere-Marichalar14} from Hershell observations. We retrieved from the LAMOST archive a spectrum of HIP\,11437A  in the 3800-9000\,\AA\,\, region. We detected the H$\alpha$ line in emission and measured an  EW = $-$0.57$\pm$0.01\,\AA.\\
 A rotation period P = 13.6828\,d was first measured by \citet{Norton07}, and subsequently confirmed by \citet{Messina11}, who reported P = 12.5\,d. 
The most recent photometric study of AG Tri was carried out by \citet{Messina15b}, who measured for the first time a rotation period P = 4.66\,d for the component B;  that paper provides more detailed information on this object.\\
 We derived a luminosity L$_{\rm A}$ = 0.27$\pm$0.03\,L$_\odot$, a radius R$_{\rm A}$ = 1.02$\pm$0.12\,R$_\odot$, and an inclination $\sin{i}_{\rm A}$ $\sim$ 1.3 for the primary (see  \citealt{Messina15b}).
 We derived L$_{\rm B}$ = 0.063$\pm$0.005\,L$_\odot$, a radius R$_{\rm B}$ = 0.68$\pm$0.09\,R$_\odot$, and an inclination $i_{\rm B}$ $\simeq$ 80$\pm$9$^\circ$ for the secondary component.\\

\item \bf \object{HIP 12545} \rm \\

S; V$-$K$_s$ = 3.301\,mag; P = 4.83\,d\\

HIP\,12545 is classified as K6Ve by \citet{Torres06}.  \citet{Torres08} reported the detection of RV variations and suggested that this star might be  a SB1 binary with the K6Ve as primary component. However, this claim was not confirmed by \citet{Bailey12} who did not find any significant RV variation. Images from
SEEDS did not detect any companion candidate within 8.5$\arcsec$ \citep{Brandt14}. 
Its membership to the $\beta$ Pictoris association was first proposed by \citet{Song03} and \citet{Zuckerman04}. Its membership was also recovered  by \citet{Torres08} and \citet{Schlieder12}. \citet{Malo14b} listed this star among the bona fide members of the association. The membership was only questioned by \citet{Elliott14} on the basis on new RV measurements. 
In the literature we found the following values of the rotational velocity: $v \sin{i}<$  8\,km\,s$^{-1}$ \citep{Favata95}; $v \sin{i}=$ 9.3\,km\,s$^{-1}$ 
        \citep{Scholz07}; $v \sin{i}=$  9\,km\,s$^{-1}$ \citep{daSilva09}; $v \sin{i}=$  9\,km\,s$^{-1}$ \citep{Weise10}; and
$v \sin{i}=$ 8.7$\pm$0.2\, km\,s$^{-1}$ \citep{Bailey12}. The only discrepant  $v \sin{i}=$ 40\,km\,s$^{-1}$  value is reported by \citet{Torres06}, which is probably unreliable. \\
A rotation period P = 4.831\,d and a light curve amplitude $\Delta$V = 0.105\,mag were measured by \citet{Kiraga12}.
We derived a luminosity  L = 0.27$\pm$0.02\,L$_\odot$ and a radius R = 0.92$\pm$0.25\,R$_\odot$. 
Our LS and CLEAN analyses revealed  the existence of two peaks in almost all seasons of ASAS: the dominant is at P = 4.83\,d, but also a second peak is almost always present with comparable power  at P = 1.26\,d, which is however its 1-d beat period. We observed HIP\,12545 at CrAO between 25 October and 24 December 2015 for a total of 11 nights
in the B, V, R, and I filters. We collected a total of  71 frames per filter. Differential magnitudes were obtained with respect to the nearby nonvariable star \object{2MASS 02410935+0557497}. Our LS and CLEAN analyses revealed the highest peak at P = 4.85$\pm$0.01\,d and a light curve amplitude $\Delta$V = 0.14\,mag (see online Fig.\,\ref{A13}).
In contrast to what was stated by \citet{Messina11}, P = 4.85\,d is the stellar rotation period that combined with the average $< v \sin{i} > = $  9\,km\,s$^{-1}$ and the stellar radius, yields an inclination of the stellar rotation axis $i$ = 70$\pm$10$^{\circ}$. \\

\item \bf \object{2MASS J03350208+2342356} \rm \\

Bc?; V$-$K$_s$ =   5.739\,mag; P = 0.4719\,d\\

2MASS\,J03350208+2342356 is classified as a M8.5 star by \citet{Gizis00}. This  star was  proposed by  \citet{Shkolnik12} as a member of the $\beta$ Pictoris association  with  quality BAA. 
Malo et al. (\citeyear{Malo13}, \citeyear{Malo14b}) considered this star a bona fide member.
\citet{Close03} did not found any physical companion within 0.1--15$\arcsec$ from high-resolution imaging. However, no RV study was carried out to confirm its single nature.
 \citet{Reid02} classified it as a likely brown dwarf with M = 0.06\,M$_\odot$ and 1 Gyr of age, finding
 clear evidence of the presence of a Li line (Li$_{\rm EW}$ = 720\,m\AA), a 
 significant level of chromospheric activity (EW$_{\rm H\alpha}$ = $-$6.5 \AA), rapid rotation ($v \sin{i}=$  30\,km\,s$^{-1}$ ), and a distance d = 23.5\,pc from photometric parallax. 
 \citet{Ribas03} found its space velocities to be similar to the average of the Castor Moving Group with an age of about 320\,Myr. We retrieved from the LAMOST archive a spectrum in the 3800-9000\,\AA\,\, region. From our analysis, we detected the H$\alpha$ line in emission and measured an EW = -2.42$\pm$0.21\,\AA.\\
We  retrieved a photometric time series from the NSVS archive and our LS  analysis revealed a rotation period P = 0.475$\pm$0.005\,d
with a light curve amplitude of $\Delta$V = 0.14\,mag, and an average photometric precision $\sigma$ = 0.14\,mag (see online Fig.\,\ref{A14}). In contrast, the CLEAN periodogram showed the most significant power peak at P = 0.66\,d .
We also retrieved a photometric time series  from CSS. From these data our LS and CLEAN analyses revealed a rotation period P = 0.4719$\pm$0.005\,d
with a light curve amplitude of $\Delta$V = 0.03\,mag, and an average photometric precision $\sigma$ = 0.06\,mag (see online Fig.\,\ref{A15}).
Combining rotation period P = 0.4719\,d, $v \sin{i}$, and stellar radius, which we estimated to be about R = 0.12$\pm$0.04\,R$_\odot$, we inferred  $\sin{i}$  $>$ 1. This unreliable value may arise from the underestimation of either the radius or the distance. \\

\item \bf  \object{2MASS J03461399+1709176} \rm \\

S; V$-$K$_s$ =   4.079\,mag; P = 1.742\,d\\

2MASS\,J03461399+1709176 is a single  M0.5 star \citep{Janson12} at a  distance d = 64\,pc. It is listed by \citet{Schlieder12} as a high-probability candidate member of the $\beta$ Pictoris association at a kinematic distance d = 58.4\,pc. This star was observed by HATNet (HAT 308-0001461). From this data,  a rotation period P = 1.742\,d was reported by \citet{Hartman10} with an amplitude for the flux rotational modulation $\Delta$r = 0.068\,mag. This star has a visual companion StKM 1-406a (V$-$K$_s$ = 3.14\,mag) at a separation  $\rho$ = 8$^{\prime\prime}$ (14$^{\prime\prime}$ as maximum observed separation; \citealt{Mason01}) with rotation period P = 0.897\,d \citep{Hartman10}. \\

\item \bf \object{GJ 3305} \rm \\

Bc; V$-$K$_s$ =   4.177\,mag; P = 4.89\,d\\

This is a visual close binary consisting of a M1 primary component and a secondary component of spectral type later than M1 with a separation of $\rho$ = 0.093$\arcsec$ ($\sim$9\,AU; \citealt{Kasper07}).  Based on 12 epochs of astrometric observations (where the observed separation ranges from 0.095$^{\prime\prime}$ to 0.307$^{\prime\prime}$), \citet{Janson14} estimated an orbital period of 30\,yr. The membership of the $\beta$ Pictoris  association was first proposed by \citet{Zuckerman01}, \citet{Zuckerman04},  then by \citet{Torres08} and confirmed by Binks \& Jeffries (\citeyear{Binks14}, \citeyear{Binks16}), and by \citet{Elliott16}. 
\citet{Scholz07} measured a projected rotational velocity $v \sin{i}=$ 5.3\,km\,s$^{-1}$. \citet{Montet15} and  \citet{Elliott16} identified the F0V star \object{51 Eri} at a separation $\rho$ = 66$\arcsec$ ($\sim$ 1940\,AU) as a likely wide companion of this system.\\
The only available rotation period in the literature is P = 6.1\,d, which was tentatively adopted by \citet{Feigelson06} and inferred from photometric data collected at the Southern African Astronomical Observatory. We analyzed the ASAS time series obtained by using the smallest aperture to minimize
any flux contribution from the nearby F0V companion. The LS and CLEAN periodograms revealed the most significant peak at P = 4.89$\pm$0.01\,d with a light curve peak-to-peak amplitude $\Delta$V = 0.05\,mag. However, we still consider this measurement to be confirmed.\\

\item \bf \object{2MASS J04435686+3723033} (V\,962 Per) \rm \\

Bw; V$-$K$_s$ =   4.179\,mag; P = 4.2878\,d\\

2MASS\,J04435686+3723033 is a M3V star proposed by \citet{Schlieder10} and Malo et al. (\citeyear{Malo14a}, \citeyear{Malo14b}) as a likely member of the $\beta$ Pictoris  association together with its nearby proper-motion companion \object{PM I04439+3723E}, which is located at a distance  $\rho$ = 9$^{\prime\prime}$ ($\sim$531\,AU). \citet{Pecaut13} listed it among the bona fide members of the association. Two  projected rotational velocity measurements are available: $v \sin{i} =$ 8$\pm$2\,km\,s$^{-1}$ by \citet{Schlieder10} and $v \sin{i} = $10.6$\pm$1\,km\,s$^{-1}$ \citep{Malo14a}. We retrieved from the LAMOST archive one spectrum in the 3800-9000\,\AA\,\,region. From our analysis we inferred a M2V spectral type and found the H$\alpha$ line in emission with EW = $-$4.60$\pm$0.21\,\AA.\\
A rotation period P = 4.2878\,d was measured by \citet{Norton07} 
from SuperWASP data. 
We derived a luminosity  L = 0.73$\pm$0.02\,L$_\odot$ and a  radius R = 0.68$\pm$0.22\,R$_\odot$. Combining radius and the  projected rotational velocity 
 $v \sin{i} =$ 8\,km\,s$^{-1}$, we estimated an inclination for the stellar rotation axis $i$ = 90$\pm$15$^{\circ}$.\\
\\

\item \bf \object{HIP 23200} (V1005 Ori) \rm \\

SB1; V$-$K$_s$ =   3.99\,mag; P = 4.43\,d\\

HIP\,23200 is classified as a M0.5e   by \citet{Torres06}. \citet{Elliott14}, collecting literature RV measurements, 
noted significant radial velocity variation and inferred that it is a SB1 system. \citet{Elliott16} inferred the masses 0.7\,M$_\odot$ for the primary and 0.3\,M$_\odot$ for the secondary, and a separation of 0.26\,AU between the two components. This system is found to have as common proper motion the star \object{2MASS  J05015665+0108429} with a separation of 81044\,AU, and therefore is classified as a hierarchical triple system \citep{Elliott16}.
\citet{Biller13} detected two nearby stars at 4.8$^{\prime\prime}$ and 7.2$^{\prime\prime}$ that were classified as background stars. The membership was first proposed by \citet{Torres08}, confirmed by \citet{Lepine09}, and recovered by \citet{Schlieder12}. \\
The determination of the stellar rotation period was very challenging. We have four different values for the rotational projected velocity:
an upper value  $v \sin{i}<$ 3\,km\,s$^{-1}$ by \citet{Reiners12}; $v \sin{i}=$ 14\,km\,s$^{-1}$ by \citet{Favata95};  $v \sin{i}=$ 8.7\,km\,s$^{-1}$ by \citet{Vogt83}, and 
$v \sin{i}=$ 14\,km\,s$^{-1}$ by \citet{Torres06}. These values span a range too large to allow us to constrain the rotation period.
The first rotation period determination was carried out by \citet{Bopp77} who reported a period P = 1.96\,d, cautioning that it could be alias of the true period. The equally probable periods P = 1.858\,d and P = 2.20\,d 
were later found by \citet{Bopp78}. 
\citet{Byrne84} reported a period P = 4.565\,d.  Later,  \citet{Amado01} reported a period P = 4.399\,d detected in two consecutive years at the South African Astronomical Observatory. 
The most recent period determination P = 4.4236\,d is reported by \citet{Dal11} and it was obtained in two consecutive years.\\
Our LS and CLEAN analyses of ASAS time series show   the following periods in all seasons (as well as in the complete
series) in order of decreasing power: 
P = 4.42\,d, P = 1.29\,d, P = 2.2\,d, and P = 1.83\,d (see online Fig.\,\ref{A16}). They all give smooth phased light curves. 
However, when we combine the ASAS and the contemporary 2005-2006 Dal \& Evren photometry, which allows us to obtain a better sampled time series, the power peak at P = 4.40\,d becomes the dominant showing it to be the true rotation period.
Similar periods, both based on the ASAS data, P = 4.414\,d and P = 4.390\,d, are measured by \citet{Kiraga12} and in the ASAS Catalog of Variable Stars, respectively. On the other hand, at some epochs this star  may have two major spots on opposite hemispheres that modulate the light curve with a period that is a half (2.2\,d) the true rotation period (4.4\,d). This spot configuration would explain the detection of the P = 2.2\,d by us and by \citet{Bopp78}.\\
Considering that this is a binary system, we cannot discard that the other detected periodicities, either P = 1.86\,d or P = 2.2\,d, are the rotation period of the secondary component.\\

\item   \bf \object{HIP 23309} \rm \\

S; V$-$K$_s$ =   3.756\,mag; P = 8.60\,d\\

HIP\,23309 is classified as M0Ve by \citet{Torres06}, whereas \citet{Pecaut13} indicate a slightly earlier K8V spectral type. 
We consider this star to be single because neither \citet{Torres06} nor \citet{Elliott14} find evidence for significant RV variations. 
Its membership was first proposed by \citet{Zuckerman01}, later by \citet{Zuckerman04},  and subsequently confirmed in many other studies (e.g., \citealt{Neuhauser03}; \citealt{Jayawardhana06};  \citealt{Torres06}; \citealt{Mentuch08}; \citealt{Lepine09}).
The available measurements of rotational velocity are $v \sin{i}=$ 5.8\,km\,s$^{-1}$ \citep{Weise10}; $v \sin{i}=$ 6\,km\,s$^{-1}$ \citep{daSilva09}; $v \sin{i}=$ 5.77\,km\,s$^{-1}$ \citep{Scholz07}; and a discrepant value of $v \sin{i}=$ 11\,km\,s$^{-1}$ \citep{delaReza04}. \\
Our LS and CLEAN periodogram analyses of ASAS time series find in all seasons (as well as in the complete series) the same rotation period, 
P = 8.6\,d, and a light curve amplitude up to $\Delta$V = 0.07\,mag (see online Fig.\,\ref{A17}). A similar period P = 8.729\,d is reported by \citet{Kiraga12}.
We derived a luminosity  L = 0.17$\pm$0.04\,L$_\odot$ and a radius R = 0.90$\pm$0.29\,R$_\odot$. Combining radius and the average projected stellar velocity 
$<v \sin{i}>   =$ 5.8\,km\,s$^{-1}$, we estimated  $\sin{i}$ = 1.1$\pm$0.15.\\

\item   \bf  \object{TYC 1281 1672 1} \rm  \\

S; V$-$K$_s$ =  3.153\,mag; P = 2.759\,d\\

TYC\,1281\,1672\,1 is a K2IV star \citep{White07}. It is listed by \citet{Schlieder12} among the candidate members of the $\beta$ Pictoris association at a kinematic distance d = 53.8\,pc. \citet{White07} measured a projected rotational velocity $v\sin{i}$ = 17.22$\pm$5.0\,km\,s$^{-1}$.\\
\citet{Kiraga12} measured a rotation period P = 2.763\,d and an amplitude for the rotational flux modulation $\Delta$V = 0.08\,mag.
We  retrieved a magnitude time series from the INTEGRAL/OMC archive (ID 1281000037). The LS periodogram revealed the same rotation period  P =  2.759$\pm$0.005\,d  with an amplitude $\Delta$V = 0.12\,mag, whereas the CLEAN periodogram showed its harmonic P = 1.38\,d to be the dominant periodicity (see online Fig.\,\ref{A18}). This circumstance may be explained with the present of two spot groups on opposite hemispheres.
We derived a luminosity  L = 0.24$\pm$0.06\,L$_\odot$ and a radius R = 0.89$\pm$0.29\,R$_\odot$. Combining radius and the projected rotational velocity, we find $\sin{i}$ = 1.05$\pm$0.10. Taking the uncertainties into account,  this star is likely seen equator on.\\

\color{black}

\item    \bf \object{2MASS J05015665+0108429} \rm  \\

S?; V$-$K$_s$ =  5.523\,mag; P = 2.08\,d\\

2MASS\,J05015665+0108429 is classified by \citet{Schlieder12} as a M4 dwarf with a projected rotational velocity 
 $v \sin{i}=$ 8$\pm$2\,km\,s$^{-1}$. We have no multiple RV measurements to establish its single/binary nature. The membership has been first proposed by \citet{Schlieder12}, recently rejected by \citet{Binks16}, and later recovered by \citet{Elliott16}, who found this star to be a common proper motion companion of the bona fide member HIP\,23200 at a distance of 81044\,AU.\\
 It was observed by ASAS and NSVS, however our periodogram analysis could not find any significant periodicity.
A first attempt to measure the photometric rotation period was carried out by us in 2012 and 2015 using the facilities at the Xinglong station. We  observed it for seven nights collecting a total of 34 frames in the V filter during the first run, and for ten nights during the second run collecting a total of 44 frames. Our LS and CLEAN periodograms revealed a photometric rotation period P = 2.08\,d and a light curve amplitude $\Delta$V = 0.06\,mag (see online Fig.\,\ref{A19}). Subsequently,
we retrieved a much longer photometric time series from the INTEGRAL/OMC archive (ID 0098000028). The LS and CLEAN periodograms
confirmed the rotation period to be P = 2.08$\pm$0.02\,d (see online Fig.\,\ref{A20}) and we also detected a number of flare events.\\

\item \bf \object{HIP 23418} \rm  \rm\\

Tc; (V$-$K$_s$)$_A$ =  4.780\,mag; P = 1.22\,d\\

HIP\,23418 is a triple system.  The primary A component is a SB2 binary with spectral type M3,
an orbital period P$_{\rm orb}$ = 11.96\,d, and eccentricity $e$ = 0.323. The secondary B component has a spectral type later than M3, an angular separation from the primary component of $\rho$ =  0.677$^{\prime\prime}$ measured by Hipparcos (ESA  1997). Subsequently,   \citet{Delfosse98} determined $\rho$ =  0.97$^{\prime\prime}$  with a magnitude difference $\Delta$K = 0.9 mag. 
\citet{Scholz07} measured the following rotational velocities:  $v \sin{i}$$_A$ = 7.67\,km\,s$^{-1}$ and $v \sin{i}$$_B$ = 21\,km\,s$^{-1}$, whereas \citet{Shulyak11} measured $v \sin{i}$$_A$ = 6.65\,km\,s$^{-1}$ and $v \sin{i}$$_B$ = 23.77\,km\,s$^{-1}$. 
\citet{Riedel14} measured a trigonometric distance d = 24.6$\pm$1.3\,pc and 
inferred the following deblended magnitudes and colors V$_{\rm AC}$ = 12.46\,mag and (V$-$I)$_{\rm AC}$ = 2.71\,mag, V$_{\rm B}$ = 13.56\,mag, and  (V$-$I)$_{\rm B}$ = 3.15\,mag. 
The membership of the $\beta$ Pictoris association was first proposed by \citet{Song03}, \citet{Zuckerman04},  and then by \citet{Torres06} and \citet{Lepine09}.
The membership  was recently confirmed by Binks \& Jeffries (\citeyear{Binks14}, \citeyear{Binks16}), \citet{Riedel14}, and \citet{Elliott16}.\\
Our LS and CLEAN analyses of ASAS time series revealed two possible rotation periods: P = 6.43\,d and P = 1.22\,d (see online Fig.\,\ref{A21}). We note that one period is the beat of the other.  \citet{Kiraga12} measured a similar period, P = 6.431\,d. 
Applying a correction of $\Delta$V = 0.44\,mag, to take care of the duplicity of the unresolved AC components, we derived  L$_A$ = 0.019$\pm$0.002\,L$_\odot$ and a radius R$_A$ = 0.41$\pm$0.06\,R$_\odot$. Combining radius and the average projected stellar velocity, 
$<v \sin{i}>  = $ 7\,km\,s$^{-1}$, we found that the rotation period must be shorter than about three days. Therefore, we inferred that the stellar rotation period is P = 1.22\,d and the other is its beat period.
This is a system where the rotation/orbital periods of the SB2 HIP\,23418A components are not synchronized, yet. The faintness of the component B, and the photometric precision $\sigma$ = 0.032\,mag,
did not allow us to detect its rotation period. Combining rotation period, radius, and $v \sin{i}$, we infer a very low inclination of the rotation axis $i$ = 10$^{\circ}$ for the A component.\\

\item \bf \object{BD -21 1074}AB \rm \\

Tw; (V$-$K$_s$)$_A$ =  4.350\,mag; P$_A$ = 9.3\,d; (V$-$K$_s$)$_B$ =  4.640\,mag; P$_B$ =5.4\,d\\

BD\,$-$21\,1074 is a triple system consisting of a M1.5V star (A) at  8.2$\arcsec$ from the secondary a M2.5V star (B) that has a nearby companion at 0.8$\arcsec$ ($\sim$21\,AU) of M5 or later spectral type \citep{Torres08}. It is listed in the \citet{Gershberg99} catalog of the UV Cet-type flare stars. \\
The most recent photometric and spectroscopic investigation of this system was carried out by \citet{Messina14} who reported periods P = 9.3\,d and P = 5.4\,d for the primary and secondary components, respectively. That paper provides more detailed information on this system.  The rotation periods, combined with their measurements of projected rotational velocities,  $v \sin{i}$$_A$ = 3.7$\pm$1\,km\,s$^{-1}$ and  $v \sin{i}$$_B$ = 4.9$\pm$1\,km\,s$^{-1}$, and the stellar radii R$_A$ = 0.80\,R$_\odot$ and R$_B$ = 0.68\,R$_\odot$ yield  inclination values $i_{\rm A}$ = 60$^{\circ}$ and $i_{\rm B}$ = 50$^{\circ}$.\\

\item \bf \object{2MASS J05082729-2101444} \rm \\

S; V$-$K$_s$ = 5.867\,mag; P = 0.2804\,d\\

2MASS\,J05082729-2101444 is classified as  M5V by \citet{Riaz06}  from the TiO-band strength at a spectroscopic distance d = 26\,pc.
They associate this star with the X-ray source 1RXS\,J050827.3$-$210130, which exhibits strong X-ray to bolometric luminosity 
L$_{\rm X}$/L$_{\rm bol}$ = $-$3.19. \citet{Bergfors10} observed this target in 2008.88 with the Lucky Imaging camera AstraLux Sur 
but did not discover any close companion. 
There are  two RV measurements (Malo et al. 2014a) that agree within their errors suggesting it is a single star. 
Malo et al. (\citeyear{Malo13}, \citeyear{Malo14b}) proposed it as member of the $\beta$ Pictoris association and its membership
was confirmed by Binks \& Jeffries (\citeyear{Binks14}, \citeyear{Binks16}) who inferred a kinematic distance d = 30\,pc against the statistical distance, d = 25\,pc, from \citet{Malo14b}. \\
We analyzed the SuperWASP magnitude time series with LS and CLEAN periodograms and found
a rotation period P = 0.2804$\pm$0.0005\,d with FAP $<$ 1\% and a peak to-peak light curve amplitude $\Delta$V = 0.07\,mag (see online Fig.\,\ref{A22}).
We derived a luminosity  L = 0.010$\pm$0.003\,L$_\odot$ and a radius R = 0.39$\pm$0.12\,R$_\odot$. Combining the radius and the projected stellar velocity 
27.6$\pm$4.3\,km\,s$^{-1}$  (Malo et al. 2014), we determined $i$ = 20$\pm$5$^{\circ}$.\\
A very interesting feature that came out of the SuperWASP data was the discovery of a megaflare at HJD = 2454445.5 with a magnitude increase of $\Delta$V = 3.29 mag.\\

\item \bf       \object{TYC 112 1486 1}  \rm \\

Tc; V$-$K$_s$ = 3.11\,mag; P = 2.18\,d\\

TYC\,112\,1486\,1 is a visual binary at a distance of d = 71\,pc consisting of a K3V primary component 
separated by $\rho$ = 0.42$^{\prime\prime}$ (29.8\,AU; PA = 253$^{\circ}$) from the secondary component,
which is $\Delta$K = 2.19\,mag fainter. The two components have estimated masses  M$_1$ = 0.66\,M$_\odot$,
and  M$_2$ = 0.15\,M$_\odot$. This system (unresolved) was first studied by \citet{Alcala00} who  detected the H$\alpha$ line in emission,  the presence of a strong Li line  (EW$_{\rm H\alpha}$ = $-$0.90\,\AA\,\, and EW$_{\rm Li}$ = 510$\pm$8\,m\AA), and measured a projected rotational velocity $v \sin{i}$ = 17$\pm$1\,km\,s$^{-1}$. They identified this system as the optical counterpart of the X-ray source RXJ\,0520.5+0616 detected by the ROSAT All Sky Survey.  \citet{Biazzo05} measured an effective temperature T = 4900\,K and a metallicity [Fe$_{\rm I}$/H] = $-$0.08\,dex. The membership of the $\beta$ Pictoris association was first proposed by \citet{Elliott14}, who collected repeated RV measurements and found no variability. Membership of the $\beta$ Pictoris association was also considered by \citet{Alonso-Floriano15}. \citet{Bell15} consider this star a member of the 32 Ori group with an estimated age of 22$\pm$4\,Myr. \citet{Elliott16} found this star to be component of a hierarchical triple system together with the common proper motion stars  TYC\,112\,917\,1 and 2MASS 05195327+0617258.\\
\citet{Kiraga12} measured a photometric rotation period P = 2.18\,d and a light curve amplitude $\Delta$V = 0.09\,mag from ASAS data.\\

\item \bf       \object{TYC 112 917 1} \rm \\

Tw; V$-$K$_s$ = 3.00\,mag; P = 3.51\,d\\

TYC\,112\,917\,1 is a K3V star at a distance d = 67.8\,pc and an estimated mass M = 0.67\,M$_\odot$. This star is found by \citet{Elliott16} to be common proper motion of TYC\,112\,1486\,1 at a separation of 39269\,AU.
This system (unresolved) was first studied by \citet{Alcala00} who  detected the H$\alpha$ line in emission,  the presence of a strong Li line  (EW$_{\rm H\alpha}$ = $-$0.20\,\AA\,\, and EW$_{\rm Li}$ = 470$\pm$5\,m\AA), and measured a projected rotational velocity $v \sin{i}$ = 8$\pm$1\,km\,s$^{-1}$. They identified this system as the optical counterpart of the X-ray source RXJ\,0520.0+0612 detected by the ROSAT All Sky Survey. \\
\citet{Kiraga12} measured a photometric rotation period P = 3.51\,d and a light curve amplitude $\Delta$V = 0.085\,mag from ASAS data.\\

\item \bf \object{2MASS J05241914-1601153} \rm \\

Bc; V$-$K$_s$ = 5.603\,mag; P = 0.4008\,d\\

2MASS\,J05241914-1601153 is classified as  M4.5V by \citet{Riaz06}  from the TiO-band strength at a spectroscopic distance d = 11\,pc. 
It is associated with the X-ray source 1RXS\,J052419.1-160117       with  L$_{\rm X}$/L$_{\rm bol}$ = $-$3.14.
\citet{Bergfors10} discovered that it is a visual binary consisting of  M4.5 + M5.0 components separated by 
$\rho$ = 0.639$^{\prime\prime}$ (PA = 69.1$^{\circ}$; $\sim$13\,AU) with magnitude differences $\Delta$z$^\prime$ = 0.36\,mag and 
$\Delta$i$^\prime$ = 0.43\,mag. Subsequently, \citet{Janson14} estimated an orbital period of 40 yr or longer and
a likely close to edge-on orbit.  Malo et al. (\citeyear{Malo13}, \citeyear{Malo14b}) suggested it is a member of the $\beta$ Pictoris association and inferred a statistical distance d = 20$\pm$5\,pc and
measured a $v \sin{i}$ = 50$\pm$4.5\,km\,s$^{-1}$. Binks \& Jeffries (\citeyear{Binks14}, \citeyear{Binks16}) confirmed it  as a member and inferred a kinematic distance d = 31.6$\pm$4.9\,pc.\\
We retrieved a photometric time series from CSS catalog and found a rotation period P = 0.4008\,d
with high confidence level (see online Fig.\,\ref{A23}). Other significant power peaks do not conciliate with radius and projected rotational velocity.\\

\item \bf \object{HIP 25486} (AF Lep) \rm \\

SB2; V$-$K$_s$ = 1.294\,mag; P = 0.9660\,d\\

HIP\,25486 is a  SB2 system with a F7V primary component and a mass ratio $q$ = 0.72 (\citealt{Nordstrom04}; \citealt{Torres08}). 
The SEEDS images did not detect any companion candidates within 7.5$^{\prime\prime}$ ($\sim$200\,AU projected) \citep{Brandt14}.
The membership of the association was first proposed by \citet{Zuckerman01}, \citet{Zuckerman04}, and then by \citet{Torres06}, \citet{Fernandez08}, \citet{Binks14}, and \citet{Mamajek14}.
\citet{Weise10} reported  $v \sin{i}$ = 71.6 km\,s$^{-1}$; \citet{White07}  $v \sin{i}$ = 55.18\,km\,s$^{-1}$; \citet{Scholz07} $v \sin{i}$ = 52\,km\,s$^{-1}$; \citet{Valenti05}  $v \sin{i}$ = 54.7\,km\,s$^{-1}$, and \citet{delaReza04}  $v \sin{i}$ = 50\,km\,s$^{-1}$.\\
 The most recent and detailed photometric investigation was carried out by \citet{Jarvinen15}
who measured a rotation period P = 0.9660\,d,  which is in agreement with the rotational velocity and a radius, R = 1.18\,R$_{\odot}$, of a middle ($i$ = 50$^{\circ}$) inclination star.\\

\item \bf  \object{2MASS J05294468-3239141} \rm \\

S?; V$-$K$_s$ = 5.474\,mag; P = 1.532\,d\\

2MASS\,J05294468-3239141 is a M4.5Ve star  (\citealt{Cutispoto96}; \citealt{Janson12})  proposed by \citet{Riedel14} as member of the association at a trigonometric distance d = 26\,pc. 
We have no multiple RV measurements to establish its single/binary nature. Our periodogram analysis of the SuperWASP magnitude time series, with a photometric accuracy
$\sigma$ = 0.03\,mag, allowed us to measure a rotation period P = 1.532$\pm$0.005\,d  and a peak-to-peak light curve amplitude $\Delta$V = 0.05\,mag (see online Fig.\,\ref{A24}). 
Unfortunately, the projected rotational velocity has never been measured to derive the inclination of the rotation axis.\\

\item \bf \object{TYC 4770 0797 1} (V1311 Ori) \rm \\

Bc; V$-$K$_s$ = 4.311\,mag; P = 4.372\,d\\

TYC\,4770\,0797\,1 is a close visual binary whose components are M2 + M3.5 with a separation $\rho$ = 0.20$^{\prime\prime}$ ($\sim$4.4\,AU) and an estimated orbital
period P = 23\,yr \citep{Janson14} at a distance d = 22\,pc. 
\citet{Hartmann86} measured $v \sin{i}$ = 12\,km\,s$^{-1}$; whereas \citet{daSilva09} also reported the same projected rotational velocity  $v \sin{i}$ = 12\,km\,s$^{-1}$. The membership was proposed by \citet{Torres06}, \citet{daSilva09}, \citet{McCarthy12}, and \citet{Malo14a}. It is, however, questioned by \citet{Janson14} who detected a significant but relatively modest orbital motion, despite the estimated 23\,yr orbital period. Since the system is seen almost edge-on, the system orbital plane may be aligned with the components equatorial planes. This configuration may be an explanation of the measured relatively modest orbital rate. The membership has been recently rejected by \citet{Binks16}. \\
The first rotation period determination, P = 4.5\,d,  was reported by \citet{Gahm93} with the warning that it needed confirmation, since the star was observed for a time span of only six days. A similar period P = 4.37\,d is found by \citet{Kiraga12} from ASAS data. Our analysis of ASAS data confirms a period P = 4.37\,d.
Moreover, we retrieved a long photometric time series from the INTEGRAL/OMC archive (ID 4770000050). The LS and CLEAN periodograms
also determined the rotation period to be P = 4.372$\pm$0.002\,d (see online Fig.\,\ref{A25}).\\

\item \bf \object{2MASS J05335981-0221325} \rm \\

S; V$-$K$_s$ = 4.725\,mag; P = 7.23\,d\\

 2MASS\,J05335981-0221325 is classified as M3V by \citet{Riaz06}  from TiO-band strength at a spectroscopic
 distance d = 25\,pc. 
It is associated with the X-ray source 1RXS\,J053359.8-022131. Its  X-ray to  bolometric luminosity 
 is L$_{\rm X}$/L$_{\rm bol}$ = $-$2.88. It was proposed by Malo et al. (\citeyear{Malo13}, \citeyear{Malo14a}) as a high-probability member of the association. They inferred a statistical distance d = 42\,pc 
and did not find evidence of RV variation.  The membership of the $\beta$ Pictoris association was confirmed by Binks \& Jeffries (\citeyear{Binks14}, \citeyear{Binks16}).\\
 \citet{Kiraga12} reported a period P = 7.25\,d and a light curve amplitude $\Delta$V = 0.17\,mag from ASAS data.
We retrieved a long photometric time series from the INTEGRAL/OMC archive (ID 4770000109). The LS and CLEAN periodograms
confirmed the rotation period P = 7.23$\pm$0.02\,d and light curve amplitude $\Delta$V = 0.10\,mag  (see online Fig.\,\ref{A26}).\\
We derived a luminosity L = 0.093$\pm$0.025\,L$_\odot$ and a radius R = 0.89$\pm$0.29\,R$_\odot$. Combining $v \sin{i}$ = 5.4$\pm$1.3\,km\,s$^{-1}$ from \citet{Malo14a}
and the stellar radius,  we derived $i$ = 60$\pm$10$^{\circ}$.\\

\color{black}
\item \bf \object{2MASS J06131330-2742054} \rm \\

Tc; V$-$K$_s$ = 5.23\,mag; P = 16.8\,d\\

2MASS\,J06131330-2742054 is a close visual binary whose components have a separation $\rho$ = 0.095$^{\prime\prime}$ ($\sim$2.7\,AU), a magnitude difference in the optical $\Delta$V = 0.59 (FGS F583W filter), and an estimated orbital period P $>$ 4\,yr  (\citealt{Bowler15} report 8.3\,yr) at a trigonometric distance d = 29.4\,pc \citep{Riedel14}. Actually, the HST observations reported by \citet{Riedel14} indicate that the A component is itself a binary.
\citet{Riaz06}  inferred a M3.5V spectral type from the TiO-band strength for the unresolved system. 
It is associated with the X-ray source 1RXS\,J061313.2-274205. Its  X-ray to  bolometric luminosity 
is L$_{\rm X}$/L$_{\rm bol}$ = $-$3.05.
\citet{Malo13} inferred a M4 spectral type and proposed  its membership to the association. The membership was also confirmed by \citet{Riedel14} who suggested it may be a common proper motion companion of 2MASS\,J06085283$-$2753583,  a brown dwarf member of the $\beta$ Pictoris association. \citet{Malo14b} concluded it is a bona fide member of the association. \\
\citet{Hosey15} reported evidence of photometric variability with a possible period of $\sim$0.4\,yr. 
Our analysis of  SuperWASP data time series allowed us to find three photometric rotation periods  P = 39\,d, P = 17.7\,d, and P = 8.6\,d, whose powers never differ by more than 30\%. We observed this system at YCO and found in all B, V, and R filters two major periods  P = 36.6\,d and P = 16.8\,d (see online Fig.\,\ref{A27}).  However, the longer periods are in disagreement with either stellar radius (R = 0.86\,R$_\odot$ as reported by Malo et al. 2014b) or the projected rotational velocity  $v \sin{i}$ = 2.4\,km\,s$^{-1}$ \citep{Malo13}. On the other hand, the shorter period P$\sim$17\,d, independently found in both SuperWASP and YCO data, conciliates with both radius and projected rotational velocity.\\

\item \bf \object{HIP 29964} (AO Men) \rm  \\

S+D; V$-$K$_s$ = 2.986\,mag; P = 2.67\,d\\

HIP\,29964  is a K4Ve single star hosting a debris disk at a distance d = 39\,pc \citep{Carpenter05}. Its membership to the association was first proposed by \citet{Zuckerman01}, \citet{Zuckerman04}, and then by \citet{Neuhauser03}, \citet{Ortega09}, \citet{Makarov07}, \citet{Torres06}, \citet{Lepine09}, and is considered a bona fide member in all subsequent studies.
\citet{delaReza04} measured a 
 $v \sin{i}$ = 13\,km\,s$^{-1}$; \citet{Jayawardhana06}  $v \sin{i}$ =16\,km\,s$^{-1}$; 
\citet{Torres06}  $v \sin{i}$ = 16.4\,km\,s$^{-1}$; \citet{Scholz07}  $v \sin{i}$ = 15.96\,km\,s$^{-1}$; \citet{Weise10}  $v \sin{i}$ = 15.9\,km\,s$^{-1}$; and \citet{Cutispoto99}  $v \sin{i}$ = 17\,km\,s$^{-1}$. \\
A rotation period P = 2.65\,d was first reported by \citet{Cutispoto99}. A period P = 2.66\,d was measured by \citet{Koen02} from the Hipparcos data. \citet{Messina10} found a rotation period P = 2.67\,d from ASAS data, in agreement with the value  P = 2.673\,d as  listed in the ACVS, and
subsequently confirmed by \citet{Kiraga12}.
We derived a luminosity L = 0.27$\pm$0.07\,L$_\odot$ and a radius R = 0.91$\pm$0.29\,R$_\odot$. Combining the average $<v \sin{i} >$ =15.7\,km\,s$^{-1}$ and the stellar radius,  we derived $i$ = 70$\pm$10$^{\circ}$.\\

\item \bf   \object{2MASS J07293108+3556003}AB \rm  \\

Bc; V$-$K$_s$ = 4.024\,mag; P = 1.94\,d\\

2MASS\,J07293108+3556003AB is a M1 + M3 visual binary whose components have a separation $\rho$ = 0.198$\arcsec$ ($\sim$8\,AU), an inferred orbital period of about 38\,yr, and  magnitude differences of $\Delta$$z^{\prime}$ = 1.15\,mag and $\Delta$$i^{\prime}$ = 1.28\,mag (Janson et al. \citeyear{Janson12}, \citeyear{Janson14}). This system was proposed by \citet{Schlieder12} as likely member of the $\beta$ Pictoris association. However, the membership has  recently been  rejected by \citet{Binks16}.\\
This star was regularly observed by SuperWASP  starting from the end of 2006. However, the components could not be resolved. From this data we detected the most significant peak in the LS and CLEAN periodograms  at P = 1.967$\pm$0.005\,d (see online Fig.\,\ref{A28}). We assume it is the stellar rotation period of the brighter M1 component.\\

\item \bf   \object{2MASS J08173943-8243298} \rm  \\

Bc?; V$-$K$_s$ = 5.032\,mag; P = 1.318\,d\\

2MASS\,J08173943-8243298 is classified as  M3.5V by \citet{Riaz06}  from the TiO-band strength at a spectroscopic  distance d = 14\,pc. It is associated with
the X-ray source 1RXS\,J081742.4-824331. Its X-ray to bolometric luminosity  is
 L$_{\rm X}$/L$_{\rm bol}$ = $-$3.01. It was proposed by \citet{Malo14b} as member of the association with a statistical distance d = 27\,pc. Furthermore,  they measured a projected rotational velocity $v \sin{i}$ = 32.2$\pm$2.6\,km\,s$^{-1}$. \\
 \citet{Kiraga12} reported a period P = 1.318\,d and 
  noted that it may also be a EB or Ell star.\\

\color{black}

\item \bf   \object{2MASS J08224744-5726530} \rm \\

Tc; V$-$K$_s$ = 5.568\,mag; P = ?\,d\\

2MASS\,J08224744-5726530  is a triple system consisting of a M4.5 primary, a secondary component with spectral type later than L0 separated by $\rho$ = 0.64$^{\prime\prime}$  ($\sim$5.2\,AU),  and a farther M6 companion 
at a distance  $\rho$ = 8.43$^{\prime\prime}$  ($\sim$67\,AU) with a magnitude difference $\Delta$$z^{\prime}$ = 1.83\,mag \citep{Bergfors10}, and at a  spectroscopic distance d = 8\,pc \citep{Riaz06}.
Similar values are measured by \citet{Janson14}. The M4.5 and M6 components are known to be common proper motion stars. This system was studied  by \citet{Malo14a} who measured  $v \sin{i}$ = 6.4$\pm$2.4\,km\,s$^{-1}$ and  statistical distance d = 6\,pc, and classified this star among the ambiguous candidate members of the $\beta$ Pictoris association. The membership has been recently rejected by \citet{Binks16}. We could not infer the rotation period from the ASAS time series.\\

\item \bf \object{2MASS J09361593+3731456}AB (GJ\,9303) \rm \\

SB2; V$-$K$_s$ = 4.100\,mag; P = 12.9\,d\\

2MASS\, J09361593+3731456AB is a M0.5 + M0.5 SB2 spectroscopic binary proposed by \citet{Schlieder12} as a likely member of the $\beta$ Pictoris association. 
However, \citet{Malo13} raised some doubt based on the discrepant distance by 40\,pc with respect to the average distance of bona fide members.  
\citet{Schlieder12} measured a projected rotational velocity  $v \sin{i}$ = 6$\pm$2\,km\,s$^{-1}$, whereas \citet{Malo14a} measured  $v \sin{i}$ $=$ 1.9\,km\,s$^{-1}$.\\ 
This star was regularly observed by SuperWASP   starting from the end of 2006. Considering this data, we found the most significant peak in the LS and CLEAN periodograms is at P = 12.9\,d (see online Fig.\,\ref{A29}). We assume it is the stellar rotation period that is in  agreement with the lower $v \sin{i}$ measurement.
 Together with other three bona fide $\beta$ Pictoris members of similar spectral type, this system is among the slowest members.
 It is interesting to note that \citet{Schlieder12}, based on the large variation of RVs of both components over timescales as small as one day, suspected that the system may have a short orbital period. If this is the case, then we may deal with a nonsynchronized system where the rotation period is longer than the orbital one. \\

\item \bf     \object{2MASS J10015995+6651278}      \rm \\

Bc?; V$-$K$_s$ = 4.159\,mag; P = 2.50\,d\\

2MASS\,J10015995+6651278 is a M3 dwarf \citep{Schlieder12}.
 Only one RV measurement is available for this star from  \citet{Schlieder12},
which does not allow us to establish its single/binary nature. 
The membership was proposed by \citet{Schlieder12}, but recently rejected by \citet{Binks16}.\\
We carried out a first attempt to measure the photometric rotation period  in 2012 using the facilities at the Xinglong station. We observed it for only five nights collecting a total of 26 frames in the V filter. Our LS and CLEAN periodograms did not reveal any significant periodicity, but we detected one likely flare event with a magnitude brightening of $\Delta$V = 0.32\,mag. 
However, this star was also observed by NSVS (ID 2549392) and in this case our LS and CLEAN analyses of this time series allowed us to derive a 
firm determination of the stellar rotation period, P = 2.49$\pm$0.02\,d, and a light curve amplitude of $\Delta$V = 0.06\,mag (see online Fig.\,\ref{A30}). 
 We found the same rotation period P = 2.50$\pm$0.01 in the LS and CLEAN 
periodograms of the INTEGRAL/OMC time series (ID 4143000079) with an amplitude for the light rotational modulation $\Delta$V = 0.04 mag (see online Fig.\,\ref{A31}).
We inferred  a luminosity L = 0.054$\pm$0.014\,L$_\odot$, a radius R = 0.58$\pm$0.19\,R$_\odot$, and $i$ = 90$\pm$15$^{\circ}$ when we combined R and P 
with the projected rotational velocity  $v \sin{i}$ = 12\,km\,s$^{-1}$ \citep{Schlieder12}.\\

\color{black}

\item \bf \object{HIP 50156} (DK Leo) \rm \\

Bc; V$-$K$_s$ = 3.819\,mag; P = 8.05\,d\\

 HIP\,50156 is a close visual binary consisting of a M0.5V primary component separated by $\rho$ = 0.09$\arcsec$ ($\sim$2.3\,AU) from the secondary of unknown spectral type, and 
at a distance d = 23.1\,pc \citep{Bowler15}. The possible membership of the $\beta$ Pictoris association was first proposed by \citet{Schlieder12}, whereas \citet{Shkolnik09} suggested that this star has an age of $\sim$ 400 Myr. Subsequently, \citet{Shkolnik12} proposed the membership of the \object{Carina association}. On the other hand, \citet{Malo13} proposed the membership of Columba, whereas \citet{Klutsch14} estimated an age similar to that of the Pleiades. Recently, Binks \& Jeffries (\citeyear{Binks14}, \citeyear{Binks16}) confirmed the membership of the $\beta$ Pictoris association. \citet{Lopez-Santiago10} measured $v\sin{i}$ $=$ 7.68$\pm$0.7\,km\,s$^{-1}$.\\
The photometric rotation period P = 8.05\,d was first measured by \citet{Busko80}. Later, \citet{Kiraga12} reported P = 7.86\,d
based on ASAS time series.\\

\item \bf \object{TWA 22} \rm \\

Bc; V$-$K$_s$ = 6.301\,mag; P = 0.84\,d\\

TWA\,22 is a visual close binary consisting of  M5 + M6  components separated by $\rho$ = 0.1$\arcsec$ ($\sim$1.7\,AU). The orbital plane has an inclination $i$ = 27.43$\pm$4.40$^{\circ}$ and the orbital period is P = 5.15$\pm$0.09 yr \citep{Bonnefoy09}. \citet{Scholz07} reported $v\sin{i}$ = 9.67 kms$^{-1}$; and \citet{Jayawardhana06} $v\sin{i}$ = 9.7 kms$^{-1}$. 
\citet{Song03} first suggested its membership to the \object{TW Hya association}, based on the high lithium content. \citet{Torres08} considered TWA 22 a candidate member of the $\beta$ Pictoris association, and its membership was confirmed with the revised kinematic data by \citet{Teixeira09}. Also \citet{Malo14b} and \citet{Elliott16} considered it to be a bona fide member of the association.\\
Our analysis of ASAS data revealed in a few seasons the period P = 0.84\,d, and P =1.54\,d in other seasons.
Our analysis of INTEGRAL/OMC time series (ID 8600000041), although the very low photometric precision  ($\sigma$ = 0.37\,mag), seems to give P = 0.84\,d as the correct rotation period (see online Fig.\,\ref{A32}).\\

\item \bf \object{BD+262161}AB \rm \\

Bw+D; (V$-$K$_s$)$_A$ = 2.61\,mag; (V$-$K$_s$)$_B$ 3.25\,mag; P =  2.022\,d;
P =  0.973\,d\\

BD+\,262161AB is a visual binary consisting of  K2 + K5 components separated by $\rho$ = 5.26$\arcsec$ ($\sim$115\,AU) \citep{Mason01}. This star was proposed by \citet{Schlieder12} as a likely member of the $\beta$ Pictoris association.
However,  its UVW space velocity components would place it right at the edge of the $\beta$ Pictoris association in all three velocity components, and furthermore it lies about 30\,pc above the bulk of the bona fide $\beta$ Pictoris members. \citet{Brandt14} considered its membership to be highly doubtful. The SEEDS images revealed no companion candidates other than the known K5 secondary component.\\
 This system was observed by SuperWASP with sparse observations since 2006 and in a regular way starting from  2007. Owing to the  $\sim$5.26$\arcsec$ separation, SuperWASP could not resolve the system's components. Nonetheless, our LS and CLEAN periodograms were able to detect two highly significant periodicities, P = 0.973$\pm$0.005\,d and  P = 2.022$\pm$0.005\,d, which likely represent the rotation periods of the two components (see online Fig.\,\ref{A33}). The only available measurement of the projected rotational velocity,  $v \sin{i}$ = 6$\pm$2\,km\,s$^{-1}$  (Schlieder et al. 2012), refers to the fainter K5 component. A comparison with \citet{Siess00} models shows that these components are below the sequence traced by the $\beta$ Pictoris members, as better fitted by isochrones corresponding to older ages.\\
This system was observed at ARIES  in the VRI filters. We observed it for only six nights. However, we  retrieved the same P = 2.02\,d period in the R and I filters.\\

\item \bf \object{2MASS J11515681+0731262} \\ \rm

SB2; V$-$K$_s$ = 4.613\,mag; P = 2.291\,d\\

2MASS\,J11515681+0731262 is a close visual binary consisting of M2 + M8 components separated by $\rho$ = 0.5$^{\prime\prime}$ ($\sim$ 18\,AU) with a magnitude difference $\Delta$J =  5.4\,mag, and  at a photometric distance d = 37$\pm$6\,pc  \citep{Bowler15}. The primary component is itself a M2.5 + M2.5 spectroscopic binary whose components A and B have projected rotational velocities  
$v \sin{i}_A$ = 14$\pm$2\,km\,s$^{-1}$ and $v \sin{i}_B$ = 12$\pm$2\,km\,s$^{-1}$, and unblended magnitudes V$_A$ = 13.01\,mag and V$_B$ = 13.57\,mag, respectively \citep{Bowler15}. We retrieved one spectrum in the 3800-9000\,\AA\,\, region from the LAMOST archive. From our analysis we inferred a M2 spectral type for the primary component and detected the H$\alpha$ line in emission with EW = $-$1.66$\pm$0.13\,\AA.\\
\citet{Schlieder12} listed this star among the high-probability candidate members of the $\beta$ Pictoris association and inferred a kinematic distance d = 33.2\,pc. However, \citet{Bowler15} concluded that this system does not belong to any specific young moving group. On the other hand, the membership of the $\beta$ Pictoris association was recently rejected by \citet{Binks16}.\\
\citet{Kiraga12} reported a photometric rotation period P = 2.291\,d and a light curve amplitude $\Delta$V = 0.13\,mag.\\

\item \bf \object{2MASS J13545390-7121476} \rm\\

S?; V$-$K$_s$ = 4.568\,mag; P = 3.65\,d\\

2MASS\,J13545390-7121476 is classified as M2.5V by  \citet{Riaz06}  from the TiO-band strength at a spectroscopic distance d = 28\,pc. 
It is associated with the X-ray source 1RXS\,J135452.3-712157 with an X-ray to bolometric luminosity L$_{\rm X}$/L$_{\rm bol}$ = $-$3.39. 
It is listed by \citet{Malo13}  among the candidate members of the association with 91\% probability and a predicted distance d = 24\,pc. 
\citet{Malo14a} measured an upper limit for the projected rotational velocity  $v\sin{i}$ $<$ 3.5 kms$^{-1}$ and re-evaluate the membership probability obtaining 99.1\% when the RV information is also considered. They also determined a smaller predicted distance  of d = 21\,pc.\\
This star was observed at the YCO  from June 14 until September 20, 2014, for a total of ten nights.
We obtained a total of 49 frames in the V and R filters and 44 frames in B filter. We analyzed all time series with LS and CLEAN
periodograms and derived the most powerful power peak at P = 3.65$\pm$0.01\,d in the V and R time series with  peak-to-peak 
light curve amplitudes $\Delta$V = 0.025\,mag and $\Delta$R = 0.020\,mag (see online Fig.\,\ref{A34}). 
We derived a luminosity L = 0.024$\pm$0.007\,L$_\odot$, a radius R = 0.43$\pm$0.14\,R$_\odot$, and an inclination $i$  $=$30$\pm$5$^\circ$.\\

\item \bf \object{HIP 69562} (MV Vir) \rm  \\

Tc; V$-$K$_s$ = 3.669\,mag; P = 0.2982\,d\\

HIP\,69562  is a visual triple system. The primary component A is a K5.5V star with a nearby component B at $\rho$ = 0.3$^{\prime\prime}$, and a farther third component C at  $\rho$ = 1.3$^{\prime\prime}$ from the primary \citep{Mason01} at a distance d = 30\,pc. \citet{Malo13} measured the projected rotational velocities of the components A and C: $v\sin{i}$ $=$ 102$\pm$9\,km\,s$^{-1}$ and $v\sin{i}$ $=$ 24$\pm$3\,km\,s$^{-1}$, respectively.  Its membership to the $\beta$ Pictoris association was suggested by Malo et al. (\citeyear{Malo13}, \citeyear{Malo14b}) and it was also considered as a member by \citet{Bell15} and \citet{Alonso-Floriano15}. \\
\citet{Kiraga12} measured a photometric rotation period P = 0.2982\,d, which likely refers to the primary A component.\\

\item \bf \object{TYC 915 1391 1} \rm \\

S; V$-$K$_s$ = 3.600\,mag; P = 4.34\,d\\

TYC\,915\,1391\,1 was classified as K4V by \citet{Stephenson86}, and later as M0V by \citet{Lepine09}.
\citet{Schlieder12} classified this star as a high-probability candidate member of the $\beta$ Pictoris association
and inferred a kinematic distance d = 51.8\,pc. We retrieved one spectrum in the 3800-9000\,\AA\,\,region from the
LAMOST archive. From our analysis, we inferred a K7V spectral type and detected the H$\alpha$ line in emission with EW = $-$3.37$\pm$0.11\,\AA.\\
In the ASAS Catalog of Variable stars \citep{Pojmanski02}
it is listed with a photometric rotation period P = 4.340\,d and a light curve amplitude $\Delta$V = 0.36\,mag. 
\citet{Hoffman09} reported a similar photometric rotation period 
P = 4.329\,d and a light curve amplitude $\Delta$V = 0.28\,mag inferred from the NSVS survey.\\

\item \bf \object{HIP 76629} (V343 Nor) \rm \\

SB1; V$-$K$_s$ = 2.118\,mag; P = 4.27\,d\\

HIP\,76629 is a triple system consisting of a SB1 K0V binary with a nearby M5Ve visual companion (\object{HD139084}B) 
        at an angular separation $\rho$ = 10$\arcsec$ ($\sim$400\,AU) and $\Delta$V = 6.8\,mag \citep{Song03}. 
        \citet{Thalmann14} derived  (see, also Guenther \& Esposito 2007; \citealt{Torres06}) 
a tentative orbital solution from RV variations for an additional companion  with minimum mass M = 0.11\,M$_\odot$ and a period of 
about 4.5 years and moderate eccentricity (0.5--0.6). The membership was first proposed by \citet{Zuckerman01}, later by \citet{Song03}, \citet{Zuckerman04}, \citet{Makarov07}, \citet{Torres06}, \citet{Lepine09}, and \citet{Malo13}.
The following values of rotational velocity were measured at $v\sin{i}$ = 11\,km\,s$^{-1}$ by \citet{delaReza04};  $v\sin{i}$ = 16.6\,km\,s$^{-1}$  by \citet{Torres06};  $v\sin{i}$ = 17\,km\,s$^{-1}$  by \citet{Scholz07};   and $v\sin{i}$ = 18.2\,km\,s$^{-1}$ by \citet{Weise10}. \\
  A rotation period P = 4.20\,d was first reported by \citet{Udalski85}; then P = 4.4\,d by \citet{LloydEvans87}; P = 4.567\,d by \citet{Cutispoto93}; P = 4.25\,d by \citet{Cutispoto96}; P = 4.32\,d 
by \citet{Cutispoto98a}; and P = 4.24\,d by \citet{Cutispoto98b}. 
\citet{Messina10} found the same period P = 4.27\,d from ASAS photometry, which does not resolve the visual companion. \\
Among SB members of the $\beta$ Pictoris association, HIP\,76629 exhibits the longest rotation period.\\

\item \bf \object{2MASS J16430128-1754274} \rm \\

S; V$-$K$_s$ = 3.951\,mag; P = 5.14\,d\\

2MASS\,J16430128-1754274 is classified as a M0.5 star by \citet{Riaz06} from the TiO-band strength and with a spectroscopic distance d = 59\,pc.  
It is associated with the X-ray source 1RXS\,J164302.3-175418 with a normalized X-ray to bolometric luminosity L$_{\rm X}$/L$_{\rm bol}$ = $-$3.11. 
We consider this star a single since \citet{Siebert11},  \citet{Kiss11}, and \citet{Binks14} find similar values of RV. 
This star was first proposed by \citet{Kiss11} as a member of the association and, subsequently, confirmed by \citet{Malo13} and Binks \& Jeffries (\citeyear{Binks14}, \citeyear{Binks16}).
\citet{Malo13} measured an average projected rotational velocity $v\sin{i}$ $=$ 8.8$\pm$2.5\,km\,s$^{-1}$ and \citet{Binks14} $v\sin{i}$ $=$ 8.7$\pm$2.7\,km\,s$^{-1}$. \\
\citet{Messina11} reported a period P = 5.14\,d inferred from the ASAS time series. We also retrieved a long photometric time series from the INTEGRAL/OMC archive (ID 6221000015). We found a rotation period P = 4.74$\pm$0.01\,d and a light curve amplitude $\Delta$V = 0.12\,mag (see online Fig.\,\ref{A35}). However, the data extraction is reported to be affected by point spread function (PSF) problems. 
 We inferred  a luminosity L = 0.10$\pm$0.023\,L$_\odot$, a radius 
R = 0.73$\pm$0.24\,R$_\odot$ and $\sin{i}$ = 1.20$\pm$20. Taking into consideration the uncertainties, we can assume this star is seen equator on.\\

\item \bf \object{2MASS J16572029-5343316} \rm  \\

S; V$-$K$_s$ = 4.646\,mag; P = 7.17\,d\\

2MASS\,J16572029-5343316 is classified as
 M3V by \citet{Riaz06}  from the TiO-band strength at a spectroscopic distance d = 28\,pc. 
It is associated with the X-ray source 1RXS\,J165719.9-534328 with a normalized X-ray to bolometric luminosity L$_{\rm X}$/L$_{\rm bol}$ = $-$3.55. 
\citet{Malo13} proposed this star as candidate member at a predicted distance d = 5\,pc and a membership 
probability of 99.5\% with no evidence of RV variations.
\citet{Malo14a} measured $v\sin{i}$ $<$ 3.5\,km\,s$^{-1}$ and revised the membership probabiliy to 91\%, which  increased to 99.9\% when the RV information was also considered.\\
This star was observed at the YCO from June 14 until September 20, 2014, for a total of  ten nights.
We  obtained a total of 49 frames in the B, 48 in the V, and 41 in the R filter. We analyzed all time series with LS and CLEAN
periodograms and derived the most powerful power peaks  at P = 33d in both periodograms and P = 7.15$\pm$0.03\,d with a peak-to-peak light curve amplitude $\Delta$V =  0.025\,mag only in the  LS periodogram (see online Fig.\,\ref{A36}).
We assume the shorter rotation period to be more consistent with the upper value of projected rotational velocity. 
We inferred  a luminosity L = 0.13$\pm$0.03\,L$_\odot$, a radius R = 1.01$\pm$0.34\,R$_\odot$, and $i$ = 30$\pm$5$^{\circ}$. \\

\item \bf \object{2MASS J17150219-3333398} \rm\\

 Bc?; V$-$K$_s$ = 3.862\,mag; P = 0.3096\,d\\

2MASS\,J17150219-3333398 is classified as M0V by \citet{Riaz06}   from the TiO-band strength  at a spectroscopic distance d = 31\,pc. 
It is associated with the X-ray source 1RXS\,J171502.4-333344
 with  L$_{\rm X}$/L$_{\rm bol}$ = $-$3.10. 
 A X-ray flare was detected by \citet{Naze10}. It was proposed by \citet{Malo13} as member of the association. \citet{Malo14a}
  measured a projected rotational velocity 
 $v\sin{i}$ $=$ 76.3$\pm$5.3\,km\,s$^{-1}$ and a statistical distance d = 23\,pc. \\
 \citet{Kiraga12} measured a rotation period P = 0.3106\,d from the ASAS time series. We retrieved a long photometric time series from the INTEGRAL/OMC archive (ID 7366000048) and found a rotation period P = 0.3096$\pm$0.0005\,d and a light curve amplitude $\Delta$V = 0.04\,mag (see online Fig.\,\ref{A37}).
 We inferred  a luminosity L = 0.063$\pm$0.016\,L$_\odot$, a radius R = 0.55$\pm$0.18 R$_\odot$, and $i$ = 58$\pm$10$^{\circ}$. 
 \\

\item \bf \object{HD\,155555}ABC  (\object{V824 Ara}) \rm  \\

Tw; (V$-$K$_s$)$_{AB}$ = 2.528\,mag; P$_{AB}$ =  1.687\,d;
(V$-$K$_s$)$_C$ 5.081\,mag; P$_B$ =  4.43\,d\\

HD\,155555 is a triple stellar system  at a distance d = 31.4$\pm$0.5 pc  \citep{vanLeeuwen07}. 
This system consists of a double-line spectroscopic binary {HD\,155555AB}, also known as V824 Ara,  and  a fainter M dwarf companion 
{HD\,155555C} at an angular distance $\rho$ = 33$^{\prime\prime}$ ($\sim$1040\,AU). 

The spectroscopic binary {HD\,155555AB} was discovered by \citet{Bennett67} to be composed of  G5IV + K0IV stars that rotate with an orbital period P = 1.687\,d. 
The most recent determination of orbital and physical properties,   and first magnetic maps \rm of both components were provided by the spectro-polarimetric study of \citet{Dunstone08}.
The membership of HD155555AB of the {$\beta$ Pictoris} association was first suggested by \citet{Zuckerman01} based on distance, UVW
velocity components, high $v\sin{i}$, and L$_{\rm X}$/ L$_{\rm bol}$ ratio, later confirmed by \citet{Torres06}, \citet{Makarov07}, and \citet{Lepine09}.\\
  The fainter \object{HD155555C} component (also named  {LDS\, 587B})  is a M3.5 dwarf  \citep{Pasquini91}.
The most recent photometric investigation of this  system was carried out by \citet{Messina15c}, who confirmed the rotation period of the spectroscopic binary and discovered
for the first time the rotation period P = 4.43$\pm$0.01\,d of HD155555C. This paper provides a more detailed description of the system.\\

\item \bf \object{TYC 8728 2262 1} \rm  \\

S; V$-$K$_s$ = 2.186\,mag; P = 1.775\,d\\

TYC\,8728\,2262\,1 is classified as K1V  star by \citet{Torres06} at a distance d = 66\,pc. Their RV measurements together with those from \citet{Song12} and \citet{Elliott14}
do not show evidence of significant variation. Therefore, we consider it a single star.
\citet{Torres06} proposed it as a member of the association and measured a $v\sin{i}$ = 35.3\,km\,s$^{-1}$. Its membership was subsequently suggested by 
\citet{Kiss11}, whereas \citet{Song12} proposed the membership to the Upper-Scorpius subgroup of the Sco-Cen complex with a photometric distance d = 72\,pc. 
\citet{Brandt14} gave a lower 50\% membership probability of the $\beta$ Pictoris association.\\
\citet{Messina10} measured a rotation period P = 1.819\,d inferred from ASAS time series. The same period was later reported by \citet{Kiraga12}. The periodogram analysis carried out by \citet{Desidera15} revealed two other significant periodicities at 
P = 2.21\,d and P = 0.686\,d. However, the longest period does not conciliate with the radius nor with the projected rotational velocity. 
The shortest period would imply an inclination $i$ $\sim$20$^{\circ}$, which is too low to allow the observed light curve amplitude of $\Delta$V=0.12\,mag. Therefore, these periodicities likely arise from aliasing. We inferred a luminosity L = 0.70$\pm$0.19\,L$_\odot$, a radius R = 1.10$\pm$0.35\,R$_\odot$, and $\sin{i}$ = 1.15$\pm$0.10 when P = 1.819\,d is assumed as the rotation period. 
We retrieved a long photometric time series from the INTEGRAL/OMC archive (ID 8728000045). The LS and CLEAN periodograms
confirmed the rotation period to be P = 1.775$\pm$0.005\,d with a light curve amplitude $\Delta$V = 0.15\,mag (see online Fig.\,\ref{A38}).\\

\item \bf \object{GSC 08350-01924} \rm \\

Bc; V$-$K$_s$ = 4.766\,mag; P = 1.906\,d\\

GSC\,08350-01924 is classified as  M3V by \citet{Riaz06} from the TiO-band strength at a spectroscopic distance d = 29\,pc. 
It is associated with the X-ray source 1RXS\,J172919.1-501454
with  L$_{\rm X}$/L$_{\rm bol}$ = $-$3.35.  It was proposed by \citet{Torres06} as member of the association. \citet{Kiss11}
also considered it a member. Malo et al. (\citeyear{Malo13}, \citeyear{Malo14a}) and Elliott et al. (\citeyear{Elliott15}, \citeyear{Elliott16}) also found it to be a high-probability member.\\
\citet{Torres08} reported on the discovery of a close companion separated by 0.8$^{\prime\prime}$ ($\sim$60\,AU). 
\citet{Malo14a} measured $v\sin{i}$ = 23.5$\pm$1.9\,km\,s$^{-1}$, inferred a distance d = 64\,pc,
and did not detect any significant RV variation. \\
We observed this target at the YCO  in the BVR filters and  measured a rotation period P = 1.906$\pm$0.005\,d and an amplitude of light curve $\Delta$V = 0.10\,mag (see online Fig.\,\ref{A39}).
We also retrieved a long photometric time series from the INTEGRAL/OMC archive (ID 8350000036). However,  all frames are reported to have bad PSF. Their LS and CLEAN periodograms
revealed a rotation period  P = 1.24$\pm$0.01\,d, which we mention to provide more complete information but we do not use it  in our analysis.\\

\item \bf \object{HD 160305} \rm \\

S+D; V$-$K$_s$ = 1.358\,mag; P = 1.336\,d\\

HD\,160305 is listed in the Hipparcos catalog as a F9V star. This star hosts a debris disk \citep{Patel14}.
It was identified as new member of the association by \citet{Kiss11} who did not find evidence that this star may reside in a tight
multiple system. \citet{Song12} suggested it is younger than the $\beta$ Pictoris association and belongs to the Scorpius-Centaurus complex. \\
\citet{Messina11} from  LS and CLEAN analyses of ASAS data detected two periodicities, P = 3.92\,d and P = 1.336\,d, which are present in almost all seasons. 
These are in 1-d beat relation, meaning that only one is the true rotation period. Unfortunately, no $v\sin{i}$ is measured to independently constrain the rotation period. 
Similar  periods P = 3.93\,d and P = 1.339\,d were also found by \citet{Kiraga12}. Based on the color-period distribution of $\beta$ Pictoris members, 
the shorter period P = 1.336\,d, also reported by \citet{Messina11}, is closer, but still longer than the periods of single bona fide members.
The periodogram of Hipparcos data gives P = 3.63\,d, but other peaks of comparable power are present. 
We retrieved a long photometric time series from the INTEGRAL/OMC archive (ID 8355000055). However, neither  P=1.33\,d nor P = 3.93\,d were found in the periodograms.
We inferred  a luminosity L = 1.97$\pm$0.54\,L$_\odot$ and a radius R = 1.27$\pm$0.39\,R$_\odot$. \\

\item \bf \object{TYC 8742 2065 1}  \rm  \\

Tc; V$-$K$_s$ = 2.164\,mag; P = 2.60\,d\\

TYC\,8742\,2065\,1 is a triple system consisting of a SB2 K0IV+? with a nearby visual companion \citep{Torres08}  physically associated \citep{Chauvin10} at $\rho$ = 0.114$^{\prime\prime}$ (PA = 232$^{\circ}$) and $\Delta$K  = 0.2\,mag. \citet{Torres06} listed this system among the members of the association, whereas \citet{Song12} suggested it is younger than the members of the $\beta$ Pictoris association and belongs to the Scorpius-Centaurus complex. \citet{Torres06} reported a $v\sin{i}$ = 10$\pm$1.4\,km\,s$^{-1}$ and \citet{Weise10}  $v\sin{i}$ = 10.1$\pm$1\,km\,s$^{-1}$. \\
The periodogram analysis of \citet{Messina10} revealed two peaks at P = 2.60\,d  and P = 1.62\,d. The period P = 1.614\,d is also reported by \citet{Kiraga12}. The shorter period may be the rotation period of the SB2. 
The longer period may be the rotation period of the visual companion.\\
 
\item \bf \object{HIP 88399} \rm\\

Bw; V$-$K$_s$ = 4.227\,mag; P = ?\\

HIP\,88399 is a F6V + M2Ve spectroscopic binary whose components are separated by $\rho$ = 6.44$^{\prime\prime}$ (PA = 89$^{\circ}$) \citep{Rameau13}. The membership of the association was first proposed by \citet{Zuckerman01}, later by \citet{Zuckerman04},   \citet{Moor06}, \citet{Torres06},  \citet{Makarov07}, and \citet{Lepine09}. \citet{Scholz07} reported $v\sin{i}$$_A$ = 22.5 km\,s$^{-1}$ and \citet{Torres06} $v\sin{i}$$_A$ = 20.0 km\,s$^{-1}$. From our analysis of ASAS time series we could not derive the stellar rotation period.\\

\item \bf \object{V4046 Sgr} \rm  \\

SB2+D; V$-$K$_s$ = 3.191\,mag; P = 2.42\,d\\

V4046\,Sgr  is a spectroscopic binary consisting of two nearly equal-mass  K5Ve + K7Ve components \citep{Stempels04} 
separated by 0.045\,AU with a nearby visual M1e companion UCAC2\,18035440 at $\rho$ = 2.82$^{\prime}$ \citep{Kastner11}.
The binary hosts a circumbinary dust disk  \citep{Jensen97} and it is actively accreting from a gaseous disk \citep{Stempels04}.
The circumbinary molecular disk is inclined at 33.5$^{\circ}$ \citep{Rodriguez10} and extends to
$\sim$ 350\,AU. The membership was proposed by \citet{Torres06} and confirmed by \citet{Kastner11}. It is also considered as bona fide member by \citet{Pecaut13},  Malo et al. (\citeyear{Malo13}, \citeyear{Malo14b}), and most recently by Elliott et al. (\citeyear{Elliott14}, \citeyear{Elliott15}).
The $v\sin{i}$$_A$ = 14.2\,km\,s$^{-1}$ and $v\sin{i}$$_B$ = 13.7\,km\,s$^{-1}$ are reported by \citet{Stempels04} and  $v\sin{i}$$_A$ = 14\,km\,s$^{-1}$ by \citet{daSilva09}. \\
The first rotation period determination was P = 1.7\,d by \citet{Busko78}. A period P = 2.43\,d was later reported by \citet{delaReza86}; P = 2.44\,d was found by \citet{Mekkaden91}. The orbital period P = 2.4213305\,d from radial velocity  measurements was determined by \citet{Quast00}, and P = 2.42537\,d by \citet{Stempels04}. The LS and CLEAN periodogram analysis by \citet{Messina10} revealed in almost all seasons two peaks at P = 2.42\,d and P = 1.71\,d. Assuming that the axial rotations of both components are synchronized with the orbital period, then we inferred that P = 1.71d is the beat period. On the other hand, we note that in the time interval JD 2453576--2453580, when ASAS observations were very highly sampled for a total of about 300 observations, the period P = 2.42\,d is absent, whereas P = 1.71\,d is highly significant. \\

\item \bf \object{UCAC2 18035440} \rm \\

SB; V$-$K$_s$ = 4.241\,mag; P = 12.05\,d\\

UCAC2\,18035440 is classified as  M1.5V by \citet{Riaz06}    from the TiO-band strength at a spectroscopic distance d = 51\,pc. 
It is associated with the X-ray source 1RXS\,J181422.6-324622   with  L$_{\rm X}$/L$_{\rm bol}$ = $-$2.69. 
\citet{Torres06} reported a  $v\sin{i}$ = 3$\pm$1.5\,km\,s$^{-1}$ and suggested it may be a spectroscopic binary. 
These authors first proposed its membership of the $\beta$ Pictoris association. \citet{Kastner11} suggested this star is likely a widely separated ($\rho$ = 2$^\prime$.82) companion of V4046\,Sgr on the basis of common proper motions and radial velocities. However, to be coeval,
the star must be a spectroscopic binary (M1+M1), as suspected by \citet{Torres06}. \citet{Elliott14} reported a larger distance d = 98.1\,pc derived from XYZ positions.
Song et al. (2012) suggested a membership to the Upper Scorpius and inferred a photometric distance d = 71\,pc.\\
A rotation period P = 12.05\,d was first reported by \citet{Nataf10} based on the HATnet photometry. After \citet{Nataf10}, we have analyzed again the ASAS photometry and recovered, with both LS and CLEAN periodograms, the same period P = 12.08\,d in two seasons.  We also retrieved a long photometric time series from the INTEGRAL/OMC archive (ID 7396000227). In the LS periodogram, we found the major peak at P = 5.96\,d and a light curve amplitude $\Delta$V = 0.14\,mag (see online Fig.\,\ref{A40}). This period is half of the rotation period measured by \citet{Nataf10}. This may arises from the presence of two major spot groups on opposite hemispheres at the epochs of INTEGRAL/OMC observations.
\\

\item \bf \object{2MASS J18151564-4927472} \rm \\

SB1; V$-$K$_s$ = 4.78\,mag; P = 0.447\,d\\

2MASS\,J18151564-4927472 is classified as  M3V by \citet{Riaz06}   from the TiO-band strength at a spectroscopic distance d = 29\,pc. 
It is associated with the X-ray source 1RXS\,J181514.7-492755  with  L$_{\rm X}$/L$_{\rm bol}$ = $-$3.17. 
\citet{Malo13} suggested its membership of the association with a probability of 91.2\% and inferred a distance d = 61$\pm$4\,pc.
\citet{Moor13} found that its kinematic properties match those of  the $\beta$ Pictoris members. However, owing to the lack of a detectable amount of Li, 
they proposed it as candidate member of the Argus association. \citet{Malo14a} measured the projected rotational velocity 
$v\sin{i}$ = 76.7$\pm$10.9\,km\,s$^{-1}$ and redetermined the membership probability to 89.4\%, which increases to 99.9\% when the RV information is considered. 
The large RV variation they measured suggests that it is a spectroscopic binary.\\
This star was observed at the YCO  from May 23 until September 29, 2014, for a total of 12 nights.
We obtained a total of 70 frames in the B and V filters and 64 frames in the R filter. We analyzed all time series with LS and CLEAN
periodograms and derived the most powerful power peak at P = 0.447$\pm$0.002\,d with a peak-to-peak light curve amplitude $\Delta$mag =  0.09\,mag by combining all filters (see online Fig.\,\ref{A41}).\\

\item \bf \object{HIP 89829} \rm \\

S; V$-$K$_s$ = 1.837\,mag; P = 0.571\,d\\

HIP\,89829 is a G1V \citep{Torres06} single star \citep{Waite11} at a distance d = 75\,pc as measured by Hipparcos. 
It was proposed by \citet{Torres08} and \citet{Malo13} as a high-probability member of the $\beta$ Pictoris 
association. Its kinematics matches those of the $\beta$  Pictoris members 
at the $<$ 2.0 sigma level \citep{Desidera15}. Moreover, the large lithium EW, high levels of activity,
and photometric variability are consistent with the $\beta$ Pictoris membership assignment, and the isochrone fitting further supports the pre-MS status (age of 15$\pm$3\,Myr) \citep{Desidera15}.
\citet{Torres06} measured  $v\sin{i}$= 114.7\,km\,s$^{-1}$, and a similar value  
$v\sin{i}$ = 114\,km\,s$^{-1}$ was measured by \citet{Waite11}. \\
The  rotation period P =  0.571\,d is reported by ACVS, \citet{Messina10}, and \citet{Kiraga12},  where all measurements are based on the ASAS time series. We also retrieved a long photometric time series from the INTEGRAL/OMC archive (ID 6856000028) and found a 
rotation period P = 0.5687$\pm$0.0005\,d and a light curve amplitude $\Delta$V = 0.07\,mag (see online Fig.\,\ref{A42}). 
We derived a luminosity L = 1.48$\pm$0.40\,L$_\odot$, a radius R = 1.38$\pm$0.43\,R$_\odot$, and an inclination $i$ = 70$\pm$12$^\circ$.\\

\item \bf \object{2MASS J18202275-1011131} (FK\,Ser) \rm \\

Bw+D; V$-$K$_s$ = 3.350\,mag; P = 4.65\,d\\

2MASS\,J18202275-1011131 is a K5Ve + K7Ve visual binary whose components have a separation $\rho$ = 1.33$^{\prime\prime}$ ($\sim$81\,AU) and a magnitude difference $\Delta$V = 0.7\,mag. 
\citet{Anthonioz15} reported the presence of a debris disk around the primary component. The membership was proposed by \citet{Torres06} and by Malo et al. (\citeyear{Malo13}, \citeyear{Malo14a}).\\
\citet{Chugainov74} first measured the photometric
rotation period P = 5.20\,d. Similar rotation period P = 5.15\,d was subsequently found by \citet{Batalha98}. However, also a period 
P = 4.89\,d by \citet{Batalha98} is found in some seasons. These two periods may refer to the two components that, having similar brightness, can both contribute
to the observed variability. \citet{Kiraga12} from ASAS data measured a rotation period P = 4.65\,d with  a light curve amplitude $\Delta$V = 0.07\,mag.
Our analysis of ASAS time series found two peaks of comparable power at P = 4.65\,d and P = 5.15\,d.
We also retrieved a long photometric time series from the INTEGRAL/OMC archive (ID 5681000021) and found a 
rotation period P = 5.00$\pm$0.02\,d and a light curve amplitude $\Delta$V = 0.07\,mag  (see online Fig.\,\ref{A43}).
 No projected rotational velocity is available to infer the inclination of the rotation axis.\\

\item \bf \object{2MASS J18420694-5554254} \rm \\

S?; V$-$K$_s$ = 4.946\,mag; P =5.40\,d\\

2MASS\,J18420694-5554254 is classified as  M3.5V by \citet{Riaz06} from the TiO-band strength at a spectroscopic distance d = 37\,pc. 
It is associated with the X-ray source 1RXS\,J184206.5-555426  with  L$_{\rm X}$/L$_{\rm bol}$ = $-$2.77. 
Malo et al. (\citeyear{Malo14a}, \citeyear{Malo14b}) suggested it is a member of the  association, measured an upper value of the projected rotational velocity $v\sin{i}$ $<$ 8.4\,km\,s$^{-1}$,
 and inferred a statistical distance d = 54\,pc. Membership is also proposed by \citet{Elliott16} who found the star 2MASS\,J18420483-5554126 to be a common proper motion companion at a separation of 1138\,AU.\\
\citet{Kiraga12}  found a rotation period P = 5.403\,d  and a light curve amplitude $\Delta$V = 0.07\,mag from ASAS data.
We retrieved a long photometric time series from the INTEGRAL/OMC archive (ID 8762000047) and found a 
rotation period P = 5.18$\pm$0.02\,d and a light curve amplitude $\Delta$V = 0.16\,mag  (see online Fig.\,\ref{A44}).
We derived a luminosity L = 0.065$\pm$0.018\,L$_\odot$, a radius R = 0.79$\pm$0.26\,R$_\odot$, and an inclination $i$ = 70$\pm$15$^\circ$.\\

\item \bf \object{TYC 9077 2489 1} \rm \\

Tc; V$-$K$_s$ = 3.204\,mag; P = 0.3545\,d\\

TYC\,9077\,2489\,1 is a triple system consisting of a close visual binary  and a
wide A7V companion (\object{HIP 92024}) at a distance of 70$^{\prime\prime}$. The close binary has a primary component with K8V spectral type \citep{Zuckerman01}, and is separated by $\rho$ = 0.18$^{\prime\prime}$ ($\sim$5\,AU) from the secondary component that is $\Delta$K = 2.3\,mag fainter \citep{Chauvin10}.
Its membership to the association was first proposed by \citet{Zuckerman01}, \citet{Zuckerman04}, and then by \citet{Moor06}, and was subsequently confirmed by \citet{Torres06}, \citet{Lepine09}, and \citet{Elliott16}. \citet{delaReza04} reported $v\sin{i}$ = 150\,km\,s$^{-1}$; \citet{Jayawardhana06} $v\sin{i}$ = 102.7\,km\,s$^{-1}$, with the caveat that this value may be affected by line blending; \citet{Torres06} measured  $v\sin{i}$= 110\,km\,s$^{-1}$ and \citet{Garcia-Alvarez11} $v\sin{i}$ = 121.3\,km\,s$^{-1}$. \\
The  rotation period P = 0.3545\,d is reported by \citet{Messina10} and \citet{Garcia-Alvarez11}, which is likely the rotation period of the brighter component K8V of the close binary. \\

\item \bf \object{TYC 9073 0762 1}\rm  \\

S; V$-$K$_s$ = 3.946\,mag; P = 5.37\,d\\

TYC\,9073\,0762\,1 is classified as an  M1Ve star by \citet{Torres06}. It is associated with the X-ray source 1RXS\,J184657.3-621037 \citep{Haakonsen09}.
\citet{Elliott14} and \citet{Malo13} did not find evidence of RV variations. 
Its membership to the association was first proposed by \citet{Torres06}, and then by \citet{Lepine09} who 
inferred a kinematic distance d = 55.8$\pm$3.5\,pc. \citet{Brandt14} proposed a lower 80\% membership probability of the $\beta$ Pictoris association.
\citet{Moor13} suggested that this star may be the wide separation ($\sim550^{\prime\prime}$) companion of the F5V star 
HD\,173167, having similar proper motions and radial velocities.
\citet{Torres06} measured a projected rotational velocity  $v\sin{i}$  = 9.9$\pm$0.6\,km\,s$^{-1}$.  \\
A rotation period P = 5.373\,d is reported in ACVS. The same rotation period P = 5.37\,d is also found by \citet{Messina10} in all ASAS seasons. We also retrieved a long photometric time series from the INTEGRAL/OMC archive (ID 9073000047) and found the same rotation period P = 5.35$\pm$0.01\,d and a light curve amplitude $\Delta$V = 0.15\,mag  (see online Fig.\,\ref{A45}).
We derived a luminosity L = 0.14$\pm$0.03\,L$_\odot$, a radius R = 0.95$\pm$0.12\,R$_\odot$, and an inclination $i$ $\sim$ 90$^\circ$.\\
We note that its light curve amplitude is significantly
larger than the amplitude exhibited by other stars sharing similar rotation period and spectral type.\\

\item \bf \object{HD 173167} \rm \\

SB1; V$-$K$_s$ = 1.144\,mag; P = 0.25\,d\\

HD 173167 is a probable SB1 binary consisting of F5V primary component. It is considered the wide companion of  TYC\,9073\,0762\,1 and, therefore, member of the $\beta$ Pictoris association (\citealt{Moor13}; \citealt{Elliott16}). \\
From ASAS data time series (ASAS 184806-6213.8), we derived a rotation period P = 0.250\,d with both LS and CLEAN periodogram analyses. 
We also retrieved a long photometric time series from the INTEGRAL/OMC archive (ID 9073000056) and found a similar rotation period P = 0.291$\pm$0.001\,d and a light curve amplitude $\Delta$V = 0.22\,mag  (see online Fig.\,\ref{A46}).
However, no projected rotational velocity has ever been measured to constrain this value.
The light curve amplitude is very large for the F5V spectral type. Even assuming an inclination  $i$ $\sim$ 90$^\circ$ that maximizes the rotational modulation, such an amplitude has never been observed in F-type dwarfs.\\

\item \bf \object{TYC 7408 0054 1}\rm \\

EB; V$-$K$_s$ = 3.66\,mag; P = 1.075\,d\\

TYC\,7408\,0054\,1 is a K8Ve star \citep{Lepine09}. It is associated with the X-ray source 1RXS\,J185044.7-314748. \citet{Torres06} first suggested its membership to the association, and later,  \citet{Lepine09} who inferred a kinematic distance d = 51.2\,pc, based on proper motion and membership of the association, and measured  $v\sin{i}$ = 50$\pm$2\,km\,s$^{-1}$. \citet{Torres06} measured $v\sin{i}$ = 49.7\,km\,s$^{-1}$. The RV measurements from \citet{Torres06}, \citet{Lepine09}, and \citet{Song12}
 range from $-$3 to $-$7.1\,km\,s$^{-1}$, suggesting it is a close binary star. 
 \citet{Malo13} found a membership probability of 92.5\%, which increases to 99\% when the RV information is also used. 
\citet{Song12} listed this star among candidate members of the Upper Scorpius.\\
In the ACVS this star is reported to be a semi-detached or contact eclipsing binary with a period P = 1.04154\,d. \citet{Messina10} detected a longer period P = 1.089\,d from their analysis of the same ASAS time series. 
A similar period P = 1.0420\,d was subsequently reported by \citet{Kiraga12}. An intriguing aspect of this star is that in most ASAS seasons only one light minimum is visible, whereas there is only evidence of two eclipses in three seasons. 
We also retrieved a long photometric time series from the INTEGRAL/OMC archive (ID 7408000058) and found a similar rotation period P = 1.075$\pm$0.005\,d and a light curve amplitude $\Delta$V = 0.10\,mag, but no evidence for eclipses (see online Fig.\,\ref{A47}).
Until spectroscopy is unavailable, we suspect that the double minima may arise from starspots on opposite hemispheres.
We derived a luminosity L = 0.201$\pm$0.05\,L$_\odot$, a radius R = 0.95$\pm$0.12\,R$_\odot$, and an inclination $i$ $\sim$ 90$^\circ$.\\

\item \bf \object{HIP 92680} (PZ Tel) \rm \\

Bw; V$-$K$_s$ = 1.924\,mag; P = 0.94\,d\\

HIP\,92680 is a K8Ve single star with a substellar $\sim$28\,M$_{Jup}$ companion \citep{Mugrauer10}. 
The membership of the association was first proposed by \citet{Zuckerman01}, later by \citet{Zuckerman04},   \citet{Torres06}, and \citet{Makarov07}, and \citet{Lepine09}.
Numerous measurements of projected rotational velocity are reported in the literature as follows:  $v\sin{i}$ = 70\,km\,s$^{-1}$ \citep{Randich93};  $v\sin{i}$ = 58\,km\,s$^{-1}$; \citep{Soderblom98};  $v\sin{i}$ = 68\,km\,s$^{-1}$ \citep{Barnes00};   $v\sin{i}$ = 70\,km\,s$^{-1}$ \citep{Cutispoto02}; 
 $v\sin{i}$ = 67\,km\,s$^{-1}$ \citep{delaReza04};  $v\sin{i}$ = 69\,km\,s$^{-1}$ \citep{Torres06};  and 
$v\sin{i}$ = 77.50\,km\,s$^{-1}$ \citep{Scholz07}. \\
The first rotation period determinations were P = 0.942\,d  by \citet{Coates80} and P = 0.943\,d by \citet{Coates82}. \citet{LloydEvans87} reported P = 0.9447\,d, and \citet{Innis90} P = 0.94486\,d. A period P = 0.9447\,d is determined by \citet{Cutispoto98a}, P = 0.94\,d by \citet{Innis07}. \citet{Kiraga12} reported a period P = 0.9457\,d. 
The most recent photometric investigation on both PZ Tel and the companion brown dwarf is reported by \citet{Maire16} together with the high-contrast imaging results obtained with SPHERE at VLT. In that work the known rotation period of PZ Tel was confirmed by new photometric observations carried out at ESO with the Rapid Eye Mount (REM) Telescope and at the YCO.\\

\item   \bf \object{TYC 6872 1011 1}\rm \\

Bw; V$-$K$_s$ = 3.762\,mag; P = 0.503\,d\\

TYC\,6872\,1011\,1 is a M0Ve single star whose membership of the association was first proposed by \citet{Torres06} who reported $v\sin{i}$  = 33.8\,km\,s$^{-1}$. 
\citet{Moor13} suggested that this star is a likely member and may form a visual binary with \object{UCAC2 19527490} at a distance of 28.3$^{\prime\prime}$ ($\sim$2210\,AU), as their RV and proper motions are consistent to each other. UCAC2 19527490 is also found by \citet{Elliott16} to be the common proper motion companion. Malo et al. (\citeyear{Malo13}, \citeyear{Malo14a},\citeyear{Malo14b}) also found it to be a high-probability member at a predicted distance  d = 76\,pc.
It is  also listed among the association members by \citet{Pecaut13} and Elliott et al. (\citeyear{Elliott15}, \citeyear{Elliott16}).\\
The rotation period P = 0.504\,d was determined by \citet{Messina10} from ASAS data. In the time interval JD 2452781--245288,  the ASAS sampling was very high, allowing a period measurement with a very high confidence level. The same period was subsequently reported by \citet{Kiraga12}. 
We also retrieved a long photometric time series from the INTEGRAL/OMC archive (ID 6872000042) and found a rotation period P = 0.343$\pm$0.005\,d and a light curve amplitude $\Delta$V = 0.27\,mag (see online Fig.\,\ref{A48}). However, this photometry is reported to suffer from bad PSF. We derived a luminosity L = 0.31$\pm$0.1\,L$_\odot$, a radius R = 1.31$\pm$0.17\,R$_\odot$, and an inclination $i$ $\sim$ 15$^\circ$.\\

\item  \bf \object{2MASS J19102820-2319486} \rm \\

S; V$-$K$_s$ = 4.985\,mag; P = 3.60\,d\\

2MASS\,J19102820-2319486  is classified as  M4V by \citet{Riaz06}  from the TiO-band strength at a spectroscopic distance d = 18\,pc. 
It is associated with the X-ray source 1RXS\,J191028.6-231934 with  L$_{\rm X}$/L$_{\rm bol}$ = $-$2.97.
\citet{Malo13} suggested its membership to the association with a probability of 99.9\% and a predicted statistical
distance d = 67$\pm$5\,pc. The membership was confirmed by Binks \& Jeffries (\citeyear{Binks14}, \citeyear{Binks16}) 
who derived a kinematic distance d = 69.2$\pm$3.4\,pc, and again it was proposed as a candidate member by Malo et al. (\citeyear{Malo14a},\citeyear{Malo14b}), who measured a projected rotational velocity 
$v\sin{i}$ = 12.2$\pm$1.8\,km\,s$^{-1}$ and found no evidence of significant RV variations.\\
 This star was observed at the YCO  from October 20 until November 11, 2014, for a total of seven nights. We  obtained a total of 51 frames in the B, V, and R filters. 
We analyzed all time series with LS and CLEAN periodograms and derived the most powerful power peak at P = 3.64$\pm$0.02\,d with a peak-to-peak light curve amplitude $\Delta$V = 0.13\,mag (see online Fig.\,\ref{A49}).
We found the same period  P = 3.60$\pm$0.01\,d (see online Fig.\,\ref{A50}) in the LS and CLEAN periodograms of SuperWASP data (1SWASP J191028.18-231948.0).
We derived a luminosity L = 0.12$\pm$0.03\,L$_\odot$ and a radius R = 1.09$\pm$0.15\,R$_\odot$.
Combining stellar radius, projected rotational velocity, and rotation period, we derived an inclination of $i$ = 55$^{\circ}$.\\

\item \bf \object{TYC 6878 0195 1} \rm \\

Bw; V$-$K$_s$ = 2.904\,mag; P = 5.70\,d\\

TYC\,6878\,0195\,1 is a wide visual binary whose primary component is a K4Ve star at 1.1$^{\prime\prime}$ ($\sim$88\,AU) from the secondary \citep{Torres08} with a magnitude V =13.80\,mag \citep{Mason01}. \citet{Torres06}, who first proposed the membership of this star in the association,  reported $v\sin{i}$  = 9.8\,km\,s$^{-1}$. It is also considered member of the $\beta$ Pictoris association  by \citet{Pecaut13} and Elliott et al. (\citeyear{Elliott14}, \citeyear{Elliott15}).\\
The rotation period  P = 5.70\,d was first determined by \citet{Messina10}, based on ASAS photometric time series. This period was subsequently confirmed by the analysis of SuperWASP time series by \citet{Messina11}.\\

\item \bf \object{2MASS J19233820-4606316} \rm \\

S; V$-$K$_s$ = 3.598\,mag; P = 3.237\,d\\

2MASS\,J19233820-4606316 is classified as  M0V by \citet{Riaz06}  from the TiO-band strength at a spectroscopic distance d = 57\,pc.
They associated this star with the X-ray source 1RXS\,J192338.2-460631, which exhibits strong X-ray to bolometric luminosity 
L$_{\rm X}$/L$_{\rm bol}$ = $-$3.24. The membership of the $\beta$ Pictoris association was first proposed by \citet{Malo13} and later by \citet{Malo14a} as a high-probability member with a predicted distance d = 70\,pc 
and a measured projected rotational velocity 
$v\sin{i}$ = 15.4$\pm$3\,km\,s$^{-1}$. Also \citet{Malo14a} did not find  evidence of significant RV variations. \citet{Moor13} measured a kinematic distance d = 71\,pc and identified it as a new 
probable candidate member of the association, measuring the same RV as Malo et al. (2013).
Therefore, we consider it a single star. \\
The rotation period P = 3.242\,d and V-band light curve amplitude $\Delta$V = 0.11\,mag were measured  by \citet{Kiraga12} by analyzing the ASAS photometric time series. We analyzed the SuperWASP time series of this target (1SWASP J192338.19-460631.5) and our LS and CLEAN periodograms confirmed the rotation period P = 3.237$\pm$0.005\,d with a light curve amplitude $\Delta$V = 0.07\,mag (see online Fig.\,\ref{A51}).
We derived a luminosity L = 0.19$\pm$0.05\,L$_\odot$, a radius R = 0.92$\pm$0.30\,R$_\odot$, and an inclination $i$ $\sim$ 90$^\circ$.\\

  \color{black}
\item \bf \object{2MASS J19243494-3442392} \rm  \\

Bc?; V$-$K$_s$ = 5.495\,mag; P = 0.7072\,d\\

2MASS\,J19243494-3442392 is classified as  M4V by \citet{Riaz06}  from the TiO-band strength at a spectroscopic distance d = 13\,pc.
They associated this star with the X-ray source 1RXS\,J192434.2-344230, which exhibits strong X-ray to bolometric luminosity 
L$_{\rm X}$/L$_{\rm bol}$ = $-$3.27. \citet{Malo14b} found it to be a high-probability member with a predicted distance d = 54 pc and a projected rotational velocity 
$v\sin{i}$ = 10.9$\pm$2.9\,kms$^{-1}$. They obtained three RV measurements that exhibit a small but significant variability
that point toward a binary nature of this star.\\
In 2015 we carried out a photometric monitoring of this object  at the  ROAD  Observatory. Our LS and CLEAN analyses of the V and I time series revealed a rotation period P = 0.7072$\pm$0.008\,d, an amplitude of the light curve $\Delta$V = 0.02\,mag and $\Delta$I = 0.015\,mag, and a positive correlation between the V and V$-$I color variations (the star is redder when it is fainter) (see online Fig.\,\ref{A52}).
We were able to retrieve a magnitude time series for this star 
(SSS\_J192435.0-344240) from the CSS survey  whose periodogram analysis provided a rotation period P = 0.74$\pm$0.01\,d (see online Fig.\,\ref{A53}). 
This target was also observed by SuperWASP (1SWASP J192434.97-344239.3). For this data the LS periodogram gives the highest periodicity at one day and a secondary period at P = 0.678$\pm$0.005\,d (see online Fig.\,\ref{A54}), which is in good agreement with the period found with our ROAD observations;  whereas the CLEAN periodogram did not reveal any significant power peak.\\

\item \bf \object{TYC 7443 1102 1} \rm\\

Tw; V$-$K$_s$ = 3.954\,mag; P = 11.3\,d\\

TYC\,7443\,1102\,1 is a M0.0V single star at a predicted distance of d = 57.7\,pc and proposed by \citet{Lepine09}, together with the common proper motion
companion \object{2MASS J195602.8-320720} at an angular distance of $\sim$ 26.3$^{\prime\prime}$ ($\sim$1450\,AU), to be  a member of the $\beta$ Pictoris association (see, also \citealt{Elliott16}). 
Also \citet{Kiss11} found these two stars to have the same kinematic distances and radial velocities,
confirming they are a physical pair in the association.
They also suggested that this pair may be physically associated with the other member, \object{2MASS J20013718-3313139}.
A high-probability membership of both components is also found by \citet{McCarthy12}, on the basis of Li EW, and by \citet{Malo14b}.
This star has been investigated by \citet{Delorme12} and \citet{Biller13} to search for substellar mass companions, but no evidence has been found.
\citet{Messina11} reported for TYC\,7443\,1102\,1 a rotation period  of P = 11.3\,d, which was found by LS and CLEAN periodograms in all the ASAS observation seasons and a period of P = 11.8\,d  in all the SuperWASP observation seasons.  The rotation period P = 11.3\,d combined with a stellar radius R = 0.96\,R$_{\odot}$ is in agreement with the rotational velocity  $v\sin{i}$ = 6$\pm$2\,km\,s$^{-1}$  measured by \citet{Lepine09}. \\
We derived a luminosity L = 0.16$\pm$0.03\,L$_\odot$, a stellar radius R = 0.94$\pm$0.09\,R$_{\odot}$ and, in combination with the rotational velocity, we inferred   $\sin{i}$ = 1.18$\pm$0.1. This high value may arise from an underestimation of the stellar radius.\\

\item \bf \object{2MASS J19560294-3207186}AB \rm \\

Tc; V$-$K$_s$ = 5.116\,mag; P = 1.569\,d\\

2MASS\,J19560294-3207186AB is classified as  M4V by \citet{Riaz06}  from the TiO-band strength. They associated this star with the X-ray source 1RXS\,J195602.8-320720, which exhibits strong X-ray to bolometric luminosity 
L$_{\rm X}$/L$_{\rm bol}$ = $-$2.91. Malo et al. (2014a) measured a projected rotational velocity $v\sin{i}$ = 35$\pm$4.9\,km\,s$^{-1}$. 
This star was discovered to be a close visual binary whose components have a separation $\rho$ = 0.2$^{\prime\prime}$ ($\sim$10\,AU), an orbital period P = 66 yr \citep{Bowler15}, and a magnitude difference $\Delta$H = 1.22\,mag.
Together  with TYC\,7443\,1102\,1 and 2MASS\,J20013718-3313139, they form a hierarchical quadruple system.\\
The membership of the association was proposed by \citet{Lepine09},  \citet{Kiss11}, and \citet{McCarthy12}.
This star was photometrically monitored at KKO. We detected with both LS and CLEAN analyses only one significant power peak at 
P = 1.569$\pm$0.008\,d, which we attribute to the brighter A component  (see online Fig.\,\ref{A55}).\\

\item \bf \object{2MASS J20013718-3313139} \rm  \\

Tw; V$-$K$_s$ = 4.056\,mag; P = 12.8\,d\\

2MASS\,J20013718-3313139 is classified as  M1V by \citet{Riaz06}  from the TiO-band strength at a spectroscopic distance d = 48\,pc.
They associate this star with the X-ray source 1RXS\,J200136.9-331307, which exhibits strong X-ray to bolometric luminosity 
L$_{\rm X}$/L$_{\rm bol}$ = $-$3.39. 
\citet{Elliott14}  did not find evidence for multiplicity from their RV measurements. 
\citet{Kiss11} first proposed it to be a likely member of the association at a distance d = 62\,pc. 
\citet{Malo13} also proposed this star as member with a probability of 99.9\%.  
\citet{Kiss11} suggested that this star is associated with TYC\,7443\,1102\,1 since they found these two stars  have the same kinematic distances and radial velocities,
confirming they are a physical pair in the association. TYC\,7443\,1102\,1 is itself associated with the visual close binary 2MASS J19560294-3207186AB.
Malo et al. (\citeyear{Malo13}, \citeyear{Malo14a}) measure an upper limit for the projected rotational velocity $v\sin{i}$ $<$ 2.6\,km\,s$^{-1}$ and find it a high probability member. Also \citet{Pecaut13} and  \citet{Elliott15} consider it a member of the association. \\
ASAS and SuperWASP photometric time series were analyzed by \citet{Messina11} who found in all seasons a rotation period 
P = 12.8$\pm$0.2\,d (ASAS) and 12.7$\pm$0.2\,d (SuperWASP), and a maximum light curve amplitude $\Delta$V = 0.13\,mag. 
Similar period P = 12.77\,d was also found by \citet{Kiraga12} in the ASAS data.
We derived a luminosity L = 0.14$\pm$0.03\,L$_\odot$, a radius R = 0.89$\pm$0.09\,R$_\odot$, and an inclination $i$ $=$ 47$\pm$5$^\circ$.
\\

\item \bf  \object{2MASS J20055640-3216591} (V5663\,Sgr) \rm \\

S; V$-$K$_s$ = 4.022\,mag; P = 8.368\,d\\

\citet{Moor13} assigned it a M2 spectral type on the basis of its effective temperature and suggested that this star
 is a probable member of the association. They derived a kinematic distance  d = 52\,pc and detected no sign for multiplicity.
 Therefore, we consider it a single star.
It is associated with the X-ray source 1RXS\,J200556.1-321651 with a X-ray to bolometric luminosity L$_{\rm X}$/L$_{\rm bol}$ = $-$3.56.\\
This star was observed by SuperWASP in three seasons (1SWASP J200556.41-321658.6). Our LS and CLEAN analyses  of this data allowed us to find the rotation period P = 8.368$\pm$0.005\,d
in both periodograms with a maximum light curve amplitude of $\Delta$V = 0.13\,mag  (see online Fig.\,\ref{A56}).
\citet{Berdnikov08} measured  a similar rotation period P = 8.307\,d analyzing the ASAS photometry time series.
We derived a luminosity L = 0.13$\pm$0.03\,L$_\odot$ and a radius R = 0.86$\pm$0.09\,R$_\odot$. \\

\item \bf \object{HD 191089} \rm \\

S+D; V$-$K$_s$ = 1.104\,mag; P = 0.488\,d\\

HD\,191089 is a F5V single star first proposed by \citet{Barrado99} as member of the $\beta$ Pictoris association. Subsequently, it was considered by \citet{Moor06}, \citet{Torres08}, \citet{Lepine09}, and \citet{Malo13} as a bona fide member. 
The star hosts a debris disk with
T$_{dust}$ = 95\,K, R$_{dust}$ = 15\,AU, and M$_{dust}$ = 3.4$\times$10$^{-2}$M$_{\oplus}$ \citep{Rhee07}.
The disk was recently spatially resolved by \citet{Churcher11}.
The  literature RVs  show  a  modest
scatter, which seems compatible with measurement errors. 
The projected rotational velocity was measured by \citet{Schroder09} $v\sin{i}$ = 37.7\,km\,s$^{-1}$
and by \citet{White07}  $v\sin{i}$ = 37.0\,km\,s$^{-1}$.\\
A tentative period P = 0.488\,d 
is derived from the Hipparcos data analysis \citep{Desidera15}.  We derived a luminosity L = 2.89$\pm$0.80 L$_\odot$, a radius R = 1.35$\pm$0.10 R$_\odot$, and an inclination $i = 15$$\pm$2$^\circ$.\\

\item \bf \object{2MASS J20100002-2801410}AB\rm \\

Bc; (V$-$K$_s$) = 4.64\,mag; P = 0.4702\,d\\

2MASS\,J20100002-2801410AB is classified as  M3V by \citet{Riaz06}  from the TiO-band strength at a spectroscopic distance d = 26\,pc.
They associate this star with the X-ray source 1RXS\,J201001.0$-$280139, which exhibits strong X-ray to bolometric luminosity 
L$_{\rm X}$/L$_{\rm bol}$ = $-$3.16. \citet{Bergfors10} observed this target in 2008.87 with the Lucky Imaging camera AstraLux Sur and discovered it is a close visual system consisting of M2.5 + M3.5 components that have a separation $\rho$ = 0.615$^{\prime\prime}$ ($\sim$16 AU), a position angle PA = 280.4$^{\circ}$, and  magnitude differences 
$\Delta z^{\prime}$ =  0.80\,mag and $\Delta i^{\prime}$ =  0.75\,mag. Janson et al. (2012) derived the following masses M$_A$ = 0.355\,M$_\odot$ and M$_B$  = 0.245\,M$_\odot$, a separation of 19.5\,AU, and an orbital period of 157\,yr.
\citet{Malo13} proposed this system to be member of the association with a probability of 99\% 
and inferred a statistical distance d = 53\,pc. \citet{Malo14a} measured a projected rotational velocity $v\sin{i}$ = 46.5$\pm$4.1\,km\,s$^{-1}$ and variable radial velocity,  and they listed this star among the bona fide members of the association. This system was investigated by \citet{Riedel14} who inferred a distance d = 48\,pc from parallax measurements and deblended magnitudes V = 13.62\,mag and I = 10.85\,mag for the A component and V = 13.86\,mag and I = 11.06\,mag for the B component confirming the membership of the association. Their Hubble-FGS observations confirmed the separation $\rho$ = 0.614$^{\prime\prime}$ and position angle PA = 281.6$^{\circ}$.\\
This system was observed by SuperWASP in two seasons (1SWASP J201000.03-280140.7). Our LS and CLEAN periodogram analyses exhibit the most powerful peak at P = 0.4702$\pm$0.0003\,d and other peaks, which are its beat periods. When we use the CLEAN periodogram, which effectively removes the aliasing effect arising from the spectral window, we find that the most powerful
peak is at P = 0.470204\,d (see online Fig.\,\ref{A57}). No evidence of a second periodicity related to the other component is found.\\

\item \bf  \object{2MASS J20333759-2556521}\rm \\

S; V$-$K$_s$ = 5.993\,mag; P = 0.71\,d\\

2MASS J20333759-2556521 is classified as  M4.5V by \citet{Riaz06}  from the TiO-band strength at a spectroscopic distance d = 13\,pc.
They associate this star with the X-ray source 1RXS\,J203336.9-255654, which exhibits strong X-ray to bolometric luminosity 
L$_{\rm X}$/L$_{\rm bol}$ = $-$3.15. Its membership was first suggested by \citet{Malo13} and then confirmed by \citet{Malo14b} who listed it among the bona fide members of the association and measured a projected rotational velocity   $v\sin{i}$ = 21\,km\,s$^{-1}$ and  a high Li EW = 504 m\AA. Their multi-epoch RV measurements ruled out its binary nature.
Membership was subsequently confirmed by  \citet{Riedel14}. \\
Our LS and CLEAN analyses of SuperWASP data (1SWASP J203337.61-255651.7) allowed us to measure a photometric rotation period P = 0.7097$\pm$0.0005\,d
and a light curve amplitude $\Delta$V = 0.05\,mag (see online Fig.\,\ref{A58}). We derived a luminosity L = 0.037$\pm$0.009\,L$_\odot$, a radius R = 0.76$\pm$0.09\,R$_\odot$, and an inclination $i  = 23$$\pm$3$^\circ$.\\

  \color{black}

\item \bf \object{HIP 102141}AB (\object{AT Mic}) \rm \\

Bc; (V$-$K$-s$)$_A$ = (V$-$K$_s$)$_B$ = 5.416\,mag; P = 1.19\,d; P = 0.78\,d\\

HIP\,102141AB  is a M4 + M4 close visual binary system ($\rho$ = 3.3$^{\prime\prime}$; \citealt{McCarthy12}) ($\sim$34\,AU). The membership of the association was first proposed by \citet{Barrado99}, and later by \citet{Zuckerman01}, \citet{Zuckerman04}, \citet{Torres06}, and \citet{Makarov07}, \citet{Lepine09}, \citet{Nakajima10}, and \citet{Riedel14}.
\citet{Torres06} reported $v\sin{i}$$_A$ = 10.1\,km\,s$^{-1}$ and $v\sin{i}$$_B$ = 15.8 \,km\,s$^{-1}$ for the components A and B, respectively; whereas \citet{Scholz07} determined $v\sin{i}$$_A$ = 10.56\,km\,s$^{-1}$ and $v\sin{i}$$_B$ = 17\,km\,s$^{-1}$. The most recent photometric analysis of this system is presented by \citet{2016c}.
Our analysis of the first season of SuperWASP data, allowed us to measure the stellar rotation periods of both components. Although the photometry could not resolve the system, both components equally contribute to the stellar variability, allowing an accurate period measurement. We find the  rotation periods P = 1.191$\pm$0.005\,d and  P = 0.781$\pm$0.002\,d. The  period  P = 0.781\,d was also found by \citet{Kiraga07} from the analysis of spatially unresolved ASAS data. The period P = 1.191\,d was also retrieved by \citet{Messina11} from the same SuperWASP photometry. For instance,  the first determination reported by \citet{Messina10} turned out to be the beat period of the faster component.
For AT Mic A we derived a luminosity L = 0.034$\pm$0.01\,L$_\odot$, a radius R = 0.61$\pm$0.09\,R$_\odot$, and an inclination $i  \sim 25$$^\circ$. 
For AT Mic B, we derived a luminosity L = 0.031$\pm$0.01\,L$_\odot$, a radius R = 0.59$\pm$0.09\,R$_\odot$, and the same inclination $i  \sim 25$$^\circ$. \\

\item \bf       \object{2MASS J20434114-2433534} \rm \\

Bc; V$-$K$_s$ = 4.971\,mag; P = 1.61\,d\\ 

2MASS\,J20434114-2433534 is a close visual binary consisting of  M3.7 + M4.1 components at a trigonometric distance d = 28\,pc \citep{Shkolnik12} with separation $\rho$ = 1.47$^{\prime\prime}$ (PA = 217$^{\circ}$) ($\sim$ 42\,AU).
\citet{Malo14a} measured a projected rotational velocity $v\sin{i}$ = 26\,km\,s$^{-1}$ (Malo et al. \citeyear{Malo13}; \citeyear{Malo14a}) and found it to be a highly probability member of the association. In earlier studies by Shkolnik et al. (\citeyear{Shkolnik09}, \citeyear{Shkolnik12}) it was first identified as candidate member of the Castor moving group  with an estimated age of about 200\, Myr (or younger)  with quality match BAA.\\
Our LS analysis of SuperWASP data (1SWASP J204341.16-243352.8) allowed us to measure the photometric rotation period P = 1.61$\pm$0.01\,d
and a V-band light curve amplitude $\Delta$V = 0.04\,mag, whereas the CLEAN algorithm found only one major power peak at one day (see online Fig.\,\ref{A59}). \\

\item   \bf \object{HIP 102409} (AU Mic) \rm \\

 S+D; V$-$K$_s$ = 4.201\,mag; P = 4.83\,d\\

HIP\,102409 is a M1Ve \citep{Maldonado15} single star \citep{Bailey12}. The SEEDS images do not detect any companions within 3.2$^{\prime\prime}$, $\sim$30 AU projected \citep{Brandt14}.
It hosts a well-known debris disk that appears nearly edge-on and extends out to 200\,AU in radius \citep{Kalas04}. The membership of the association was first proposed by \citet{Barrado99}, and later by \citet{Zuckerman01},  \citet{Zuckerman04}, \citet{Torres06}, \citet{Makarov07},  \citet{Lepine09}, and confirmed in several subsequent studies.
\citet{Weise10} reported $v\sin{i}$ = 8.2\,km\,s$^{-1}$; \citet{Browning10} $v\sin{i}$ $<$ 8.5\,km\,s$^{-1}$; \citet{Scholz07} $v\sin{i}$ = 8.49\,km\,s$^{-1}$;  and \citet{Torres06} $v\sin{i}$ = 9.3\,km\,s$^{-1}$. \\
The rotation period P = 4.865\,d was discovered by \citet{Torres73}. A new determination P = 4.854\,d was performed by \citet{Busko78}. A period P = 4.852\,d is reported by \citet{Kiraga12} and P = 4.822\,d is listed in ACVS. 
A period P = 4.837\,d is found by \citet{Hebb07}. Our LS and CLEAN analyses find P = 4.86\,d from ASAS, P = 4.83\,d from SuperWASP time series, and  P = 4.890\,d from Hipparcos data.
We derived a luminosity L = 0.105$\pm$0.03\,L$_\odot$, a radius R = 0.82$\pm$0.08\,R$_\odot$, and an inclination $i = 80$$\pm$10$^\circ$.\\

\item \bf \object{HIP 103311} \rm \\

Bc; V$-$K$_s$ = 1.539\,mag; P = 0.3558\,d\\

HIP\,103311 is a close visual binary consisting of a F8V + M0V at 1.1$^{\prime\prime}$  ($\sim$ 49 AU) \citep{Kaisler04}. 
Its membership to the association was proposed by \citet{Zuckerman01}, \citet{Zuckerman04}, \citet{Torres06},  \citet{Lepine09}, and later confirmed by other studies.
It is one of the fastest rotating members:  $v\sin{i}$ = 115.5\,km\,s$^{-1}$ was measured by \citet{Garcia-Alvarez11}; $v\sin{i}$ = 127.5\,km\,s$^{-1}$ by \citet{Torres06}; and $v\sin{i}$ = 160\,km\,s$^{-1}$ by \citet{Weise10}. \\
The stellar rotation period P = 0.3558\,d was first determined by \citet{Messina10} and subsequently confirmed by \citet{Garcia-Alvarez11} who inferred a radius R = 2.20\,R$_{\odot}$ and an inclination $i  = 21$$^\circ$.\\

\item   \bf \object{TYC 6349 0200 1} \rm (AZ Cap) \\

Bw; V$-$K$_s$ = 3.541\,mag; P = 3.40\,d\\ 

TYC\,6349\,0200\,1 is a wide visual binary consisting of  K6Ve + M2 components at 2.2$^{\prime\prime}$ ($\sim$105 AU) 
and with a magnitude difference $\Delta$K = 1.6\,mag (Neuhauser et al. \citeyear{Neuhauser02}; \citeyear{Neuhauser03}). 
The system has the  same proper motion as HIP\,103311 and is located nearby, therefore  that it
most certainly has the same distance as HIP\,103311 (\citealt{vandenAncker00}; \citealt{Elliott16}). Its membership was proposed by \citet{Zuckerman01}, \citet{Zuckerman04}, \citet{Torres06},  and \citet{Lepine09}. The value
 $v\sin{i}$ = 15.6\,km\,s$^{-1}$ is reported by \citet{Torres06}; $v\sin{i}$ = 14.6\,km\,s$^{-1}$ by \citet{Jayawardhana06}; $v\sin{i}$ = 12\,km\,s$^{-1}$ 
by \citet{delaReza04},  and $v\sin{i}$ = 20$\pm$2\,km\,s$^{-1}$ by \citet{Lepine09}. 
 \\
\citet{Messina10} found a period P = 3.4\,d from 
the ASAS photometry analysis. A period P = 3.403\,d is subsequently reported by \citet{Kiraga12} based on the same ASAS photometry, where 
the two components are unresolved. Similar period P = 3.4082\, is listed in ACVS. 
We derived a luminosity L = 0.28$\pm$0.08\,L$_\odot$, a radius R = 1.11$\pm$0.10\,R$_\odot$, and an inclination $i$ = 70$\pm$10$^\circ$. \\

\item \bf  \object{2MASS J21100535-1919573} \rm \\

S; V$-$K$_s$ = 4.344\,mag; P = 3.71\,d\\ 

2MASS\,J21100535-1919573 is classified as  M2V by \citet{Riaz06}  from the TiO-band strength at a spectroscopic distance d = 24\,pc.
They associate this star with the X-ray source 1RXS\,J211004.9$-$192005, which exhibits strong X-ray to bolometric luminosity 
L$_{\rm X}$/L$_{\rm bol}$ = $-$2.99. The RV measurements by Malo et al. (\citeyear{Malo13}, \citeyear{Malo14a}) and \citet{Moor13} do not differ significantly,
indicating it is likely a single star.
Its membership to the $\beta$ Pictoris association was investigated by \citet{Moor13} who found  the Galactic space motion and space 
motion to be consistent with those of other association members. However, their measured Li EW upper limit (EW$_{\rm Li} < 40$ m\AA) 
did not enable the application of their age diagnostics. Malo et al. (\citeyear{Malo13}, \citeyear{Malo14a})  confirmed it as a high-probability candidate member and measured
 $v\sin{i}$ = 9.7$\pm$1.2\,km\,s$^{-1}$ and a predicted distance d = 32\,pc. Also \citet{Elliott16} found it was a member.\\
This star was observed  by ASAS with a photometric precision $\sigma$ = 0.024\,mag. Considering this data, our LS and CLEAN periodogram analyses allowed us to find a rotation period P = 3.71$\pm$0.02\,d in the complete series as well as in 
its yearly seasons with a maximum light curve amplitude of $\Delta$V = 0.29\,mag. This star was also observed in the NSVS
 (ID 17204671) from June to October 1999 with a better photometric precision  $\sigma$ = 0.011\,mag. Our LS and CLEAN periodogram analyses allowed us 
 to find  the same rotation period P = 3.71$\pm$0.05\,d\ with a light curve amplitude of  $\Delta$V = 0.09\,mag (see online Fig.\,\ref{A60}).
In addition, we observed this target at CASLEO from September 3 untill October 22, 2014 for a total of nine nights. Our LS analysis revealed a period P = 3.71$\pm$0.01\,d with high 
confidence level and an amplitude $\Delta$R = 0.26\,mag. The CLEAN algorithm revealed its beat period at P=0.78\,d as its most significant peak (see online Fig.\,\ref{A61}). 
However, it would mean an inclination $i$ $<$ 10$^{\circ}$, and in contrast with the high amplitude of light variation.
We derived a luminosity L = 0.09$\pm$0.025\,L$_\odot$, a radius R = 0.79$\pm$0.08\,R$_\odot$, and an inclination $i$ = 73$\pm$7$^\circ$. \\
Among the single members of the association, this star exhibits a different behavior. Although it is the fastest star among four equal-mass stars (V$-$I = 1.97--2.00) in our sample, 
it shows the smallest content of lithium. On the contrary, we generally observe a positive correlation between Li content and rotation rate (see \citealt{Messina16a}). Moreover, this star exhibits a light curve amplitude that is significantly larger than the average observed among the other $\beta$ Pictoris members. \\

  
  \item \bf  \object{2MASS J21103147-2710578} \& \object{2MASS J21103096-2710513} \rm \\ 
  
 Bw;  V$-$K$_s$ = 5.60\,mag; P = 1.867\,d\\ 
  
This is a visual binary consisting of two M-type components at an angular distance of 9.5$^{\prime\prime}$ ($\sim$390\,AU).
\citet{Riaz06}  inferred a M4.5V spectral type  from the TiO-band strength and a spectroscopic distance d = 16\,pc 
for the primary component J21103147-2710578, and a M5V spectral type and a spectroscopic distance d = 22\,pc
for the secondary component  J21103096-2710513. Both components, owing to their small angular separation,
are associated with the same X-ray source
1RXS\,J211031.2$-$271046. Their X-ray to bolometric luminosities are
 L$_{\rm X}$/L$_{\rm bol}$ = $-$3.00 for the primary and L$_{\rm X}$/L$_{\rm bol}$ = $-$2.66 for the secondary \citep{Riaz06}.
A first position measurement was made by \citet{Cutri03} who found  a separation
$\rho$ = 9.4$^{\prime\prime}$  and a position angle PA = 313$^{\circ}$. Subsequently, this visual binary
was observed with the Lucky Imaging camera AstraLux Sur
that measured $\rho$ = 9.501$^{\prime\prime}$ (152 AU), position angle PA = 313.2$^{\circ}$, and magnitude differences 
$\Delta z^{\prime}$ =  1.07\,mag and $\Delta i^{\prime}$ =  1.20\,mag \citep{Bergfors10}. In their study,  \citet{Bergfors10} noticed 
that the two components form a common proper motion pair.
The components of this system were proposed as candidate members of the $\beta$ Pictoris association by \citet{Malo13} who inferred a membership probability of 99.9\%, a statistical distance d = 41 pc,  and a  projected rotational velocity for the primary component  of $v\sin{i}$ = 15.8$\pm$1.3\,km\,s$^{-1}$ and  $v\sin{i}$ = 14.6$\pm$2.4\,km\,s$^{-1}$ for the secondary \citep{Malo14a}. Membership was confirmed by Malo et al. (\citeyear{Malo14a}, \citeyear{Malo14b}).\\
This system was observed photometrically by SuperWASP (1SWASP J211031.38-271056.7) in two consecutive seasons. Owing to low angular resolution, both components were included in the aperture photometry. 
Our LS and CLEAN analyses detected a rotation period P = 0.650$\pm$0.007\,d in the complete series and in the first season. Owing to the faintness of the star, the achieved average photometric precision $\sigma$ = 0.14\,mag is very poor, whereas the light curve amplitude is $\Delta$V = 0.04\,mag. We note the presence of a secondary peak at P = 1.867\,d in the LS periodogram, which is the 1-d beat period, and it is removed by CLEAN (see online Fig.\,\ref{A62}).
For the primary, we derived a luminosity L = 0.017$\pm$0.005\,L$_\odot$, a radius R = 0.47$\pm$0.05\,R$_\odot$, and ${i}$ = 26$\pm$3$^{\circ}$.\\

  \color{black}
\item \bf \object{HIP 105441} (V390 Pav) \&  \object{TYC 9114 1267 1} \rm  \\

 Bw; (V$-$K$_s$)$_A$ = 2.370\,mag; (V$-$K$_s$)$_B$ = 3.581 \,mag; P$_A$ = 5.50\,d; P$_B$ = 20.5\,d\\

This is a visual binary consisting of  K2V + K7V components separated by 26$^{\prime\prime}$ ($\sim$785\,AU). The primary component A, designed as V390 Pav in the 74th special name-list of variable stars
\citep{Kazarovets99},  was proposed by \citet{Zuckerman01} 
as a possible member of the \object{Tucana association}. Subsequently, \citet{Ortega09} proposed it as a dynamical member of the 
$\beta$ Pictoris association.\\
 \citet{Torres06} did not detect the presence of the  Li line. 
The Li line was not detected also by \citet{Song03}. Doubts on the membership has also been raised by \citet{Elliott16} since the position of the primary in the M$_{\rm V}$-T$_{\rm eff}$ diagram is below the 24-Myr isochrone that fits the other members of the $\beta$ Pictoris association. This circumstance suggests that HIP 105441 may be older.\\
\citet{Kiraga12} reported for the primary component a rotation period P = 5.30\,d and a light curve amplitude $\Delta$V = 0.05\,mag.
We derived a luminosity L = 0.32$\pm$0.09\,L$_\odot$, a radius R = 0.80$\pm$0.11\,R$_\odot$, and an inclination  $i$  $=$ 53$\pm$5$^\circ$. \\
The secondary component B was first proposed as a member of the $\beta$ Pictoris association by \citet{Torres06} who 
measured a $v\sin{i}$ = 4.5$\pm$1.2\,km\,s$^{-1}$. Its membership was questioned by \citet{daSilva09} who
considered it as an intruder, owing to its very low Li content (EW$_{\rm Li}$ = 15 m\AA), and its membership was also rejected by \citet{Schlieder10}.
We derived a luminosity L = 0.11$\pm$0.03\,L$_\odot$, a radius R = 0.71$\pm$0.23\,R$_\odot$ for  component B.
Malo et al. (2013) found this system to be a high-probability member of the association. They inferred a statistical distance  d = 32\,pc for  the secondary component TYC\,9114-1267-1, which is in fair agreement with the trigonometric distance d =  30.2\,pc inferred by \citet{vanLeeuwen07}. 
However, the RV measurements in the literature are uncertain and exhibit  large scatter, which suggests
that the primary component may be an unresolved spectroscopic binary \citep{Torres06}.

Both stars were observed at KKO from October 10 until\ December 1, 2014 for a total of 24 nights  and at CASLEO from September 4 until October 20, 2014 for a total of 9 nights. From this data, we derived the rotation period P = 5.50$\pm$0.02\,d for V390 Pav (see online Fig.\,\ref{A63}) and P = 20.5$\pm$1.\,d for 
TYC\,9114\,1267\,1 (see online Fig.\,\ref{A64}). In the latter case, the rotation period is however
in disagreement with the projected rotational velocity and stellar radius; the expected period should be  P $<$ 8\,d.
The disagreement may derive from distance. In this case the star should be more distant than d = 32\,pc and, therefore, have a larger radius. However,
 both stars seem to have the same proper motions and are likely be physically bound at the trigonometric distance measured from V390 Pav.\\

\item \bf \object{TYC 9486 927 1} \rm \\

 Bc;  V$-$K$_s$ = 4.36\,mag; P = 0.54\,d\\
  
TYC\,9486\,927\,1 is a M1V star  with uncertain single/binary nature and a projected rotational velocity  $v\sin{i}$ = 43.5$\pm$1.2\,km\,s$^{-1}$ \citep{Torres06}. 
The membership was proposed by \citet{Torres06}, and then TYC\,9486\,927\,1 was considered  a candidate member by \citet{Malo13} and proposed to be member of the Carina association by \citet{Elliott15}. A recent study by \citet{Deacon16} concludes that this is likely a single star and suggests membership to the $\beta$ Pictoris association.\\
 \citet{Kiraga12} measured a photometric rotation period P = 0.5419\,d and a peak-to-peak light curve amplitude $\Delta$V = 0.19\,mag.\\

\item \bf   \object{2MASS J21374019+0137137}AB \rm    \\

 Bc;  V$-$K$_s$ = 5.476\,mag; P = 0.201\,d\\ 

2MASS\,J21374019+0137137AB is a M5V star \citep{Mochnacki02} at a predicted distance d =  39\,pc \citep{Schlieder12}.
This is a highly active star that exhibits strong X-ray, NUV, and FUV emission, and a fast rotation rate of 
 $v\sin{i}$ = 55\,km\,s$^{-1}$ \citep{Mochnacki02}, $v\sin{i}$ = 45$\pm$5\,km\,s$^{-1}$ \citep{Schlieder12},
and $v\sin{i}$ = 66$\pm$9\,km\,s$^{-1}$ \citep{Malo14a}.
The available RV measurements show evident variability that is, however, attributed to uncertainties arising from fast rotation.
We retrieved one spectrum in the 3800-9000\,\AA\,\, region from the LAMOST spectroscopic survey archive. From our analysis, we inferred a M7V spectral type and detected the H$\alpha$ line in emission with EW = $-$10.1$\pm$0.6\,\AA\,\, \citep{Zhang16}.\\
It was discovered by \citet{Bowler15} to be a close visual binary with components separated by $\rho$ = 0.44$^{\prime\prime}$ ($\sim$17\,AU), magnitude difference $\Delta$K$_s$ = 0.8\,mag, and orbital period of 40 yr.
This star was proposed by \citet{Schlieder12}  and by \citet{Malo14b} as a likely member of the $\beta$ Pictoris association.\\
\citet{Kiraga12} measured a rotation period P = 0.213086\,d based on the ASAS photometry. 
This star was also observed by NSVS (ID 14441065) and our LS analysis revealed a period P = 0.2037\,d 
(P = 0.1932\,d from CLEAN periodogram) (see online Fig.\,\ref{A65}). 
We also observed this target at CASLEO from September 4 untill October 20, 2014 for a total of nine nights. Our LS analysis revealed a period P = 0.2015$\pm$0005\,d in agreement with ASAS and NSVS determinations,  and an amplitude $\Delta$R = 0.13\,mag (see online Fig.\,\ref{A66}).\\

\item \bf  \object{2MASS J21412662+2043107}  \rm \\

Bc?;   V$-$K$_s$ = 4.891\,mag; P = 0.899\,d\\ 

2MASS\,J21412662+2043107 is a M3V star \citep{Alonso-Floriano15}. It was suggested to be a high-probability member of the $\beta$ Pictoris association by \citet{Schlieder12} at a kinematic distance d = 52.7\,pc. We retrieved one spectrum in the 3800-9000\,\AA\,\, region from the LAMOST spectroscopic survey archive.  We inferred a M3V spectral type and detected the H$\alpha$ line in emission with EW = $-$1.35$\pm$0.02\,\AA\ from our analysis.\\
No multiple RV measurements are available to assess its single/binary nature.\\
Our LS and CLEAN analyses of SuperWASP data (ID 1SWASP J214126.63+204311.0) allowed us to measure with very high confidence
level a photometric rotation period P = 0.899$\pm$0.001\,d with an amplitude of the light curve $\Delta$V = 0.03\,mag, and a second period P = 8.94$\pm$0.02\,d with lower power, but it is still highly significant (see online Fig.\,\ref{A67}). The second period is close to but different than the 1-d beat periods. In fact, the CLEAN periodogram, which effectively removes such aliases, also exhibits this second period. Then, we may suspect that this star is an unresolved close binary and the detected rotation periods refer to the two components. 
No projected rotational velocity has been measured to infer the inclination of the rotation axis. \\

\item \bf \object{TYC 2211 1309 1} \rm \\

SB;  V$-$K$_s$ = 3.666\,mag; P = 1.109\,d\\ 

TYC\,2211\,1309\,1  is a M0Ve  star  at a predicted distance d = 45.6\,pc \citep{Lepine09}. \citet{Lepine09} found this star to be magnetically active (the H$\alpha$ line was found in emission) and measured a projected rotational velocity $v\sin{i}$ = 30$\pm$2\,km\,s$^{-1}$.
This star is associated with the X-ray source 1RXS\,J220042.0+271520 and it was identified by Fuhrmeister \& Schmitt (2003) as a variable X-ray source with flaring behavior. \citet{Lepine09} identified this star as a highly probable member of the $\beta$ Pictoris association. However, \citet{McCarthy12} found no evidence of lithium, suggesting it is much older. \citet{Brandt14} found no optical companions and assigned a lower probability membership (50\%) to it because of the mentioned lack of lithium.
We retrieved one spectrum in the 3800-9000\,\AA\,\, region from the LAMOST spectroscopic survey archive. From our analysis, we inferred a M0V spectral type and detected the H$\alpha$ line in emission with EW = $-$1.52$\pm$0.08\,\AA.\\
In the literature, we found two values of RV as follows: $-$13.3$\pm$2.4\,km\,s$^{-1}$ from \citet{Lepine09}, $-$0.3$\pm$0.5\,km\,s$^{-1}$ from \citet{Malo14a}, and a new value RV = 27.2\,km\,s$^{-1}$  from the LAMOST spectrum, which are significantly different, indicating this star may be a binary system.\\
\citet{Norton07} find a rotation period P = 0.5235\,d based on the first SuperWASP season.  A similar period P = 0.476\,d is reported by \citet{Messina10}  based on the unique ASAS 2002  observation season. \citet{Messina11} analyzed all three SuperWASP seasons and reported a rotation period P = 0.5229\,d. However, we note that in all periodograms there is a peak at P = 1.109\,d that has the same (and in a few seasons a larger) power as the quoted period P = 0.52\,d.\\
Since we do not know the real single/binary nature, we can only make an estimate of the luminosity L = 0.13$\pm$0.04\,L$_\odot$ and a radius R = 0.78$\pm$0.07\,R$_\odot$. In the case of P = 0.52\,d, we infer an inclination $i$ = 23$\pm$4$^\circ$, whereas, in the case of P = 1.10\,d, we infer an inclination $i$ = 57$\pm$5$^\circ$. Owing to the very large amplitude of the light curve ($\Delta$V = 0.10\,mag), the longer period (and higher inclination) seems to be more likely. In this case, the star may have two major active regions at opposite
hemispheres that modulate the observed flux with half the rotation period.\\

\item \bf \object{TYC 9340 0437 1} \rm (CPD$-$72 2713) \\

 S;  V$-$K$_s$ = 3.706\,mag; P = 4.46\,d\\

TYC\,9340\,0437\,1 is classified as K7Ve by \citet{Torres06} at a distance d = 36\,pc. \citet{Biller13} and \citet{Elliott14} did not find evidence for binarity
from RV measurements. \citet{Torres06}  and \citet{Weise10} reported  $v\sin{i}$ $=$ 7.5\,km\,s$^{-1}$ and $v\sin{i}$ = 6.6\,km\,s$^{-1}$, respectively. 
 The membership of the  $\beta$ Pictoris association was proposed by \citet{Torres06}, \citet{Lepine09}, \citet{Malo14b}, and \citet{Brandt14}. \\ 
\citet{Messina10}, using LS and CLEAN periodograms, found the rotation period P = 4.46\,d in eight out of nine seasons of ASAS photometry. The same period was subsequently found by \citet{Kiraga12}.\\
We derived a luminosity L = 0.17$\pm$0.05\,L$_\odot$, a radius R = 0.91$\pm$0.08\,R$_\odot$, and an inclination $i$ = 45$\pm$4$^\circ$.\\

\item \bf \object{HIP 112312} (WW Psa) \rm \\

Bw;  V$-$K$_s$ = 5.168\,mag; P = 2.35\,d\\ 

WW Psa is a M4Ve star with a fainter visual companion TX Psa at about $\rho$ = 33$^{\prime\prime}$ ($\sim$780\,AU).
The membership of the $\beta$ Pictoris association was first proposed by \citet{Song03}, and later by \citet{Zuckerman04}  and \citet{Torres06}.
\citet{Torres06} reported $v\sin{i}$= 12.1\,km\,s$^{-1}$;   \citet{Jayawardhana06} reported $v\sin{i}$ = 14$\pm$1.73\,km\,s$^{-1}$.\\
Our analysis of ASAS (ASAS J224458-3315.1) and SuperWASP (1SWASP J224457.83-331500.6) data shows P = 2.35\,d to be the rotation period. P = 2.358\,d is reported by \citet{Kiraga12}, and P = 2.3546\,d is listed in ACVS.
We observed WW Psa at Siding Spring Observatory (Australia), at CASLEO, and at the PEST Observatory,  and measured a rotation period P = 2.37$\pm$0.01\,d. The results of our analyses are reported by Messina et al. \citet{2016d}.
Adopting a  distance d = 23.6\,pc (from Hipparcos), we inferred a luminosity L = 0.059$\pm$0.005\,L$_\odot$, a stellar radius R = 0.82$\pm$0.08\,R$_\odot$, and an inclination of the stellar rotation axis  $i$ $\simeq$ 43$\pm$7$^\circ$. \\

 \bf \object{TX Psa} \rm   \\
 
 Bw;  V$-$K$_s$ = 5.567\,mag; P = 1.080\,d\\ 
   
TX Psa is a M5Ve single star that is the fainter optical companion of WW Psa \citep{Torres08}. 
The membership of the $\beta$ Pictoris association was first proposed by \citet{Song03}, and later by \citet{Zuckerman04}  and \citet{Torres06}.
\citet{Torres06} report  v$\sin$i = 16.8\,km\,s$^{-1}$; \citet{Jayawardhana06} report  $v\sin{i}$ = 24.3$\pm$4.93\,km\,s$^{-1}$.
Unfortunately, both ASAS and SuperWASP aperture photometry are often contaminated by the flux coming from the brighter WW Psa. 
We observed it at Siding Spring Observatory (Australia), at CASLEO,  and at the PEST Observatory, and measured a rotation period P = 1.086$\pm$0.005\,d. The results of our analyses are reported by Messina et al. \citet{2016d}.
We assume for TX Psa the same distance d = 23.6\,pc as for WW Psa. We inferred a luminosity L = 0.027$\pm$0.006\,L$_\odot$ and a stellar radius R = 0.59$\pm$0.09\,R$_\odot$.  We inferred the inclination of the stellar rotation axis  $i$ $\simeq$ 48$\pm$7$^\circ$ from these values and stellar rotation period.\\
Therefore, the two components of this physical pair are found to have similar inclinations of their rotation axes.\\

\item \bf \object{2MASS J22571130+3639451} \rm \\

S;   (V$-$K$_s$) = 3.862\,mag; P = 1.22\,d\\ 

2MASS\,J22571130+3639451 is a M3V star and proposed as likely member of the $\beta$ Pictoris association at a distance d = 68.7\,pc \citep{Schlieder12}.
They measured a projected rotational velocity   $v\sin{i}$ = 20$\pm$2\,km\,s$^{-1}$ and inferred a M3
spectral type, which was later confirmed by \citet{Frith13}. However, they found a discrepancy between the
observed and predicted RV, that led them to led them to change their minds about this membership.\\
We retrieved one spectrum in the 3800-9000\,\AA\,\, region from the LAMOST spectroscopic survey archive. We detected the H$\alpha$ line in emission with EW = $-$2.13$\pm$0.08\,\AA\ from our analysis.\\
We retrieved a time series from the SuperWASP archive (1SWASP J225712.91+363928.2). However, the photometry is contaminated by a nearby
star ($\Delta$J = 1.8\,mag fainter) whose dilution effect determines  a smaller amplitude of photometric variability. We determined with both LS and CLEAN periodograms 
a photometric rotation period P = 1.22$\pm$0.01\,d with very high confidence level and a light curve amplitude $\Delta$V = 0.04\,mag (see online Fig.\,\ref{A68}).
Adopting a distance d = 68.7\,pc, we derived a luminosity L = 0.12$\pm$0.03\,L$_\odot$, a radius R = 0.80$\pm$0.08\,R$_\odot$, and an inclination $i$  $\sim$37$\pm$7$^\circ$.\\

\item \bf \object{TYC 5832 0666 1} \rm (BD$-$13 6424 )\\

S;   V$-$K$_s$ = 3.971\,mag; P = 5.68\,d\\ 

TYC\,5832\,0666\,1 is classified as  M0V by \citet{Riaz06}, M0.8V by \citet{Shkolnik09}, and M0Ve by \citet{Torres06}. \citet{Torres06}, \citet{Lepine09},  \citet{Weise10}, and \citet{Malo14a} measure similar values of projected rotational velocities, $v\sin{i}$ = 8.8\,km\,s$^{-1}$, $v\sin{i}$ = 7$\pm$2\,km\,s$^{-1}$,  
$v\sin{i}$ = 9.6 km\,s$^{-1}$, and  $v\sin{i}$ $<$ 9\,km\,s$^{-1}$, respectively. \citet{Riaz06} measured a spectroscopic distance d = 25\,pc, whereas \citet{Lepine09} a 
kinematic distance d = 27.3$\pm$0.4\,pc. No evidence for binarity is found either by the Lucky Imaging survey \citep{Bergfors10} nor by \citet{Delorme12} or by \citet{Brandt14} or from RV measurements (\citealt{Malo13}; \citealt{Elliott14}; \citealt{Torres06}). All these studies indicate that it is a single star. 
The membership of the $\beta$ Pictoris association was proposed by \citet{Torres06}, \citet{Shkolnik09}, \citet{Lepine09}, \citet{Malo14b}, \citet{Brandt14}, and \citet{Elliott15}.\\
\citet{Messina10} found the rotation period  P = 5.68\,d in five out of eight ASAS seasons. Similar periods P = 5.667\,d and P = 5.6945\,d are reported by \citet{Kiraga12} and in the ACVS, respectively.
We derived a luminosity L = 0.13$\pm$0.04\,L$_\odot$ and a radius R = 0.86$\pm$0.8\,R$_\odot$, and an inclination $i$  = 90$\pm$15$^\circ$ considering P = 5.68\,d.\\

\item \bf  \object{2MASS J23500639+2659519} \rm \\

 Bc?;  V$-$K$_s$ = 4.957\,mag; P = 0.287\,d\\ 

2MASS\,J23500639+2659519 is classified as  M3.5V by \citet{Riaz06},  \citet{Malo14a}, and  by \citet{Shkolnik09}. 
Whereas \citet{Riaz06} reported a spectroscopic distance d = 47$\pm$3\,pc, \citet{Reid07} and \citet{Malo13} determined similar values of photometric distances d = 26.6$\pm$4\,pc, and d = 24$\pm$2\,pc, respectively.
On the other hand, Shkolnik et al. (2012) inferred a possible membership to the \object{Castor Moving Group}
with match quality BAA. 
We have included this target in the present study since
\citet{Malo13} reported a probability that this target is member of the $\beta$ Pictoris association of  92.8\% when the star is assumed to be a binary, and 61.6\% when it is assumed to be a single star. 
Subsequently, \citet{Malo14a} redetermined the probability that it is a member of the $\beta$ Pictoris association to 82.4\%, which increases to 96.8\%  including radial velocity information. \citet{Klutsch14} found it to be a likely member of the 200 Myr Castor MG. We obtained the first RV  measurement for this target, RV = 0.4$\pm$0.6\,km\,s$^{-1}$, and $v\sin{i}$ = 36\,km\,s$^{-1}$ from ESPaDOnS spectra (Malo et al. in preparation).\\
This target was observed by SuperWASP (1SWASP J235006.40+265952.1) in two seasons (2004 and 2006). We selected only the higher precision data 
($\sigma <$ 0.05\,mag) and searched for periodicities in both seasons with the LS and CLEAN periodograms. With both methods, we found a very significant (FAP $<$ 0.1\%) power peak at the period  P = 0.259$\pm$0.002\,d, which we consider the stellar rotation period and an amplitude of the light curve up to $\Delta$V = 0.05\,mag (see online Fig.\,\ref{A69}). 
The average  photometric precision is $\sigma$ = 0.040\,mag.
This target was also observed by MEarth \citep{Berta12} with a higher photometric precision ($\sigma$ = 0.003\,mag). From this data  we found a similar period P = 0.287$\pm$0.005\,d (see online Fig.\,\ref{A70}).\\

\item \bf \object{2MASS J23512227+2344207}  \rm (G\,68-46) \\

 S;  V$-$K$_s$ = 5.285\,mag; P = 3.20\,d\\ 

2MASS\,J23512227+2344207 is classified as  M4V by \citet{Riaz06} and \citet{Shkolnik09} who 
reported similar  spectroscopic  and  photometric  distances, d = 15$\pm$3\,pc and d = 14\,pc, respectively.
A more recent  spectroscopic distance d = 18\,pc was measured by \citet{Newton14}, whereas \citet{Malo14a} inferred a statistical distance d = 16\,pc.
\citet{Shkolnik12} did not detect RV variations and inferred a possible membership of the Chamaeleon-Near Moving Group (MG) with match quality BBB and an estimated age in the range 35--300\,Myr.  
Malo et al. (\citeyear{Malo13}, \citeyear{Malo14a}, \citeyear{Malo14b}) found a probability that the target is member of the $\beta$ Pictoris association of 81.6\%, which increases to 97.8\%  when the radial velocity information is included. 
On the other hand, the most recent investigation by \citet{Bowler15} does not find any match with any young MG, suggesting an age in the range 35--300 Myr, in agreement with \citet{Shkolnik09}. However, \citet{Klutsch14} found this star to be  likely a member of the 200 Myr Castor MG. The membership of the $\beta$ Pictoris association has been recently rejected by \citet{Binks16}.
\\
This target was included in the following three photometric surveys:  SuperWASP (ID\,1SWASP J235122.30+234421.2), the NSVS  (ID\,6277459), and the MEarth survey (ID\,LSPMJ2351+2344\_tel08\_2011-2013).
In all three time series our LS and CLEAN periodograms found a highly 
significant (FAP $<$ 0.1\%) power peak at the period  P = 3.2079$\pm$0.0039\,d, which we consider the stellar 
rotation period, and we found light curve amplitudes up to $\Delta$V = 0.06\,mag (see online Fig.\,\ref{A71}--\ref{A72}). 
The LS periodogram of the MEarth time series, which is the most accurate, shows two other power peaks  at P = 1.4\,d and P = 0.7\,d. These peaks are alias of the rotation period and arise from the sampling interval of about 1 day imposed by the rotation of the Earth and the fixed longitude of the observation site. These peaks  are effectively removed by the CLEAN algorithm (see online Fig.\,\ref{A73}).
We obtained one RV measurement, RV = $-$2.0$\pm$0.3\,km\,s$^{-1}$, which is comparable with the literature values, and $v\sin{i}$ = 4\,km\,s$^{-1}$ from ESPaDOnS spectra (Malo et al., in preparation). \\
Adopting a distance d = 18\,pc, we inferred a luminosity L = 0.006$\pm$0.001\,L$_\odot$, a stellar radius R = 0.25$\pm$0.08\,R$_\odot$, and an inclination $i$  = 83$\pm$15$^\circ$.\\

\end{itemize}
\end{appendix}

\end{document}